\documentclass[12pt,superscriptaddress,tightenlines,nofootinbib,preprint]{revtex4}
\pdfoutput=1

\usepackage[a4paper,text={16.8cm,22.4cm}]{geometry}

\usepackage{graphicx,color}  %
\usepackage{bm}  %
\usepackage{amssymb, graphics, amsmath}
\usepackage{cancel}

\def\slashchar#1{\setbox0=\hbox{$#1$}           
  \dimen0=\wd0                                    
  \setbox1=\hbox{/} \dimen1=\wd1                  
  \ifdim\dimen0>\dimen1                           
    \rlap{\hbox to \dimen0{\hfil/\hfil}}            
    #1                                             
  \else                                          
    \rlap{\hbox to \dimen1{\hfil$#1$\hfil}}        
    /                                           
 \fi}                                           %

\newcommand{\Tr}{{\textrm{Tr}}}
 
\newcommand{\slashT}{\cancel{T}}
\newcommand{\CP}{C\hspace{-.5mm}P}

\newcommand{\tb}{\bar \theta}
\newcommand{\mpi}{m_\pi}

\newcommand{\boldtau}{\mbox{\boldmath $\tau$}}
\newcommand{\boldpi}{\mbox{\boldmath $\pi$}}
\newcommand{\boldSigma}{\mbox{\boldmath $\Sigma$}}

\begin{document}
\preprint{LA-UR-15-23860}
\preprint{NT@WM-15-07}
\preprint{JLAB-THY-15-2061}

\title{\textbf{Baryon mass splittings and strong $\CP$ violation\\
 in $SU(3)$ Chiral Perturbation Theory}
 \vspace*{0.2cm}}

\author{J. de Vries \vspace*{0.2cm}}
\affiliation{Institute for Advanced Simulation, Institut f\"ur Kernphysik, 
and J\"ulich Center for Hadron Physics, Forschungszentrum J\"ulich,
D-52425 J\"ulich, Germany
\vspace*{0.4cm}}

\author{E. Mereghetti}
\affiliation{Theoretical Division, Los Alamos National Laboratory,
Los Alamos, NM 87545, USA
\vspace*{0.4cm}}

\author{A. Walker-Loud}
\affiliation{Department of Physics, The College of William \& Mary, 
Williamsburg, VA 23187-8795, USA \vspace*{0.4cm}}
\affiliation{Thomas Jefferson National Accelerator Facility, 
12000 Jefferson Avenue, Newport News, VA 23606, USA
\vspace*{0.4cm}
}

\date{\today}

\begin{abstract}
We study $SU(3)$ flavor-breaking corrections to the relation between the octet baryon masses and the nucleon-meson $\CP$-violating interactions induced by the QCD $\bar\theta$ term.
We work within the framework of  $SU(3)$ chiral perturbation theory and work through next-to-next-to-leading order in the $SU(3)$ chiral expansion, which is $\mathcal{O}(m_q^2)$.
At lowest order, the $\CP$-odd couplings induced by the QCD $\bar\theta$ term are determined by mass splittings of the baryon octet, the classic result of Crewther et al.
We show that for each isospin-invariant $\CP$-violating nucleon-meson interaction there exists one relation which is respected by loop corrections up to the order we work, while other leading-order relations are violated. With these relations we extract a precise value of the pion-nucleon coupling $\bar g_0$ by using recent lattice QCD evaluations of the proton-neutron mass splitting. 
In addition, we derive semi-precise values for $\CP$-violating coupling constants between heavier mesons and nucleons with  $\sim 30\%$ uncertainty and discuss their phenomenological impact on electric dipole moments of nucleons and nuclei.

\end{abstract}

\maketitle

\section{Introduction}

Violation of time reversal ($T$), or, equivalently, violation of the product of charge conjugation and parity ($\CP$)
is one of the ingredients \cite{Sakharov:1967dj} needed to explain the matter-antimatter asymmetry of the visible universe.
The Standard Model (SM) of particle physics contains two sources of $\CP$ violation, the phase of the CKM matrix and the QCD $\bar\theta$ term.
The phase of the CKM matrix explains the observed $\CP$ violation in $K$ and $B$ decays \cite{Agashe:2014kda}, but appears to be too small 
for electroweak baryogenesis (see \cite{Morrissey:2012db} and references therein). The QCD $\bar\theta$ term is severely constrained by the non-observation of the neutron electric dipole moment (EDM).
The current limit on the neutron EDM, $|d_n| < 2.9 \cdot 10^{-13} \, e$ fm \cite{Baker:2006ts}, bounds  $\bar\theta$ to be small,
$\bar\theta < 10^{-10}$, the well known strong $\CP$ problem.

The viability of attractive, low-scale baryogenesis mechanisms such as electroweak baryogenesis thus requires new sources of $\CP$ violation.
With the assumption that new physics is heavier than the electroweak scale and that there are no new light degrees of freedom, 
new sources of $\CP$ violation appear as higher-dimensional operators in the SM Lagrangian, suppressed by powers of the scale $M_{\slashT}$ at which $T$ violation arises.
These operators involve SM particles and respect the SM gauge symmetry. In the quark sector, they are at least dimension six,
and are suppressed by two powers of $M_{\slashT}$ \cite{Buchmuller:1985jz,Grzadkowski:2010es}. 
EDMs of the nucleon, light nuclei, atomic, and molecular systems are extremely sensitive probes of such flavor-diagonal $T$-violating ($\slashT$) 
operators, for reviews see Refs.~\cite{Pospelov:2005pr, Engel:2013lsa, Roberts:2014bka}. The current generation of experiments probes scales of TeV (or higher), and provides powerful constraints on new physics models,  
complementary to direct searches of new physics at colliders.
Furthermore, a vigorous experimental program is under way \cite{Kumar:2013qya} to improve current bounds on the neutron EDM  \cite{Baker:2006ts}  by two orders of magnitudes, 
to measure EDMs of light nuclei at the same or even higher level of accuracy \cite{Pretz:2013us, Anastassopoulos:2015ura}, and to improve the bounds on EDMs of diamagnetic atoms, like $^{199}$Hg \cite{hgbound}, $^{129}$Xe \cite{Xebound} and $^{225}$Ra \cite{Parker:2015yka}.

The extraction of robust information on possible new sources of $\CP$ violation from  EDM measurements involves dynamics 
on a large variety of scales, from the new physics scale $M_{\slashT}$ to the electroweak (EW) and QCD scales, down to the atomic scale. The step to hadronic and nuclear scales involves nonperturbative strong matrix elements which are often poorly known  leading to large uncertainties \cite{Engel:2013lsa}. In recent years, lattice QCD has made progress in computing the nucleon EDM induced by the QCD $\bar\theta$ term \cite{Shintani:2014zra,Shindler:2014oha,Guo:2015tla}, while the study of the nucleon EDM generated by higher-dimensional operators is still in its infancy (an exception is the quark EDM \cite{Bhattacharya:2015esa}). Another important ingredient for the study of EDMs are $\slashT$ pion-nucleon couplings, which determine the leading non-analytic  contribution to the nucleon EDM
\cite{Crewther:1979pi}. In addition, they generate $\slashT$ long-range nucleon-nucleon potentials, contributing to EDMs of light nuclei \cite{deVries:2011an,Bsaisou:2014zwa},
and diamagnetic atoms \cite{Pospelov:2005pr, Engel:2013lsa}. 

The direct calculation of  $\slashT$ pion-nucleon couplings on the lattice, both from $\bar\theta$ and dimension-six operators, is difficult. Some information can be gained by the study of the momentum dependence of the electric dipole form factor (EDFF) \cite{Ottnad:2009jw, Mereghetti:2010kp}, but the most recent lattice calculations are performed at too large momenta for a reliable extraction \cite{Shintani:2014zra,Guo:2015tla,Mereghetti:2015rra}. 
Fortunately, in some cases other methods exist to extract the values of the pion-nucleon interactions. 
For chiral-symmetry-breaking sources, like the QCD $\bar\theta$ term, or the light-quark chromo-electric dipole moments (qCEDMs),
the pion-nucleon couplings are intimately related 
to $\CP$-even, chiral-symmetry-breaking effects. In the case of the QCD $\bar\theta$ term this was realized in  Ref. \cite{Crewther:1979pi},
that expressed the isoscalar $\slashT$ coupling $\bar g_0$ in terms of mass splittings of the octet baryons. In particular, in $SU(2)$ Chiral Perturbation Theory ($\chi$PT) it
is possible to relate $\bar g_0$ to the neutron-proton mass difference induced by the quark mass difference $m_d - m_u$, which we denote by $\delta m_N$. All the information on non-perturbative dynamics entering 
$\bar g_0$ can thus be extracted by computing a property of the baryon spectrum, the nucleon mass splitting, a task for which lattice QCD is particularly well suited.  
Indeed, existing calculations of the nucleon mass splitting allow a determination of $\bar g_0$ with $10\%$ accuracy \cite{Walker-Loud:2014iea,Borsanyi:2014jba}, if one considers only lattice uncertainties.
Similar relations between $\slashT$ pion-nucleon couplings and modifications to the meson and baryon spectrum
can be derived in the case of the qCEDM operators~\cite{Pospelov:2005pr,deVries:2012ab}, and provide a viable route to improve the determination of couplings that, at the moment, are only known at the order-of-magnitude level \cite{Pospelov:2001ys, Pospelov:2005pr, Engel:2013lsa}.

The relations between $\slashT$ couplings and baryon masses strictly hold at leading order (LO) in $\chi$PT.
Furthermore, if one considers the strange quark as light, and extends the chiral group to $SU(3) \times SU(3)$, more LO relations can be written, \textit{e.g.}  $\bar g_0$ can be expressed in terms of the mass difference of the $\Xi$ and $\Sigma$ baryons. Using the two LO relations leads to values of $\bar g_0$ that differ by about $50\%$, well beyond the lattice QCD uncertainty. 
Is this large difference due to an inherent uncertainty in the relation between the spectrum and the $\slashT$ couplings? Does this imply that the relations to the baryon spectrum can only be used for order-of-magnitude estimates of the $\slashT$ couplings? 

In this paper we investigate these questions and seek to quantify the $SU(3)$ flavor-breaking corrections between the baryon masses and $\slashT$ couplings induced by the QCD $\bar\theta$ term.
We work in the framework of $SU(3)$ heavy baryon $\chi$PT~\cite{Jenkins:1990jv,Bernard:1992qa} and compute higher-order corrections in the chiral expansion.
We show that most LO relations are badly violated, already at next-to-leading order (NLO) and cannot be used for reliable extractions of the meson-nucleon couplings. 
However, for all isospin-invariant $\slashT$ couplings there exists exactly one relation that is preserved by all loop corrections up to next-to-next leading order (N${}^2$LO).
By using the relations that are not violated by $SU(3)$ flavor breaking, a precise extraction of the couplings is possible irrespective of the convergence of $SU(3)$ $\chi$PT through this order.
In the case of $\bar g_0$, the preserved relation is to $\delta m_N$, while the relation to the mass difference of the $\Xi$ and $\Sigma$ baryons receives large NLO and N${}^2$LO corrections, 
which show little sign of convergence. 
Expressing  $\bar g_0$  in terms of the $\Xi$ and $\Sigma$ masses overestimates the coupling by about $50\%$, well outside the uncertainty which is determined with $\delta m_N$.

For isospin-breaking couplings, such as the isovector pion-nucleon coupling $\bar g_1$,  we were not able to identify any robust relation that does not receive large violations already at NLO. We are forced to conclude that $SU(3)$ heavy baryon $\chi$PT does not provide a reliable method to extract this important coupling from known matrix elements. 

In this work we only focus on $\CP$ violation from the SM QCD $\tb$ term, leaving higher-dimensional operators arising from possible BSM physics for future work. However, our results are also relevant for scenarios of BSM physics where the strong $\CP$ problem is solved by a Peccei-Quinn mechanism. In this case, an effective $\bar\theta$ term can be induced proportional to any appearing higher-dimensional $\CP$-odd sources \cite{Pospelov:2000bw, Pospelov:2005pr}. Other BSM scenarios involve cases where parity  is assumed to be an exact symmetry at high energies, requiring $\tb=0$ \cite{Mohapatra:1978fy}, while a calculable contribution to $\bar\theta$ is induced at lower energies once parity is spontaneously broken, see for instance Ref.~\cite{Maiezza:2014ala}. In any case, a quantitive understanding of the low-energy consequences of the $\tb$ term is necessary to unravel the underlying source of $\CP$ violation once a nonzero EDM is measured \cite{Dekens:2014jka} and to test scenarios, such as the one in Ref.~\cite{Graham:2015cka}, where a small but nonzero $\tb$ term is expected. 
Our values of the $\CP$-odd pion-nucleon couplings can also be used for more precise limits on axion searches \cite{Stadnik:2013raa,Roberts:2014cga}.

This paper is organized as follows. In Section \ref{Sec2} we review the closely related chiral symmetry breaking and $\slashT$ sectors of the $\chi$PT Lagrangian. 
In Section \ref{relations} we discuss baryon masses and $\slashT$ couplings at tree level, and identify the relations between masses and couplings imposed by $SU(3)$ symmetry.
In Section \ref{NLO} we study NLO corrections to masses and $\slashT$ couplings, and identify which relations are respected by NLO loop corrections.
In Section \ref{n2lo1} we discuss in detail N${}^2$LO corrections to the nucleon mass splitting, including, for the first time, decuplet corrections. 
The expressions of N${}^2$LO corrections to the mass splittings of the $\Xi$ and $\Sigma$ baryons and to the octet baryon average masses are relegated to Appendices \ref{AppA} and \ref{AppB}.
In Section \ref{n2lo2} we examine N${}^2$LO corrections to $\bar g_0$, and show that all the loops at this order are related to contributions to the nucleon mass splitting. In Sections  \ref{n2log1} and \ref{n2lo3}
we discuss the remaining $\slashT$ nucleon couplings.
In Section \ref{Numerics} we use the conserved relations to determined the value of the $\slashT$ couplings induced by the QCD $\bar\theta$ term,
and discuss the impact of our analysis on the nucleon electric dipole form factor and on the $\slashT$ nucleon-nucleon potential. We conclude in Section \ref{conclusion}.

\section{QCD and EFT Lagrangian}\label{Sec2}

At the QCD scale, $\mu \sim 1$ GeV, heavy gauge bosons, the Higgs and the heavy quarks can be integrated out, and the SM Lagrangian involves gluons, photons, and three flavors of quarks. 
\begin{equation}\label{eq:1.1}
\mathcal L_{\textrm{QCD}} = - \frac{1}{4} F_{\mu \nu} F^{\mu \nu}- \frac{1}{4} G^{a}_{\mu \nu} G^{a\, \mu \nu} + \bar q i \slashchar D q  - e^{i \rho} \bar q_L \mathcal M q_R - e^{-i \rho} \bar q_R \mathcal M q_L 
 - \theta \frac{g_s^2}{64 \pi^2} \varepsilon^{\mu \nu \alpha \beta} G^a_{\mu\nu} G^a_{\alpha \beta}\,\,\,.
\end{equation}
$q$ is a triplet of quark fields $q = (u, d, s)$, $F_{\mu\nu}$ and $G^a_{\mu \nu}$ are the photon and gluon field strengths, and $D_{\mu}$ is the $SU(3)_c \times U(1)_{\textrm{em}}$
covariant derivative. 
The first three terms in Eq. \eqref{eq:1.1} are the photon, gluon, and quark kinetic terms. Without loss of generality the quark mass matrix can be expressed in terms of a real, diagonal matrix 
$\mathcal M = \textrm{diag}(m_u,m_d,m_s)$, and a common phase $\rho$. The last term in Eq. \eqref{eq:1.1}
is the QCD $\bar\theta$ term. Despite being a total derivative, it contributes to physical observables through extended field configurations, the instantons \cite{'tHooft:1976fv}.
The two $\CP$-violating parameters in Eq. \eqref{eq:1.1}, $\theta$ and the phase $\rho$, are not independent, and $\CP$ violation is proportional to the combination $\bar\theta = \theta - n_f \rho$,
where $n_f = 3$ is the number of flavors of light quarks. This can be explicitly seen by  performing an anomalous $U_A(1)$ axial rotation. With an appropriate choice of phase, the $\theta$ term
can be completely eliminated, in favor of a complex mass term.
The residual freedom of performing non-anomalous $SU(3)_A$ axial rotations can be used to align the vacuum in presence of $\CP$ violation to the original vacuum of the theory.
If the complex mass term is the only $\slashT$ operator in the theory, vacuum alignment is accomplished by  making the complex mass term isoscalar  \cite{Baluni:1978rf}. 
At the level of the meson Lagrangian, the condition of vacuum alignment is equivalent to setting the leading-order coupling of the pion and $\eta$ meson to the vacuum to zero.

After vacuum alignment, the QCD Lagrangian in the presence of the $\bar\theta$ term reads
\begin{equation}\label{eq:1.2}
\mathcal L_{\textrm{QCD}} = - \frac{1}{4} F_{\mu \nu} F^{\mu \nu} - \frac{1}{4} G^{a}_{\mu \nu} G^{a\, \mu \nu} + \bar q i \slashchar D q  - \bar q \left(\mathcal M - i \gamma_5 m_* \bar\theta \right)q\,\,\,,
\end{equation} 
where we denote
\begin{equation}\label{eq:1.3}
m_{*} = \frac{m_u m_d m_s}{m_s (m_u + m_d) + m_u m_d} = \frac{\bar m (1-\varepsilon^2)}{2  + \frac{\bar m}{m_s} (1 - \varepsilon^2)}\,\,\,,
\end{equation}
with $2\bar m = m_u + m_d$ and $\varepsilon = (m_d - m_u)/(m_d+m_u)$.
When providing numerical results, we take the values of these quantities from the most recent lattice average by FLAG (quoted in $\overline{\textrm{MS}}$ scheme at $\mu=2$~GeV when relevant)~\cite{Aoki:2013ldr}:
\begin{align}\label{eq:mq_flag}
&\bar{m} = 3.42 \pm 0.09 \textrm{ MeV}\,\,\,,&
&\frac{m_s}{\bar{m}} = 27.46 \pm 0.44\,\,\,,&
&\varepsilon = 0.37 \pm 0.03\, \,\,\,.&
\end{align}

The QCD Lagrangian is approximately invariant under the global chiral group $SU(3)_L \times SU(3)_R$. 
Chiral symmetry  and its spontaneous breaking to the vector subgroup $SU(3)_V$  lead to the emergence of an octet of pseudo Nambu-Goldstone (pNG) bosons, the pion, kaon, and $\eta$ mesons,
whose interactions are dictated by chiral symmetry.
The quark mass and the QCD $\bar\theta$ term break chiral symmetry explicitly.
Chiral invariance can be formally recovered by assigning the mass term the transformation properties 
\begin{equation}\label{eq:1.4}
\mathcal M + i m_* \bar\theta \rightarrow R (\mathcal M + i m_* \bar\theta ) L^{\dagger}\,\,\,, \qquad
\mathcal M - i m_* \bar\theta \rightarrow L (\mathcal M - i m_* \bar\theta ) R^{\dagger}\,\,\,,
\end{equation}
under a $SU(3)_L \times SU(3)_R$ rotation. The QCD $\bar\theta$ term thus induces $\slashT$ interactions  between pNG bosons, and pNG bosons and matter fields,
that can be constructed using the same spurion fields employed in the construction of the meson and baryon mass terms.
We refer to Refs. \cite{Crewther:1979pi,deVries:2012ab,Gasser:1984gg,Borasoy:2000pq,Mereghetti:2010tp,Bsaisou:2014oka} for more details.
In the next Sections we give the meson and baryon $\chi$PT Lagrangians relevant to the calculation of $\slashT$ baryon-pNG couplings at N${}^2$LO.

\subsection{Meson Sector}\label{Sec2Meson}
The constraints imposed by chiral symmetry and its spontaneous and explicit breaking on the interactions of pNG bosons can be formulated in an effective Lagrangian,
$\chi$PT \cite{Weinberg:1968de,Coleman:1969sm,Callan:1969sn,  Gasser:1983yg, Gasser:1984gg}, whose construction is well known. We adopt here the notation of Ref. \cite{Gasser:1984gg}.
In the absence of explicit chiral symmetry breaking, the interactions of pNG bosons are proportional to their  momentum, $q$, which guarantees that low-momentum observables can be computed as a perturbative
expansion in $q/\Lambda_\chi$, where $\Lambda_\chi$ is a typical hadronic scale, $\Lambda_\chi \sim 1$ GeV.
The quark masses explicitly break chiral symmetry, giving masses to the pNG bosons and inducing non-derivative couplings. However, the breaking is small and can be incorporated in the expansion by counting each insertion of the quark mass as $q^2$.

We assign each term in the $\chi$PT Lagrangian an integer index, that counts the powers of momentum or of the quark mass. 
The LO meson Lagrangian contains two derivatives or one light quark mass insertion, and is given by  
\begin{equation}\label{eq:2.1}
\mathcal L^{(2)}_{\pi} = \frac{F^2_{0}}{4} \textrm{Tr} \left( \partial_{\mu} U \partial^{\mu} U^{\dagger} \right)  + \frac{F^2_{0}}{4}\textrm{Tr} \left[ U^{\dagger} \chi  + U \chi^{\dagger}\right]\,\,\,,
\end{equation}
where $F_{0}$ is the pion decay constant in the chiral limit.
Beyond LO, $SU(3)$ breaking corrections break the degeneracy of the pion, kaon, and $\eta$ decay constants. We denote by $F_{\pi}$ and $F_{K}$ the empirical pion and kaon decay constant,
$F_{\pi} = 92.2$ MeV and $F_{K} = 113$ MeV  \cite{Agashe:2014kda}. $F_\eta$ can be expressed in terms of $F_K$ and $F_{\pi}$, and we use $F_{\eta} = 1.3 F_{\pi}$ \cite{Gasser:1984gg}.
In Eq. \eqref{eq:2.1} we introduced the unitary matrix 
\begin{equation}\label{eq:2.2}
U(\pi) = u(\pi)^2= \exp \left( \frac{2 i \pi}{F_0}\right)\,\,\,,
\end{equation}
where $\pi$  are the pNG boson fields
\begin{equation}\label{eq:2.3}
\pi =\pi^a t^a_{i j} =\frac{1}{\sqrt{2}} \left( \begin{array}{c c c} 
\frac{\pi_3}{\sqrt{2}} + \frac{\pi_8}{\sqrt{6}}  & \pi^+ 						& K^+ \\
\pi^-						& - \frac{\pi_3}{\sqrt{2}} + \frac{\pi_8}{\sqrt{6}} 	& K^0 \\
K^-						& \overline{K}^0					& - \frac{2}{\sqrt{6}}\pi_8 
\end{array} \right)\,\,\,.
\end{equation}
$(t^a)_{i j}$ are the generators of $SU(3)$, $a = 1,\ldots,8$ is the octet index and $i,j = 1,\ldots,3$ are indices of the fundamental representation of $SU(3)$.
Under a $SU(3)_L \times SU(3)_R$ transformation, the pNG field has a complicated non-linear transformation, while $U$ transforms simply as $U \rightarrow R U L^{\dagger}$.
The first term in Eq. \eqref{eq:2.1} is chirally invariant.  
The second term, with  $\chi = 2 B ( \mathcal M + i m_* \bar\theta)$, is the realization of the quark mass term which, with the transformation properties in Eq. \eqref{eq:1.4}, is also formally invariant.

Eq. \eqref{eq:2.1} induces the leading contribution to the pion, kaon, and $\eta$ meson masses.
\begin{eqnarray}\label{eq:2.4}
m^2_{\pi^{\pm}} &=&   2 B \bar m\,\,\,, \nonumber \\
m_{\pi^0}^2 &=& 2 B \bar m   - B \frac{\bar m^2 \varepsilon^2}{m_s - \bar m}\,\,\,, \nonumber \\
m_{\eta}^2 &=& \frac{2}{3} B (2 m_s + \bar m)  + B \frac{\bar m^2 \varepsilon^2}{m_s - \bar m}\,\,\,, \nonumber \\
m^2_{K^{\pm}}  &=& B (m_s + \bar m - \bar m \varepsilon)\,\,\,,  \nonumber \\  
m^2_{K^{0}}  &=& B (m_s + \bar m + \bar m \varepsilon)\,\,\,. 
\end{eqnarray}
When working in the isospin limit, we will denote $m_K^2 = B (m_s + \bar m)$.
At the order we are working, we  need the meson masses only at LO, and, for numerical evaluations, we will use the PDG values
$m_{\pi^{\pm}}  = 139.6$ MeV, $m_{K^+} = 493.7$ MeV, $m_{K^0} = 497.6$ MeV, $m_{\eta} = 547.9$ MeV \cite{Agashe:2014kda}. The experimental error on the meson masses is always negligible compared to other uncertainties in the calculations, 
and we can neglect it. 

The relation between the physical $\pi_0$ and $\eta$ and the pNG bosons $\pi_3$ and $\pi_8$ is determined, at LO, by  the $\pi$ -- $\eta$ mixing angle $\phi$,
\begin{eqnarray}\label{eq:2.5}
\pi_0 &=& \cos\phi \, \pi_3 + \sin\phi \, \pi_8 \,\,\,, \nonumber \\
\eta &=& -\sin\phi \, \pi_3 + \cos\phi \, \pi_8 \,\,\,,
\end{eqnarray}
with
\begin{equation}\label{eq:2.6}
\frac{\phi}{\sqrt{3}} =  
\frac{\bar m \varepsilon}{ 2 (m_s  - \bar m)  }\,\,\,. 
\end{equation}
Beyond lowest order, $\eta$ -- $\pi$ mixing cannot simply be described by a mixing angle \cite{Gasser:1984gg,Ecker:1999kr}.

In the $\slashT$ sector, vacuum alignment eliminates pion and $\eta$ tadpoles in LO. 
In $SU(2)$ $\chi$PT, vacuum alignment eliminates all LO three-pion vertices. On the other hand, in $SU(3)$ $\chi$PT
the meson mass term  induces a three-pNG vertex of the form 
\begin{equation}\label{eq:2.7}
\mathcal L_{\pi \pi \pi}= - \frac{B}{3 F_0}  m_{*}\bar\theta \,  d^{a b c} \pi_a \pi_b \pi_c\,\,\,,
\end{equation}
where $d^{a b c}$ are the constants determined by the anticommutator of $SU(3)$ generators 
\begin{equation}
\left\{ t^a, t^b \right\} = \frac{1}{3}\delta^{a b} + d^{a b c} t^c\,\,\,.
\end{equation}
The interaction in Eq. \eqref{eq:2.7} involves one $\eta$ and two pions, one $\eta$ and two kaons, or  one pion and two kaons,
and induces the $\CP$-odd decay $\eta \rightarrow \pi \pi$. Limits on this branching ratio allow to put a bound on $\bar\theta$, though several orders of magnitudes less stringent than the bound from the neutron EDM  \cite{Ambrosino:2004ww}.
Three-pion interactions also arise at LO, but they are proportional to the $\eta$ -- $\pi$ mixing angle, and vanish for large $m_s$.

The $\mathcal O(q^4)$ meson Lagrangian is well known \cite{Gasser:1984gg}.
At the order we are working, we only need the terms
\begin{eqnarray}
\mathcal L^{(4)}_{\pi} &\supset&
L_4\,   \textrm{Tr}( \partial_{\mu} U^{\dagger} \, \partial^{\mu} U^{})  \, \textrm{Tr} \left(\chi^{\dagger} U +  U^{\dagger} \chi \right) + 
L_5  \textrm{Tr} \left( \partial_{\mu} U^{\dagger} \, \partial^{\mu} U^{} \left( \chi^{\dagger} U +  U^{\dagger} \chi \right) \right) \nonumber \\
& & 
+ L_7\, \left(  \textrm{Tr}( U \chi^{\dagger} - \chi\, U^{\dagger})  \right)^2 + 
L_8  \textrm{Tr} \left( U \chi^{\dagger} U \chi^{\dagger} + \chi U^{\dagger } \chi U^{\dagger}\right)\,\,\,.
\end{eqnarray}
$L_4$ and $L_5$  contribute to the pNG wave function renormalization and to the renormalization of $F_0$.
$L_7$ and $L_8$  generate pion and $\eta$ tadpoles, that contribute to $\slashT$ pion-nucleon couplings at N${}^2$LO.
For $\slashT$ baryon-pNG couplings, the dependence on $L_4$ and $L_5$ cancels between the wave function renormalization and the corrections to $F_0$. $L_7$ and $L_8$ have been determined 
from global fits to meson data \cite{Bijnens:2011tb,Bijnens:2014lea}. We use the NLO fits in Ref. \cite{Bijnens:2014lea}, which give $L_7 = ( -0.3 \pm 0.2 ) \cdot 10^{-3}$ and $L_8 = ( 0.5 \pm 0.2 ) \cdot 10^{-3}$. $L_8$ is scale dependent, and it is evaluated at the scale $\mu = 770$ MeV.

\subsection{Baryon Sector}

The inclusion of baryons in $\chi$PT has been derived in a large number of papers, for instance Refs. \cite{Gasser:1987rb,Jenkins:1990jv,Jenkins:1991es,Bernard:1992qa,Bernard:1995dp,Hemmert:1996xg}. 
The baryon octet can be included in a way consistent with the chiral expansion by working in the non-relativistic limit and removing the large, inert octet mass $m_B$ \cite{Jenkins:1990jv,Bernard:1992qa}. 
The mass splittings of octet states vanish in the chiral limit, and scale as $\mathcal O(q^2)$.
$\chi$PT can be extended to include the decuplet baryons at the price of introducing a new scale $\Delta$, the decuplet-octet splitting, which does not vanish in the chiral limit nor can it be rotated away \cite{Jenkins:1990jv,Hemmert:1996xg}.
This octet-decuplet splitting scales as $1/N_c$ in the large $N_c$ expansion~\cite{Jenkins:1993zu,Dashen:1993jt,Jenkins:1995gc}.  The explicit inclusion of the decuplet is necessary for the chiral expansion to respect the $1/N_c$ counting rules~\cite{FloresMendieta:2000mz}, and the predictions from a combined $SU(3)$--$1/N_c$ expansion are phenomenologically well satisfied in lattice QCD calculations~\cite{Jenkins:2009wv,WalkerLoud:2011ab}.

In the heavy baryon formalism, the lowest-order chiral-invariant octet and decuplet baryon Lagrangian is given by 
\begin{eqnarray}\label{eq:3.1}
\mathcal L^{(1)} &=&  \textrm{Tr} \left(i\bar B  v \cdot \mathcal D B\right) + F\, \textrm{Tr} \left( \bar B  S_{\mu} \left[u^{\mu}, B\right]\right)
+ D\, \textrm{Tr} \left( \bar B S_{\mu} \left\{ u^{\mu}, B\right\}\right)  \nonumber \\
& & - i \overline{T}^{\mu} v \cdot \mathcal D T_{\mu} + \Delta \overline{T}^{\mu} T_{\mu} +  \frac{\mathcal{C}}{2} \left( \overline{T}^{\mu} u_{\mu} B + \bar{B} u_{\mu} T^{\mu}\right) +  \mathcal H \overline{T}^{\mu} S^{\nu} u_{\nu} T_{\mu} \,\,\,,
\end{eqnarray} 
where $v_{\mu}$ and $S_{\mu}$ denote the heavy baryon  velocity  and spin. 
$D$ and $F$ are the octet axial couplings, $D\simeq 0.8$ and $F \simeq 0.45$, and $D+F = g_A = 1.27$. $\mathcal C$ is the decuplet-octet axial coupling, $\mathcal{C}\simeq 1.5$. $\mathcal H $ is the decuplet axial coupling, which does not play a role in our discussion.
The matrix $B$ denotes the octet baryon field
\begin{equation}\label{eq:3.0}
B = B^a t^a_{i j} = \left( \begin{array}{c c c}
\frac{1}{\sqrt{2}}\Sigma^0 + \frac{1}{\sqrt{6}}\Lambda & \Sigma^+ 							& p \\
\Sigma^-					       & -\frac{1}{\sqrt{2}}\Sigma^0 + \frac{1}{\sqrt{6}}\Lambda 	& n \\
\Xi^-						       & \Xi^0								& -\frac{2}{\sqrt{6}} \Lambda
\end{array} \right)\,\,\,.
\end{equation}
The decuplet field $T^{\mu}_{i j k}$ carries three completely symmetrized fundamental indices. 
The pNG bosons appear through the combinations 
$u_{\mu} = u^{\dagger} i\partial_{\mu} u^{} - u^{} i\partial_{\mu} u^{\dagger} $ and, in the covariant derivatives,
$V_{\mu} = \frac{1}{2} ( u \partial_{\mu} u^{\dagger} +  u^{\dagger} \partial_{\mu} u^{} )  $.
The chiral covariant derivatives are
\begin{eqnarray}
\mathcal D_{\mu} B  &=& \partial_{\mu} B + \left[ V_{\mu}, B \right]\,\,\,, \\
\mathcal D_{\nu} T_{i j k}^{\mu}  &=& \partial_{\nu} T^{\mu}_{i j k} + \left( V_{\nu} \right)_{i l} T^{\mu}_{l j k} + \left( V_{\nu} \right)_{j l} T^{\mu}_{i l  k} + \left( V_{\nu} \right)_{k l} T^{\mu}_{i j l}\,\,\,.  
\end{eqnarray}
Invariant terms involving the octet and decuplet baryons are constructed using the contractions 
\begin{equation}
\bar{B} u_{\mu} T^{\mu} \equiv \bar B_{i l} u_{\mu\, j m} T^{\mu}_{k l m} \varepsilon^{i j k}\,\,\,, \qquad
\bar{T}^{\mu} u_{\mu} B \equiv \bar T^{\mu}_{k l m}   u_{\mu\, m j} B_{l i} \varepsilon^{i j k}\,\,\,.
\end{equation}

Baryon mass  terms and $\slashT$ couplings appear in the $SU(3)$ Lagrangian at $\mathcal O(q^2)$, and they are given by \cite{Jenkins:1990jv}
\begin{eqnarray}\label{eq:L2}
\mathcal L^{(2)} &=& b_0 \textrm{Tr} \left(\bar B B\right) \textrm{Tr} \chi_+  +  b_D \textrm{Tr}\left(\bar B \{ \chi_+, B \}  \right) + b_F \textrm{Tr}\left(\bar B [ \chi_+, B ]  \right) \nonumber \\
 &  & + b_C\, \overline{T}^{\mu} \chi_+ T_{\mu} + b_{\Delta}\, \textrm{Tr} (\chi U^{\dagger} + \chi^{\dagger} U ) \, \overline{T}^{\mu} T_{\mu}\,\,\,,
\end{eqnarray}
where $\chi_+ = u^{\dagger} \chi u^{\dagger} + u \chi^{\dagger} u$.
$b_0$ and $b_{\Delta}$ denote  common shifts to all octet and decuplet masses, and do not give rise to $\slashT$ nucleon couplings with one pNG boson. 
$b_D$, $b_F$, and $b_C$ induce  splittings between the different octet and decuplet states, and give rise to $\slashT$ baryon-pNG interactions.
The LECs $b_0$, $b_D$, and $b_F$ scale as $\Lambda_\chi^{-1}$, and do not depend on the quark masses. When including decuplet corrections $b_0$, $b_D$, and $b_F$ must be interpreted as series expansions in the octet-decuplet splitting $\Delta$~\cite{Bernard:1998gv, WalkerLoud:2004hf,Tiburzi:2004rh,Tiburzi:2005na}
\begin{equation}
b_i = \frac{1}{\Lambda_\chi} \left( b^{(0)}_i + b^{(1)}_i \frac{ \Delta }{\Lambda_\chi} + b^{(2)}_i \frac{ \Delta^2 }{\Lambda_\chi^2} + \ldots  \right)\,\,\,,
\end{equation}
where the higher orders in $\Delta/\Lambda_\chi$ 
arise from finite contributions and are needed to absorb the divergences arising from diagrams with decuplet intermediate states. 
There is no sense in keeping track of the finite $\Delta$ dependence in the LECs as these corrections are quark mass independent, and thus not discernible with present lattice QCD calculations at fixed $N_c=3$.

Baryon mass splittings and $\slashT$ couplings receive $\mathcal O(q^3)$ corrections from one-loop diagrams involving the chiral-invariant interactions in  Eq. \eqref{eq:3.1}, and  chiral-breaking interactions from Eq. \eqref{eq:2.1}.
This is distinct from $SU(2)$ $\chi$PT in which the $\mathcal{O}(q^3)$ corrections cancel in the isospin mass splitting.
At $\mathcal O(q^4)$,  one has to consider one-loop diagrams involving operators  in the $\mathcal O(q^2)$ Lagrangian, and tree-level diagrams with one insertion of the $\mathcal O(q^4)$ Lagrangian.
Besides the mass terms in Eq. \eqref{eq:L2}, $\mathcal L^{(2)}$ contains relativistic corrections to the interactions in Eq. \eqref{eq:3.1}, and baryon-pNG interactions with two derivatives.
The operators are listed in Ref.  \cite{Borasoy:1996bx} and here we give only those relevant to our discussion.
There are four operators containing two derivatives of pNG fields.
\begin{eqnarray}\label{eq:L2pipi}
\mathcal L^{(2)}_{\pi\pi} &=& b_1 \textrm{Tr} \left( \bar B \left[u_{\mu}, \left[ u^{\mu}, B\right]\right] \right) + b_2 \textrm{Tr} \left( \bar B \left[u_{\mu}, \left\{ u^{\mu}, B \right\}\right]\right)
+ b_3 \textrm{Tr} \left( \bar B \left\{  u_{\mu} ,\left\{ u^{\mu}, B \right\} \right\}\right) \nonumber \\ & &  + b_8 \mathrm{Tr}[\bar B B]  \mathrm{Tr}[u_\mu u^\mu] \,\,\,.
\end{eqnarray}
The effects on the baryon masses of operators similar to $b_1$, \ldots, $b_8$, but with $u_{\mu}$ replaced by $v \cdot u$, can be accounted for by a redefinition of $b_{i}$
and of the $\mathcal O(q^4)$ LECs. For this reason, we do not include these operators explicitly.
The relativistic corrections are
\begin{eqnarray}\label{eq:L2rel}
\mathcal L^{(2)}_{\textrm{rel}} &=& - \frac{ D}{2 m_B} \textrm{Tr} \left( \bar B S_{\mu} \left[i D^{\mu}, \left\{ v\cdot u, B \right\}\right] \right)
- \frac{ F}{2 m_B} \textrm{Tr} \left( \bar B S_{\mu} \left[i D^{\mu}, \left[ v\cdot u, B \right] \right] \right) \nonumber \\
& & -\frac{F}{2 m_B} \textrm{Tr} \left( \bar B S_{\mu} \left[ v\cdot u, \left[ i D^{\mu}, B\right] \right]\right)
- \frac{D}{2 m_B} \textrm{Tr} \left( \bar B S_{\mu} \left\{ v\cdot u, \left[i D^{\mu}, B\right] \right\} \right) \nonumber \\
& & + \frac{D^2 - 3 F^2}{24 m_B} \textrm{Tr} \left(\bar B  \left[ v\cdot u, \left[v\cdot u, B\right]  \right]\right) - \frac{D^2}{12 m_B} \textrm{Tr}(\bar B B) \textrm{Tr}(v\cdot u \, v \cdot u) \nonumber \\
& &- \frac{1}{2 m_B} \textrm{Tr}\left(\bar B \left[D_{\mu}, \left[D^{\mu}, B \right]\right]\right) 
+ \frac{1}{2 m_B} \textrm{Tr}\left(\bar B \left[ v \cdot D , \left[v \cdot D, B \right]\right]\right) \nonumber \\ & & - \frac{D F}{4 m_B} \textrm{Tr} \left(\bar B \left[ v \cdot u, \left\{ v\cdot u, B\right\}\right] \right) \,\,\,.
\end{eqnarray}
We find that the contribution of recoil corrections to $D$ and $F$ to the baryon masses are small.
The relativistic corrections to the octet-decuplet coupling $\mathcal C$ can be removed using the LO equations of motion~\cite{WalkerLoud:2004hf,Tiburzi:2004rh,Tiburzi:2005na}.  

The $\mathcal O(q^4)$ Lagrangian is
\begin{eqnarray}\label{eq:L4}
\mathcal L^{(4)} &=& d_1 \Tr \left( \bar B \left[\chi_+, \left[\chi_+, B \right] \right]\right) + d_2 \Tr \left( \bar B \left[\chi_+, \left\{ \chi_+, B \right\} \right]\right) + d_3 \Tr \left( \bar B \left\{ \chi_+, \left\{ \chi_+, B \right\}  \right\} \right) \nonumber \\
& &  + d_4 \Tr( \bar B \chi_+) \textrm{Tr}(\chi_+ \bar B) + d_5 \Tr \left(\bar B \left[\chi_+, B\right]\right) \, \Tr(\chi_+)
+ d_6 \Tr \left(\bar B \left\{\chi_+, B\right\}\right) \, \Tr(\chi_+) \nonumber \\
& & + d_7 \Tr(\bar B B) \Tr(\chi_+) \Tr(\chi_+) + d_8 \Tr(\bar B B)\, \Tr(\chi_+^2) \nonumber \\
& & +  d_9 \textrm{Tr} \left(\bar B \left[\chi_-, \left[\chi_-, B\right]\right]\right) + d_{10} \textrm{Tr} \left(\bar B \left[\chi_-, \left\{\chi_-, B\right\}\right]\right) \nonumber \\ &&
+ d_{11} \textrm{Tr} \left(\bar B \left\{\chi_-, \left\{\chi_-, B\right\}\right\}\right) + d_{12} \textrm{Tr} \left(\bar B \chi_- \right)  \textrm{Tr} \left(  \chi_-  B\right)  \nonumber \\
& & + d_{13} \textrm{Tr} \left( \bar B \left[\chi_-, B\right]\right) \textrm{Tr}(\chi_-) +
 d_{14} \textrm{Tr} \left( \bar B \left\{\chi_-, B\right\}\right) \textrm{Tr}(\chi_-) \nonumber \\
& & + d_{15} \textrm{Tr}(\bar B B) \textrm{Tr}(\chi_-) \textrm{Tr}(\chi_-) + d_{16} \textrm{Tr}(\bar B B) \textrm{Tr}(\chi^2_-)\,\,\,. 
\end{eqnarray}
$d_{1}, \ldots, d_{8}$ were constructed in Ref. \cite{Borasoy:1996bx}, and contribute to baryon masses and splittings.
The operators $d_9, \ldots, d_{16}$ involve  two insertions of  $\chi_- =  u^{\dagger} \chi u^{\dagger} - u \chi^{\dagger} u$. 
The $\CP$-even parts of these operators do not contribute to baryon masses and mass splittings, but do contribute to pion-nucleon scattering.
The $\CP$-odd components give $\mathcal O(q^4)$ corrections to $\slashT$ baryon-pNG couplings.

\section{Octet baryon masses and $\slashT$ couplings at tree level}\label{relations}
The Lagrangian \eqref{eq:L2}
realizes the leading effects of the light quark masses in the baryon sector. The light quark masses induce splittings between the octet and decuplet states, and, in the presence of the QCD $\bar\theta$ term,
cause the appearance of $\slashT$ couplings between baryon and pNG bosons. 
The LO corrections to the baryon masses are well known (see, for example, Refs. \cite{Jenkins:1991ts,Bernard:1993nj}),  we give them here in order to make the connection with $\slashT$ couplings explicit.

The  nucleon, $\Xi$ and $\Sigma$ mass splittings are given by
\begin{eqnarray}
\delta^{(0)} m_N &=& m_n - m_p  = - 8 B \bar m \varepsilon\,  (b_F + b_D)\,\,\,, \nonumber \\
\delta^{(0)} m_{\Xi} &=& m_{\Xi^-} - m_{\Xi^0} = - 8 B \bar m \varepsilon\, (b_F - b_D)\,\,\,, \nonumber \\
\delta^{(0)} m_{\Sigma} &=& m_{\Sigma^+} - m_{\Sigma^-} =  16 B \bar m  \varepsilon\,  b_F \label{tree}\,\,\,,
\end{eqnarray}
where we introduced the superscript $(0)$ to denote that these are the leading contributions.
The three mass splittings are not independent, but are related by the Coleman-Glashow relation \cite{Coleman:1961jn} 
\begin{equation}\label{rel1}
\delta^{(0)} m_N + \delta^{(0)} m_{\Xi} + \delta^{(0)} m_{\Sigma} = 0\,\,\,.
\end{equation}
Neglecting $\mathcal O(\varepsilon^2)$ corrections to the $\Sigma^0$ and $\Lambda$ masses, the isospin-averaged masses of the nucleon, $\Sigma$, $\Lambda$, and $\Xi$ baryon are
\begin{eqnarray}
\Delta^{(0)} m_N &=& \frac{m_n + m_p}{2} - m_B  = - 4 B ( m_s (b_0 + b_D - b_F) + \bar m ( 2 b_0 + b_D + b_F) ) \,\,\,, \nonumber \\
\Delta^{(0)} m_{\Xi} &=& \frac{m_{\Xi^-} + m_{\Xi^0}}{2} - m_B = - 4 B ( m_s (b_0 + b_D + b_F) + \bar m ( 2 b_0 + b_D - b_F) )\,\,\,, \nonumber \\
\Delta^{(0)} m_{\Sigma} &=& \frac{ m_{\Sigma^+} + m_{\Sigma^0} + m_{\Sigma^-} }{3} - m_B = - 4 B (  ( m_s + 2 \bar m) b_0  + 2 \bar m b_D )\,\,\,, \nonumber  \\
\Delta^{(0)} m_{\Lambda} &=& m_{\Lambda} - m_B = - 4 B \left(  (m_s + 2 \bar m ) b_0 + \frac{2}{3}  (\bar m + 2 m_s) b_D \right)\,\,\,, 
\label{tree1}  
\end{eqnarray}
where all the masses are measured with respect to $m_B$, the common octet mass in the chiral limit.
Finally, $m_B$ gets a corrections proportional to $m_s + 2 \bar m$
\begin{equation}
\Delta^{(0)} m_{B} =  \frac{2 \Delta^{(0)} m_N  + 2 \Delta^{(0)} m_\Xi +   3 \Delta^{(0)} m_\Sigma + \Delta^{(0)} m_\Lambda}{8}=  - 4 B  (m_s + 2 \bar m ) \left(b_0 + \frac{2}{3} b_D \right)\,\,\,.
\label{tree2}  
\end{equation}

In the presence of a $\bar\theta$ term, the operators in Eq. \eqref{eq:L2} induce $\slashT$ baryon-pNG couplings. The couplings of the greatest phenomenological interest are pion-nucleon couplings.
Besides giving a LO contribution to the nucleon EDM, the isoscalar non-derivative pion-nucleon coupling $\bar g_0$ 
induces a $\slashT$ nucleon-nucleon potential, which is expected to give a sizeable, when not  dominant, contribution to EDMs of light nuclei with $N \neq Z$ \cite{deVries:2011an,Maekawa:2011vs,Bsaisou:2012rg}.
Furthermore, EDMs of heavier systems, like $^{199}$Hg, are commonly computed in terms of three non-derivative pion-nucleon couplings \cite{Engel:2013lsa}.

Introducing the nucleon doublet, $N = (p\, n)^T$,  we write the  $\slashT$ pion-nucleon couplings as
\begin{equation}
\mathcal L_\pi = - \frac{\bar g_0}{2 F_{\pi}} \bar N \boldtau\cdot \boldpi N  - \frac{\bar g_1}{2 F_{\pi}} \pi_0 \bar N  N  - \frac{\bar g_2}{2 F_{\pi}} \pi_0 \bar N  \tau^3 N + \ldots\,\,\,,
\end{equation}
where $\boldtau$  are the Pauli matrices, and $\ldots$ include terms with more derivatives. Notice that we defined the couplings in terms of the physical pion decay constant $F_{\pi}$, rather than $F_0$. The difference between 
$F_{\pi}$ and $F_0$ is an N${}^2$LO correction.

At tree level, the $\slashT$ pion-nucleon couplings are expressed in terms of the LECs $b_D$ and $b_F$. 
$b_0$ does not generate tree-level $\slashT$ couplings with only one pNG boson, but does induce couplings with at least three pNG, which are relevant at one loop.  
$\bar g_{0}$, $\bar g_1$, and $\bar g_2$ are given by
\begin{eqnarray}\label{FFlo}
\bar g^{(0)}_0 &=&  -8  B (b_D + b_F) m_{*} \bar\theta\,\,\,, \\
\bar g^{(0)}_1 & =&   8 B (b_D - 3 b_F) \frac{\phi}{\sqrt{3}} m_{*} \bar\theta\,\,\,, \label{FFloB} \\
\bar g^{(0)}_2 & = &  4 B (b_D + b_F) \phi^2 m_* \bar\theta\,\,\,, \label{FFloC}
\end{eqnarray}
where $\phi$ is the $\eta$ -- $\pi$ mixing angle defined in Eq. \eqref{eq:2.6}.
From Eqs. \eqref{tree} and \eqref{FFlo}, we see that $\bar g_0$ is related to the tree-level contribution to the nucleon mass splitting.
\begin{equation}\label{isorel1a}
\bar g^{(0)}_0 = \delta^{(0)} m_N \frac{m_* \bar\theta}{\bar m \varepsilon} = \delta^{(0)} m_N  \frac{1-\varepsilon^2}{2\varepsilon} \bar\theta + \mathcal O\left( \frac{\bar m}{m_s}\right)\,\,\, .
\end{equation}
For $m_s \gg \bar m$, this is the same relation that holds in $SU(2)$ \cite{Mereghetti:2010tp}.
In $SU(3)$ $\chi$PT, at tree level one can also write \cite{Crewther:1979pi}
\begin{equation}\label{isorel1b}
\bar g^{(0)}_0 = (\Delta^{(0)} m_{\Xi} - \Delta^{(0)} m_{\Sigma})\frac{2 m_*}{ m_s - \bar m}  \bar\theta\,\,\,.
\end{equation}
We will  show that both Eq. \eqref{isorel1a} and \eqref{isorel1b} are violated at N${}^2$LO. 
However, Eq. \eqref{isorel1a} is only violated by finite terms  and by new LECs appearing at $\mathcal O(q^4)$, while it is respected by all loop diagrams. On the other hand, Eq. \eqref{isorel1b}
is already violated at NLO and receives much larger corrections. 

An isoscalar operator like $\bar\theta$ can generate the isospin-breaking couplings $\bar g_1$ and $\bar g_2$ only in the presence of some source of isospin violation.
In $SU(2)$ $\chi$PT this implies that $\bar g_1$ and $\bar g_2$ are suppressed and appear at $\mathcal O(q^4)$ and $\mathcal O(q^6)$, respectively.
In $SU(3)$ $\chi$PT, the $\eta$ -- $\pi$  mixing angle $\phi$ appears at LO which means that $\bar g_{1,2}$ are formally  LO as well. However, numerically they are suppressed by powers of $\bar m/m_s \sim 0.04$.

The coupling $\bar g_1$ is particularly important for EDMs of nuclei with $N=Z$ such as the deuteron. At LO, the combination of 
 LECs $b_D - 3 b_F$ can be expressed in terms of baryon-mass splittings as
\begin{eqnarray}\label{eq:g1su3}
\bar g^{(0)}_1&=&  -(\delta^{(0)} m_{\Sigma} - \delta^{(0)} m_{\Xi})\frac{\phi}{\sqrt{3}}\frac{ m_* }{ \bar m \varepsilon} \bar\theta \nonumber \\
&=& - ( \Delta^{(0)} m_{N}   - \Delta^{(0)} m_{B} )\frac{\phi}{\sqrt{3}} \frac{6 m_{*} }{ ( m_s - \bar m)} \bar\theta \nonumber \\
&=& - \left( \left(\frac{d}{d m_s} - \frac{d}{2 d \bar m }\right) \Delta^{(0)} m_N \right)\frac{\phi}{\sqrt{3}} 4 m_* \bar\theta\,\,\,,
\end{eqnarray}
where the masses and splittings are given in Eq. \eqref{tree1}. 
Eqs. \eqref{FFloB} and \eqref{eq:g1su3} were used in Ref.~\cite{Lebedev:2004va} to estimate $\bar g_1$ and its contribution to the deuteron EDM. 
Contributions to $\bar g_1$ that are not suppressed by $\bar m/m_s$ only appear at N${}^2$LO. 
However, as we discuss in Sect. \ref{n2log1}, they can be as large as Eq. \eqref{eq:g1su3}.

In the case of the QCD $\bar\theta$ term, the coupling $\bar g_2$ is suppressed by $\bar m^2/m_s^2$, and because it is of little phenomenological consequence, we neglect $\bar g_2$ henceforth.

Non-analytic LO contributions to the nucleon EDM in $SU(3)$ $\chi$PT involve other $\slashT$ nucleon couplings \cite{Ottnad:2009jw}. 
Introducing the isospin doublet $K = (K^+, K^0)$, and an isospin triplet $\Sigma = (\Sigma^+, \Sigma^0, \Sigma^-)$, we can write the isospin-invariant $\slashT$ couplings between
the nucleon and $\eta$ meson, and the nucleon, kaon and $\Sigma$ or $\Lambda$ baryon as 
\begin{eqnarray}
\mathcal L_{} &=& - \frac{\bar g_{0\, \eta}}{2 F_\eta} \eta \bar N N  - \frac{\bar g_{0\, N\Sigma K}}{2 F_{K}} \bar N \boldtau \cdot \boldSigma K   - \frac{\bar g_{0\, N\Lambda K}}{2 F_{K}} \bar N  K \Lambda^0   + \textrm{h.c.}\label{othertslash1}
\end{eqnarray}
Isospin-breaking couplings also arise at LO, but are suppressed by $\bar m \varepsilon/m_s$, and we neglect them in the following.
At LO, the couplings in Eq. \eqref{othertslash1} are given by
\begin{eqnarray}
\bar g^{(0)}_{0\, \eta} &=& 8 B \frac{b_D - 3 b_F}{\sqrt{3}}  m_* \bar\theta\,\,\,, \label{treeg0eta}\\
\bar g^{(0)}_{0\, N\Sigma K} &=& 8 B (b_F - b_D) m_* \bar\theta  \,\,\,,  \label{treeg0K1}\\
\bar g^{(0)}_{0\, N\Lambda K} &=& 8 B \frac{(b_D + 3 b_F)}{\sqrt{3}} m_* \bar\theta\,\,\,. \label{treeg0K2}
\end{eqnarray}
These $\slashT$ couplings can be expressed in terms of various combinations of baryon masses. We list some of them 
\begin{eqnarray}
\sqrt{3} \bar g^{(0)}_{0\, \eta} &=&  -(\delta^{(0)} m_{\Sigma} - \delta^{(0)} m_{\Xi})\frac{ m_* \bar\theta}{ \bar m \varepsilon}  \nonumber \\
&=& - \left( \left(\frac{d}{d m_s} - \frac{d}{2 d \bar m }\right) \Delta^{(0)} m_N \right) 4 m_* \bar\theta\,\,\,,
\label{relnew1}\\
\bar g^{(0)}_{0\, N\Sigma K}  &=&  - \delta^{(0)} m_{\Xi} \frac{ m_* \bar\theta}{\bar m \varepsilon }\nonumber \\
&=&  - (\Delta^{(0)} m_{\Sigma} - \Delta^{(0)} m_N) \frac{2 m_* \bar\theta}{m_s - \bar m}\,\,\,,  \label{relnew2}\\
\sqrt{3} \bar g^{(0)}_{0\, N\Lambda K} &= &   -(\delta^{(0)} m_N - \delta^{(0)} m_{\Sigma})\frac{ m_* \bar\theta}{ \bar m \varepsilon}\nonumber \\
& =&  -  (\Delta^{(0)}  m_{\Lambda}  - \Delta^{(0)} m_N) \frac{6 m_* \bar\theta}{m_s -\bar m} \,\,\,, \label{relnew3}
\end{eqnarray}
where the approximate $SU(3)$ symmetry enforces many other LO relations between corrections to the octet masses.
The second equation in Eq. \eqref{relnew1} is particularly interesting, since it relates $\bar g_{0\, \eta}$ not directly to the nucleon mass, but to its derivatives with respect to $m_s$ and $\bar m$. 
Through the Feynman-Hellmann theorem, these derivatives can be related to the nucleon sigma terms
\begin{equation}\label{eq:FH}
\sigma_{N  q} = m_q \langle N| \bar q q | N \rangle  = m_q \frac{\partial \Delta m_N }{\partial m_q}\,\,\,.
\end{equation}
We can thus write 
\begin{eqnarray}
\sqrt{3} \bar g^{(0)}_{0\, \eta} &=&  - \left( \frac{\sigma^{(0)}_{N s} }{m_s} - \frac{\sigma^{(0)}_{N l}}{2  \bar m }\right) 4 m_* \bar\theta\,\,\,,
\end{eqnarray}
where $\sigma_{N l} = \sigma_{N u} + \sigma_{N d}$.

In the rest of the paper we show that in most cases the relations between baryon mass splittings and $\slashT$ baryon-pNG couplings break down already at NLO. The slow convergence of $SU(3)$ baryon $\chi$PT 
then renders the usefulness of these relations to be qualitative only.
However, for each isospin invariant coupling, there exist one relation that survives NLO and most of the N${}^2$LO corrections.
These relations thus provide a powerful method to extract $\slashT$ couplings from well-known $T$-even matrix elements.
The most important example is the link between $\bar g_0$ and $\delta m_N$, which, as we discuss in Sections \ref{NLO} and \ref{N2LO}, 
receives  particularly small corrections.

\section{Octet baryon masses  and $\slashT$ couplings at NLO}\label{NLO}

\begin{figure}
\center
\includegraphics[width=16cm]{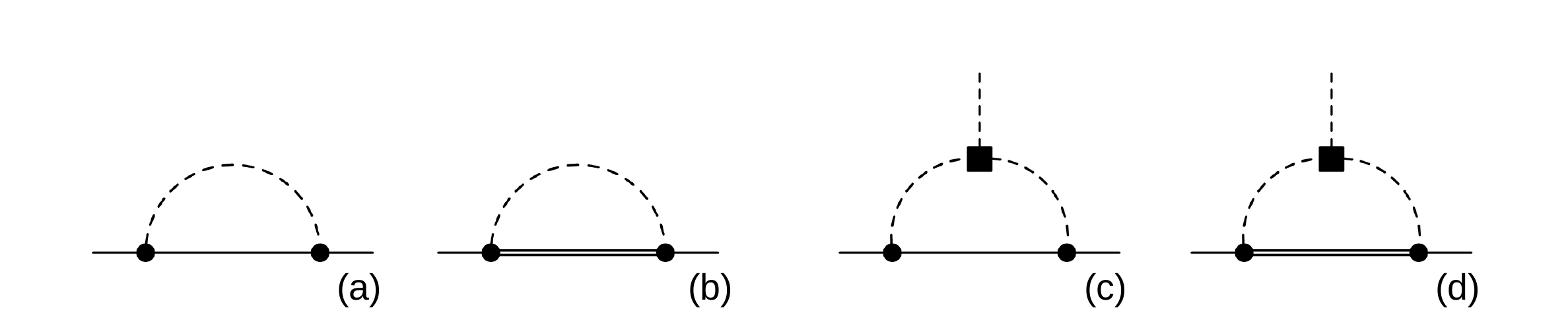}
\caption{$\mathcal O(q^3)$ corrections to the baryon masses and baryon-pNG $\slashT$ couplings. 
Plain, double, and dashed lines denote octet baryons, decuplet baryons, and pNG bosons, respectively.
Dotted vertices denote $\CP$-even couplings, \textit{i.e.} the octet-pNG axial couplings $D$ and $F$ and the decuplet-octet-pNG
coupling $\mathcal C$. A square denotes a $\slashT$ coupling.}\label{Diag2}
\end{figure}

In Fig.~\ref{Diag2} we represent one-loop corrections to the baryon masses (diagrams \ref{Diag2}(a) and (b)), and to $\slashT$ octet-pNG couplings (diagrams \ref{Diag2}(c) and (d)). 
In dimensional regularization, diagram \ref{Diag2}(a) is finite, and contributes to the octet masses and mass splittings at $\mathcal O(q^3)$. 
In particular, it affects the mass splittings $\delta m_N$, $\delta m_{\Xi}$, and $\delta m_{\Sigma}$ through the kaon mass difference $m^2_{K^0} - m^2_{K^+}  = B (m_d - m_u)$,
and the $\eta$ -- $\pi$ mixing angle $\phi$. At the same order, diagrams with intermediate decuplet states contribute via diagram \ref{Diag2}(b). These diagrams have UV poles that are linear in the octet-decuplet splitting $\Delta$, and are absorbed by $b_D$, $b_F$, and $b_0$.
Diagram \ref{Diag2}(b) contributes to the nucleon mass splitting only through the kaon mass difference because of the vanishing of the octet-decuplet-$\eta$ axial coupling. The mass splittings of the $\Xi$ and $\Sigma$ baryons receive contributions from both the kaon mass splitting and $\eta$ -- $\pi$ mixing.

All together we find
\begin{eqnarray}
\delta^{(1)} m_N &=&  \frac{ ( D^2  - 6 D F  - 3 F^2) }{ 48 \pi F_0^2} \left( m^3_{K^0} - m^3_{K^+} \right) +\frac{(D-3 F)(D + F)}{8\pi F_0^2} \frac{\phi}{\sqrt{3}} (m_{\eta}^3 - m_{\pi}^3) 
 \nonumber \\
 & & + \frac{\mathcal C^2}{144 \pi^2 F^2_{0}} \left( f(m_{K^0},\Delta) - f(m_{K^+},\Delta) \right)\,\,\,, \label{Loop1a} \\
\delta^{(1)} m_{\Xi} &=&  - \frac{ ( D^2  + 6 D F  - 3 F^2) }{ 48 \pi F_0^2} \left( m^3_{K^0} - m^3_{K^+} \right)
-\frac{(D+3 F)(D - F) }{8\pi F_0^2} \frac{\phi}{\sqrt{3}} (m_{\eta}^3 - m_{\pi}^3)
 \nonumber \\
 & & -\frac{7 \mathcal C^2}{144 \pi^2 F^2_{0}} \left( f(m_{K^0},\Delta) - f(m_{K^+},\Delta) \right)
- \frac{\mathcal C^2}{12 \pi^2 F_0^2} \frac{\phi}{\sqrt{3}} \left( f(m_{\eta},\Delta) - f(m_{\pi},\Delta) \right), \label{Loop1b} \\ 
\delta^{(1)} m_{\Sigma} &=&  \frac{  D F    }{ 4 \pi F_0^2} \left( m^3_{K^0} - m^3_{K^+} \right) + \frac{D F}{2 \pi F_0^2} \frac{\phi}{\sqrt{3}} (m_{\eta}^3 - m_{\pi}^3) \nonumber \\
& & + \frac{\mathcal C^2}{24 \pi^2 F^2_{0}} \left( f(m_{K^0},\Delta) - f(m_{K^+},\Delta) \right) + 
 \frac{\mathcal C^2}{12 \pi^2 F_0^2} \frac{\phi}{\sqrt{3}} \left( f(m_{\eta},\Delta) - f(m_{\pi},\Delta) \right).
  \label{Loop1}	
\end{eqnarray}
The loop function appearing in the decuplet diagrams is given by
\begin{eqnarray}\label{decuLO}
f(m_K,\Delta) &=& \Delta  \left(  - \Delta^2 + \frac{3}{2}m_K^2 \right) \, L  + \frac{\Delta}{6} \left(12 m_K^2 - 10 \Delta^2 + 3 \left( 3 m_K^2 - 2 \Delta^2\right) \log \frac{\mu^2}{m_K^2}\right)\nonumber  \\ && + 2 (m_K^2 - \Delta^2)^{3/2}  \textrm{arccot} \frac{\Delta}{\sqrt{m_K^2-\Delta^2}} \,\,\,. 
\end{eqnarray}
$L$ encodes the UV divergence, and is defined as	
\begin{equation}
L = \frac{1}{\varepsilon} + \log 4\pi - \gamma_E\,\,\,,
\end{equation}
where $\gamma_E$ is the Euler constant. 
For the spin projector for decuplet fields in $d = 4 - 2 \varepsilon$ dimensions, we used the definition of Ref.~\cite{Hemmert:1996xg}.
The poles are absorbed by defining the renormalized couplings $b^r_D$ and $b^r_F$. We work in the $\overline{\textrm{MS}}$ scheme, and define
\begin{equation}
b^r_D = b_D  - \Delta \frac{\mathcal C^2}{64 \pi^2 F^2_{\pi}} L\,\,\,, \qquad b^r_F = b_F  + \Delta \frac{5 \mathcal C^2}{384 \pi^2 F^2_{\pi}} L\,\,\,.
\end{equation}
In the limit $\Delta \rightarrow 0$, the divergence disappears and $f$ assumes the same form as the octet corrections
\begin{equation}
\lim_{\Delta\rightarrow 0} f(m_K,\Delta) =  \pi m_K^3\,\,\,.
\end{equation}
The NLO corrections in Eqs. \eqref{Loop1a}, \eqref{Loop1b}, and \eqref{Loop1} that do not involve the decuplet agree with Ref.~\cite{Frink:2004ic}. Both octet and decuplet corrections respect the Coleman-Glashow relation. 

In addition to the mass splittings, the tree-level relations between baryon masses and $\slashT$ couplings involve the octet isospin-averaged masses. 
Baryon masses in the isospin limit were computed at NLO in Refs.~\cite{Jenkins:1991ts,Bernard:1993nj} and at N${}^2$LO in Refs.~\cite{Borasoy:1996bx,Jenkins:1991ts,Frink:2004ic,WalkerLoud:2004hf,Ren:2012aj}. At NLO
\begin{eqnarray}
\Delta^{(1)} m_N &=& - \frac{1}{96 \pi F^2_0} \left( 2 (5 D^2 - 6 D F + 9 F^2)  m_{K}^3  + 9 (D + F)^2 m_{\pi}^3 + (D-3F)^2 m_{\eta}^3 \right) 
 \nonumber \\ &&   - \frac{\mathcal C^2}{48 \pi^2 F^2_0}  \left(f(m_{K},\Delta)  + 4 f(m_{\pi},\Delta)
\right)\,\,\,, \\
\Delta^{(1)} m_{\Xi} &=& - \frac{1}{96 \pi F^2_0} \left( 2 (5 D^2 + 6 D F + 9 F^2) m_{K}^3  + 9 (D - F)^2 m_{\pi}^3 + (D+ 3F)^2 m_{\eta}^3   \right) \nonumber\\
& &   - \frac{\mathcal C^2}{48 \pi^2 F^2_0}  \left(  3 f(m_{K},\Delta)   +  f(m_{\pi},\Delta) +  f(m_{\eta},\Delta)  \right)\,\,\,, \label{sigmaXi}
\\
\Delta^{(1)} m_{\Sigma} &=& - \frac{1}{96 \pi F^2_0} \left(  12  (D^2 + F^2) m_{K}^3  + 4 (D^2 + 6 F^2) m_{\pi}^3 + 4 D^2 m_{\eta}^3  \right)  \nonumber \\ && 
- \frac{\mathcal C^2}{144 \pi^2 F^2_0}  \left(  10 f(m_{K},\Delta)   + 2 f(m_{\pi},\Delta) + 3 f(m_{\eta},\Delta) )
\right) \,\,\, , \label{sigmaSigma} \\
\Delta^{(1)} m_{\Lambda} &=& 
- \frac{1}{24 \pi F^2_0} 
\left(    (D^2 + 9 F^2) m_{K}^3 +  D^2 ( 3 m_{\pi}^3 +  m_{\eta}^3  )  \right) \nonumber \\ & & 
- \frac{\mathcal C^2}{48 \pi^2 F^2_0} \left(  2 f(m_{K},\Delta)  + 3 f(m_{\pi},\Delta)  
\right)\,\,\,,  
\end{eqnarray}
where the decuplet loop function $f$ is given in Eq. \eqref{decuLO}.

NLO corrections to $\bar g_i$ are induced by the $\slashT$ three-pNG coupling in Eq. \eqref{eq:2.7}. This coupling is fixed at LO by the meson masses and does not involve a free coefficient.
The relevant loop diagrams with octet and decuplet intermediate states are shown in Fig.~\ref{Diag2}. Diagram \ref{Diag2}(c)  is finite, while \ref{Diag2}(d) is UV divergent. Both diagrams contribute to $\bar g_0$,  $\bar g_1$, and $\bar g_2$, although the last two couplings are suppressed by $\bar m/m_s$ and $\bar m^2/m_s^2$, respectively. The corrections to $\bar g_0$ and $\bar g_1$ are given by
\begin{eqnarray}
\bar g^{(1)}_0 &=& B m_* \bar\theta \left\{  \frac{D^2 - 6 D F - 3 F^2}{24 \pi F^2_0} \frac{m_{K^+}^2 + m_{K^0}^2 + m_{K^+} m_{K^0} }{m_{K^0}+ m_{K^+}} \right. \nonumber \\
& & \left. + \frac{(D-3 F)(D+F)}{12  \pi F_0^2} \left(\frac{m_{\eta}^2 + m_{\eta} m_{\pi} + m_{\pi}^2}{m_{\eta}+ m_{\pi}}\right) +   \frac{\mathcal{C}^2}{72 \pi^2 F^2_0}  \frac{f(m_{K^0},\Delta) - f(m_{K^+},\Delta) }{m^2_{K^0} - m^2_{K^+}}\right\}, \label{g01NLO} \\
\bar g^{(1)}_1 &=& B m_* \bar\theta \left\{ \frac{5 D^2 - 6 D F + 9 F^2}{32  \pi F^2_0} \left( m_{K^0} - m_{K^+} + (m_{K^0} + m_{K^+}) \frac{\phi}{\sqrt{3}} \right) \right. \nonumber\\ && \left.  
+ \left( 3\frac{(D-3F)^2 m_{\eta} -5(D+F)^2 m_{\pi}  }{16 \pi F^2_0 } + \frac{(D^2 + 6 D F - 3 F^2)}{6 \pi F^2_0} \frac{m_{\eta}^2 + m_{\eta} m_{\pi} + m_{\pi}^2}{m_{\eta}+ m_{\pi}} \right) \frac{\phi}{\sqrt{3}} 
\right. \nonumber \\  & &  \left.
+ \frac{\mathcal{C}^2}{32 \pi^2 F_0^2} \left( f^{\prime} (m_{K^0}, \Delta) - f^{\prime} (m_{K^+}, \Delta)  + \frac{\phi}{\sqrt{3}} (f^{\prime} (m_{K^0}, \Delta) + f^{\prime} (m_{K^+}, \Delta))
\right) \right. \nonumber \\ & & \left.  
- \frac{\phi}{\sqrt{3}} \frac{\mathcal{C}^2}{6 \pi^2 F^2_0} \left(  3 f^{\prime} (m_{\pi}, \Delta) - 2 \frac{f(m_{\eta},\Delta) - f(m_{\pi},\Delta)) }{m^2_{\eta} - m^2_{\pi}} \right)
\right\}\,\,\,. \label{g11NLO}
\end{eqnarray}
The function $f^\prime$ entering the decuplet corrections to $\bar g_1$ is
\begin{equation}
f^{\prime}(x, y) =  \frac{1}{2 x}\frac{\partial}{\partial x} f(x,y)\,\,\,.
\end{equation}
Notice that loops with only pions do not contribute to $\bar g_0$ at NLO, in accordance with the $SU(2)$ result of Ref. \cite{Mereghetti:2010tp}.
The piece  proportional to $(D+F)^2 \mpi$ contributing to $\bar g^{(1)}_1$ is the same as found in $SU(2)$ $\chi$PT \cite{Bsaisou:2014zwa} once the LO identifications $D+F = g_A$ and $\phi/\sqrt{3} = (\delta \mpi^2)/(2 B \bar m \varepsilon)$, with $\delta \mpi^2 = m^2_{\pi^{\pm}} -m_{\pi^0}^2$, are made. In $SU(2)$ $\chi$PT this contribution  appears at N${}^3$LO.

The one-loop diagrams in Fig. \ref{Diag2} give also the isospin-invariant nucleon-pNG couplings defined in Eq.~\eqref{othertslash1}. For these couplings we work in the isospin limit and find
\begin{eqnarray}
\sqrt{3}\, \bar g^{(1)}_{0 \, \eta} &=&  \frac{B m_* \bar\theta}{16 \pi F^2_0} \Big\{ m_K \left( 5 D^2 - 6 D F + 9 F^2 \right) - 9 (D + F)^2 m_{\pi} + (D - 3 F)^2  m_{\eta}   \nonumber \\ & &  
+ \frac{2 \mathcal C^2}{3} \left( f^{\prime}(m_K,\Delta) - 8 f^{\prime} (m_{\pi},\Delta) \right)
\Big\}\,\,\,, \label{g0etaNLO} \\
\bar g^{(1)}_{0\, N\Sigma K} &=& B m_* \bar\theta \left\{ \frac{5 D^2 + 18 D F - 15 F^2}{48 \pi F^2_0} \frac{ m_K^2 + m_K m_{\pi} + m^2_{\pi} }{ m_K + m_{\pi}} \right. \nonumber \\ & &  \left. + \frac{(D - F)(D + 3 F)}{48 \pi F^2_0} \frac{m^2_{\eta} + m_{\eta} m_{K} + m^2_K}{m_{K} + m_{\eta}}  \nonumber \right. \\ & & \left. + \frac{\mathcal C^2}{288 \pi^2 F^2_0} 
\left( 10 \frac{ f(m_{K},\Delta) - f(m_\pi, \Delta)  }{m^2_K - m^2_{\pi}} + \frac{ f(m_{\eta},\Delta) - f(m_K, \Delta)  }{m^2_{\eta} - m^2_{K}} \right) 
\right\}\,\,\,, \label{g0nksNLO} \\
\sqrt{3}\, \bar g^{(1)}_{0\, N\Lambda K} &=& B m_* \bar\theta \left\{ - \frac{3 (D^2 - 6 D F - 3 F^2)}{16 \pi F^2_0} \frac{m_K^2 + m_{\pi} m_K + m_{\pi}^2}{m_K + m_{\pi}} 
\right.  \nonumber \\ & & \left. + \frac{(D-F) (D + 3 F)}{16 \pi F^2_0} \frac{m^2_{\eta}+ m_{\eta} m_K + m^2_K}{m_K + m_{\eta}}   
+ \frac{\mathcal C^2}{8 \pi^2 F^2_0} \frac{ f(m_{K},\Delta) - f(m_{\pi}, \Delta)  }{m^2_{K} - m^2_{\pi}}
\right\}\,\,\,. \nonumber \\  \label{g0nklNLO}
\end{eqnarray}

\subsection{Testing the relations at NLO}
Armed with the NLO expressions for the baryon masses and the $\slashT$ nucleon-pNG couplings, we investigate the relations found in Section~\ref{relations}. We start with  $\bar g_0$,  which is of the largest phenomenological interest. We repeat the relations we want to test 
\begin{eqnarray}\label{testg0}
\bar g_0 &=&\delta m_N \frac{m_* \bar\theta}{\bar m \varepsilon} \nonumber\\
&=&  (\Delta m_{\Xi} - \Delta m_{\Sigma})\frac{2 m_*}{ m_s - \bar m}  \bar\theta\,\,\,.
 \end{eqnarray}
 As these relations hold at LO, it is sufficient to test the relation for the NLO corrections themselves. A comparison of Eq.~\eqref{Loop1a} and Eq.~\eqref{g01NLO} shows that it is possible to write
 \begin{equation}
\bar g^{(1)}_0  = 2 B \left( \frac{\delta^{(1)} m_N^{K}}{m_{K^0}^2 - m_{K^+}^2}  + \frac{1}{3} \left(\frac{\phi}{\sqrt{3}}\right)^{-1} \frac{\delta^{(1)} m_N^{\eta-\pi}}{m_{\eta}^2 - m_{\pi}^2}  \right) \;  m_* \bar\theta\,\,\,,
\end{equation}
where $\delta^{(1)} m_N^{K}$ ($\delta^{(1)} m_N^{\eta-\pi}$) denotes the pieces of Eq. \eqref{Loop1a} induced by the kaon mass splitting  ($\eta$ -- $\pi$ mixing).
Using the LO expression for the meson masses and mixing angle this simplifies into
\begin{equation}\label{isorel2}
\bar g^{(1)}_0  =  \delta^{(1)} m_N \, \frac{m_*}{\bar m \varepsilon} \bar\theta \,\,\,  .
\end{equation}
Thus, NLO corrections, both with octet and decuplet intermediate states, conserve the relation between $\bar g_0$ and $\delta m_N$.

Next we consider the second equality in Eq.~\eqref{testg0}. Using Eqs. \eqref{sigmaXi} and \eqref{sigmaSigma}, and expanding for simplicity the decuplet contributions in the limit $\Delta \rightarrow 0$, 
the NLO corrections to $\bar g_0$ can be expressed as
\begin{equation}\label{violation}
\bar g_0  = \left(  \left( \Delta^{(1)} m_{\Xi} - \Delta^{(1)} m_{\Sigma} \right)  + 
 \frac{3 (D^2 - 6 D F - 3 F^2) + \mathcal C^2}{288 \pi F_0^2}   ( m_K - m_{\pi})^2 ( m_K + 2 m_{\pi})  \right)  \frac{2 m_*}{m_s - \bar m} \bar\theta\,\,\,.
\end{equation}
So in addition to a term proportional to $\Delta^{(1)} m_{\Xi} - \Delta^{(1)} m_{\Sigma}$,
there is a second term that violates the relation. 
This second  term vanishes in the $SU(3)$ limit, $m_s = \bar m$, and is non-analytic in the quark masses.
The severity of the breaking is best illustrated by plugging in numerical values. Up to NLO it is possible to write
\begin{eqnarray}
\frac{\bar g_0}{ \Delta m_{\Xi} - \Delta m_{\Sigma} } &=& \left( 1 + \frac{3 (D^2 - 6 D F - 3 F^2) + \mathcal C^2}{288 \pi F_0^2} \frac{( m_K - m_{\pi})^2 ( m_K + 2 m_{\pi})}{\Delta m_{\Xi} - \Delta m_{\Sigma}} \right) \frac{2 m_*}{m_s - \bar m} \bar\theta\nonumber\\
&=&\left( 1 - 0.7 + 0.2  \right )\frac{2 m_*}{m_s - \bar m} \bar\theta\,\,\,,
\end{eqnarray}
where the second and third contributions in the second line are the octet and decuplet corrections, respectively.
We used the observed value of the $\Xi - \Sigma$ mass splitting, $\Delta m_{\Xi} - \Delta m_{\Sigma} = 124$ MeV \cite{Agashe:2014kda},   
and $F_{0} = F_{\pi}$, the difference being higher order.
We see that the tree-level relation is violated by a $50\%$ correction and thus it is unsuitable for a precise determination of $\bar g_0$.

Next we look at the $\slashT$ couplings in Eq.~\eqref{othertslash1}. Using the NLO results for octet masses, we conclude that the following relations survive NLO corrections:
\begin{eqnarray}
\sqrt{3} \bar g^{(1)}_{0\, \eta} &=& - \left(  \frac{\sigma_{N s}^{(1)}}{m_s} - \frac{\sigma^{(1)}_{N l}}{2 \bar m}\right)  4 m_* \bar\theta\,\,\,, \label{RelationsHeavy1} \\
\bar g^{(1)}_{0\, N \Sigma K}    &=& - \left( \Delta^{(1)} m_{\Sigma} - \Delta^{(1)} m_N  \right) \frac{2 m_* \bar\theta}{m_s - \bar m}\,\,\,, \label{RelationsHeavy2}\\
\sqrt{3}\bar g^{(1)}_{0\, N \Lambda K}    &=& - \left( \Delta^{(1)} m_{\Lambda} - \Delta^{(1)} m_N  \right) \frac{6 m_* \bar\theta}{m_s - \bar m}\,\,\,. \label{RelationsHeavy3}
\end{eqnarray}
The remaining LO relations in Eqs. \eqref{relnew1}, \eqref{relnew2} and \eqref{relnew3} 
are violated. 
We observe that NLO corrections do not spoil the relations if the baryons that enter the $\slashT$ vertices are the same as those appearing in the mass combinations, while relations to masses of baryons that are not involved in the $\slashT$ vertices are violated.

Finally we discuss $\bar g_1$. At tree level, $\bar g_1$ is closely related to $\bar g_{0\, \eta}$, $\bar g^{(0)}_1 = \bar g^{(0)}_{0\, \eta} \phi$ as can be seen from  Eqs.~\eqref{eq:g1su3} and \eqref{relnew1} . This can be understood because at this order $\bar g_1$ is induced by the emission of a $\eta$ meson by the nucleon and the consequent mixing of the $\eta$  with a neutral pion. As was the case for $ \bar g^{(1)}_{0\, \eta}$, the first two  relations of Eq.~\eqref{eq:g1su3} are violated at NLO, however, in this case also the third relation is violated. At this order, $\bar g_1$ receives a contribution identical to  $\bar g^{(1)}_{0\,\eta}$, but in addition there are contributions from the kaon mass splitting and from $\eta$ -- $\pi$
mixing in the internal pion and $\eta$ propagators. Neglecting in this discussion the decuplet correction, we find
\begin{eqnarray}\label{g1new}
\bar g^{(1)}_{1} &=& \bar g^{(1)}_{0\, \eta}\phi + B m_*\bar\theta \, \left( \frac{5 D^2 - 6 D F + 9 F^2}{32 \pi F^2_0}(m_{K^0}- m_{K^+}) 
\right. \nonumber \\ & & \left. + \frac{1}{24 \pi F^2_0} \frac{ (D^2 + 6 D F - 3 F^2) (m_{\eta} - m_{\pi})^2 + 6 (D^2 + 3 F^2) (m^2_{\eta} - m^2_{\pi})  }{m_{\eta} + m_{\pi}}  \frac{\phi}{\sqrt{3}}
\right)\,\,\,.
\end{eqnarray}
The first piece, proportional to $\bar g_{0\,\eta}$, respects the relation to the nucleon sigma term. On the other hand, we were not able to find any useful relation respected by the remaining piece of Eq. \eqref{g1new}. As numerically the violation of the relation is of similar size as the $\bar g^{(1)}_{0\, \eta}\phi$ part, the tree-level relation is of little use.

To assess the importance of corrections to the tree level value of $\bar g_1$, we evaluate Eqs. \eqref{FFloB} and \eqref{g11NLO}
using $b_D = 0.068$ GeV$^{-1}$ and $b_F = -0.209$ GeV$^{-1}$ (these values are discussed in Sect.~\ref{bEDMs}). We find
\begin{equation}\label{g1viol}
\frac{\bar g_1}{2 F_{\pi}} =  \left( 0.85   + 1.0 + 0.85 \right) \cdot 10^{-3} \, \bar\theta\,\,\,.
\end{equation}
The first number is the LO contribution. The second and third numbers are the octet and decuplet contributions to the NLO corrections, Eq. \eqref{g11NLO}.
We  see that NLO corrections are large, as big as the leading term, and the inclusion of the decuplet makes them even larger.
Since there is no surviving relation between $\bar g_1$ and baryon masses or sigma terms, and the $\chi$PT corrections show no sign of convergence, 
we conclude that in $SU(3)$ $\chi$PT there is no safe way to extract $\bar g_1$ from the baryon spectrum.

\section{Octet baryon masses and $\slashT$ couplings at N${}^2$LO}\label{N2LO}

In Section \ref{NLO} we have seen that NLO corrections affect the octet baryon mass splittings and nucleon $\slashT$ couplings in such a way that most LO tree-level relations beween 
 $\slashT$ couplings and baryon masses are violated. The exceptions are the relations between $\bar g_0$ and the nucleon mass splitting $\delta m_N$ and similar relations for the couplings $\bar g_{0\, N \Sigma K}$
and  $\bar g_{0\, N \Lambda K}$  to $\Delta m_{\Sigma} - \Delta m_{N}$ and $\Delta m_{\Lambda} - \Delta m_{N}$.  Furthermore, the link between $\bar g_{0\, \eta}$
and the nucleon sigma term also survives NLO corrections.
Of these couplings, $\bar g_0$ has the largest phenomenological impact as it contributes to the nucleon electric dipole form factor at LO and gives rise to the dominant piece of  the $\bar \theta$-induced $\slashT$ nucleon-nucleon potential. In Sections \ref{n2lo1} and  \ref{n2lo2} we therefore focus on $\delta m_N$ and $\bar g_0$. 
We discuss $\bar g_1$ in Section \ref{n2log1}, and the nucleon couplings involving $\eta$ and kaons in Section \ref{n2lo3}.
  
We show that again all loop corrections affect $\delta m_N$ and $\bar g_0$ in the same way, except for finite contributions that are quadratic in the isospin-breaking parameter $\varepsilon$ and thus numerically suppressed. All LECs from $\mathcal L^{(4)}$ that are needed to absorb divergences in the loops appear in the same way in $\bar g_0$ and $\delta m_N$. The relation between $\bar g_0$ and $\delta m_N$ is broken by additional, finite contributions to $\bar g_0$ from operators in $\mathcal L^{(4)}$, which do not contribute to $\delta m_N$. These contributions, however,  scale as 
$B^2 \bar m^2$ and not as $B^2  m_s \bar m$. Even though the values of these LECs are currently not known, they should not affect the $\bar g_0-\delta m_N$ relation in a significant way. 
Finally, $SU(3)$-breaking effects induce subleading pion and $\eta$ tadpoles, which contribute to $\bar g_0$ but not to $\delta m_N$, thus violating the relation.    
We estimate these violations and show that our results allow for a precise extraction of $\bar g_0$ from available lattice evaluations of the strong part of the nucleon mass splitting.

In what follows we calculate N${}^2$LO correction to the nucleon mass splitting (the nucleon average mass and the masses and mass splittings of the other octet baryons are given in Appendices \ref{AppA} and \ref{AppB}) and $\bar g_0$ including corrections due to the baryon decuplet. We keep terms linear in the quark mass difference, $\varepsilon$, neglecting $\mathcal O(\varepsilon^2)$ contributions. We comment on these corrections briefly at the end of the section. 

\subsection{Corrections to the nucleon mass splitting}\label{n2lo1}

\begin{figure}
\includegraphics[width=17cm]{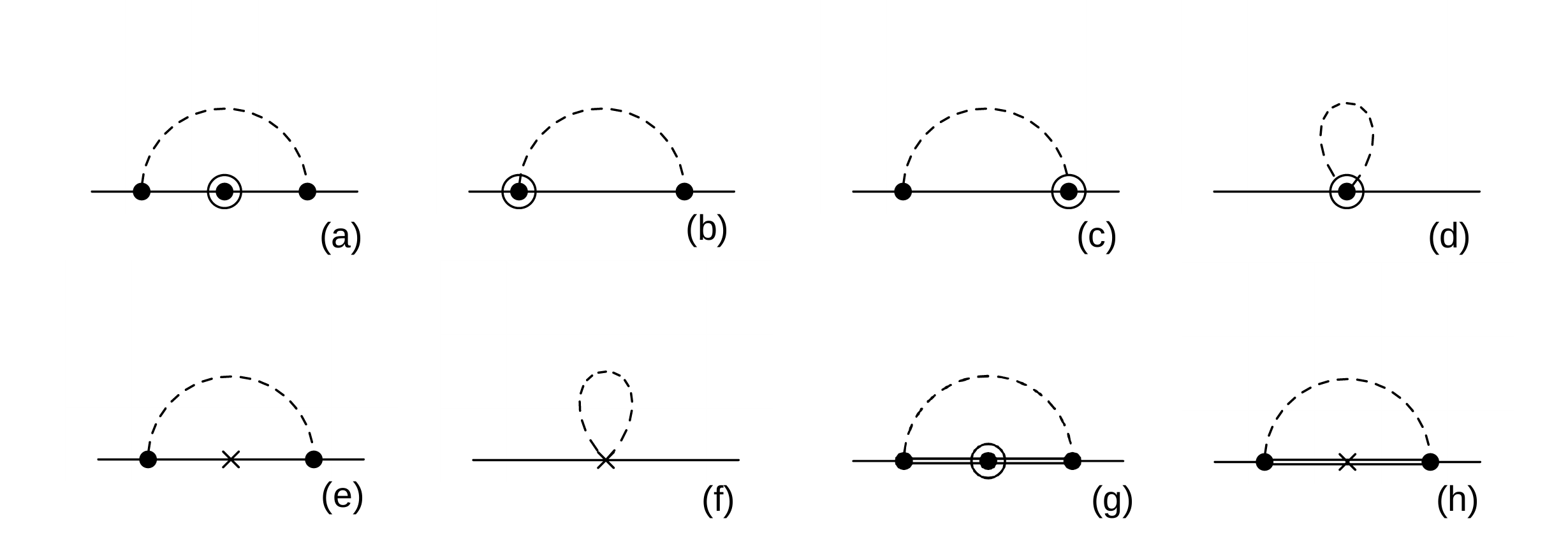}
\caption{$\mathcal O(q^4)$ corrections to the baryon octet mass splittings. Circled dotted vertices denote $SU(3)$ invariant couplings of Eqs.~\eqref{eq:L2pipi} and \eqref{eq:L2rel}. Crosses denote insertions of the octet and decuplet baryon mass terms of Eq. \eqref{eq:L2}. 
Other notation as in Fig.~\ref{Diag2}.}\label{Diag3}
\end{figure}

N${}^2$LO corrections to the nucleon and Delta mass splittings in $SU(2)$  $\chi$PT were considered in Ref. \cite{Tiburzi:2005na}.
In $SU(3)$ $\chi$PT N${}^2$LO octet masses and mass splittings were considered in Ref. \cite{Frink:2004ic}, in the infrared regularization scheme. 
Here we repeat the calculation in the heavy baryon formalism, and include corrections from the baryon decuplet.
At $\mathcal O(q^4)$, baryon masses receive corrections from loops  involving vertices in the Lagrangians $\mathcal L^{(2)}$, $\mathcal L^{(2)}_{\pi \pi}$, and $\mathcal L^{(2)}_{\textrm{rel}}$ 
in Eqs. \eqref{eq:L2}, \eqref{eq:L2pipi}, and \eqref{eq:L2rel},
and from tree-level insertions from the Lagrangian  $\mathcal L^{(4)}$ in Eq. \eqref{eq:L4}. The relevant loop diagrams are shown in Fig. \ref{Diag3}. 
Diagrams \ref{Diag3}(a) -- (f) show the contributions of octet intermediate states.
Diagram \ref{Diag3}(a) includes the correction to the propagator. It contributes to the mass splittings in two ways, through the kaon mass splitting or $\eta$ -- $\pi$ mixing,
and through the on-shell relation, which relates $v\cdot p$ to the mass of the external baryon in the diagram.
Diagrams \ref{Diag3}(b,c) contain recoil corrections to the axial couplings $D$ and $F$, and are proportional to $1/m_B$.
Diagram \ref{Diag3}(d) has a piece given by a recoil correction, and a piece proportional to the LECs $b_1$, $b_2$, and $b_3$. The LEC $b_8$ does not contribute to mass splittings, but only to isospin-averaged masses.
Diagrams \ref{Diag3}(e,f) have an insertion of the $SU(3)$ breaking couplings $b_D$, $b_F$, and $b_0$.

The diagrams in Fig. \ref{Diag3} are UV divergent and the divergences are absorbed by the counterterms in  $\mathcal L^{(4)}$. 
Of the operators defined in Eq. \eqref{eq:L4}, only $d_{1}$, $d_{2}$, $d_{3}$, $d_{5}$, and $d_{6}$ are relevant for mass splittings. Furthermore, $d_{i}$ satisfy the relation in Eq.~\eqref{rel1}, 
implying that there are only four independent counterterms.
We write
\begin{eqnarray}\label{N2LOct}
\delta m^{\textrm{ct}}_N &=&  \phantom{+} (4 B)^2 (2\bar m \varepsilon) ( \bar m \tilde{d}_1 + m_s \tilde d_2 ) \,\,\, ,\nonumber \\
\delta m^{\textrm{ct}}_{\Xi} &=& \phantom{+} (4 B)^2 (2\bar m \varepsilon)  ( \bar m \tilde{d}_3 + m_s \tilde d_4 )\,\,\,,  \nonumber  \\
\delta m^{\textrm{ct}}_{\Sigma} &=& - (4 B)^2 (2\bar m \varepsilon)  \left( \bar m (\tilde{d}_3 + \tilde d_1)+ m_s \left( \tilde d_4 + \tilde d_2 \right)\right)\,\,\, ,  
\end{eqnarray}
where we neglected terms of order $\varepsilon^2$, and $\tilde d_i$ are defined as
\begin{eqnarray}
\tilde d_1 &=&  - 2 (  d_1 + d_2  + d_3 + d_5 + d_6)\,\,\,,  \qquad \tilde d_2 =   \phantom{-} 2 d_1 - 2 d_3 - d_5 - d_6\,\,\, , \nonumber \\
\tilde d_3 &=&  \phantom{+} 2 (  d_1 - d_2  + d_3 - d_5 + d_6)\,\,\,, \qquad \tilde d_4 = - 2 d_1 + 2 d_3 - d_5 + d_6 \,\,\,.
\end{eqnarray}
Because the counterterms satisfy the Coleman-Glashow  relation,  the divergences of the diagrams in Fig. \ref{Diag3} must do so as well. We have explicitly checked that this holds and that, at N$^2$LO, Eq. \eqref{rel1} is only violated by the finite term \cite{Lebed:1994gt}
\begin{eqnarray}\label{CGviolation}
& & \delta^{} m_N + \delta m_{\Xi} + \delta^{} m_{\Sigma}  =     \frac{B ( b_F D^2 + 2 b_D D F )  }{ \pi^2 F_0^2} \left( (m_s- \bar m) m_K^2 \log \frac{m^2_{K^0}}{m^2_{K^+}} - 2 \bar m \varepsilon \, m^2_{\pi}  \log \frac{m^2_{K}}{m^2_{\pi}}    \right) \nonumber \\ 
& & - \frac{\mathcal C^2 \, B ( 9 b_D + 12 b_F + 7 b_C  )}{36 \pi^2 F^2_{0}} \left( (m_s - \bar m)  f^-_2(m_K,\Delta)  -  \bar m \varepsilon\, ( f^+_2 (m_K,\Delta) - 2 f_2(m_{\pi},\Delta)  )  \right)\,\,\,,
\end{eqnarray}
where the decuplet loop function $f^{}_2(x,y)$ is defined below in Eq. \eqref{fdecu}.

We now present our results for the nucleon mass splitting, after having subtracted the UV divergences  in the $\overline{\textrm{MS}}$  scheme.
In order to facilitate the comparison to loop corrections to $\bar g_0$, we split the N${}^2$LO corrections to $\delta m_N$ in three contributions.   
We start from diagrams \ref{Diag3}(b,c,d) and the piece of \ref{Diag3}(a) proportional to $m_B^{-1}$.
\begin{eqnarray} \label{N2LOabcd}
\delta m^{\textrm{(a,b,c,d)}}_N  & = &  \left(b_1 + b_2 + b_3\right)\frac{1}{8  \pi^2 F_0^2 } \left( m_{K^0}^4 - m_{K^+}^4  + m_{K^0}^4 \log \frac{\mu^2}{m^2_{K^0}} - m_{K^+}^4  \log \frac{\mu^2}{m^2_{K^+}} \right) \nonumber \\ & &
-\left(3 b_1 + b_2 - b_3\right)\frac{\phi}{\sqrt{3}} \frac{1}{4  \pi^2 F_0^2 } \left( m_{\pi}^4 - m_{\eta}^4  + m_{\pi}^4 \log \frac{\mu^2}{m^2_{\pi}} - m_{\eta}^4  \log \frac{\mu^2}{m^2_{\eta}} \right) \nonumber \\ & &
 - \frac{D^2 - 6 D F - 3 F^2}{96 \pi^2 F_0^2 m_B} \left(  m_{K^0}^4 \log \frac{\mu^2}{m^2_{K^0}} -  m_{K^+}^4  \log \frac{\mu^2}{m^2_{K^+}} \right) 
 \nonumber \\ & & 
 + \frac{(D - 3 F)(D + F)}{16 \pi^2 F_0^2 m_B} \frac{\phi}{\sqrt{3}} \left(  m_{\pi}^4 \log \frac{\mu^2}{m^2_{\pi}} -  m_{\eta}^4  \log \frac{\mu^2}{m^2_{\eta}}  \right) 
 \label{relativistic.1}\,\,\,. \end{eqnarray}
Here and in the following we omit the superscript $(2)$.
Diagrams \ref{Diag3}(a,e) give  
\begin{eqnarray}\label{N2LOae}
\delta m_{N}^{(\textrm{a,e})}  &=&  - 8 B \bar m \varepsilon\, (b_D + b_F) \frac{1}{16 \pi^2 F^2_0} 
\left\{ 
 (D+F)^2    m_{\pi}^2  \left(    1 +  3 \log \frac{\mu^2}{m_{\pi}^2}  \right) \nonumber \right. \nonumber  \\ 
& & \left.     
+   \frac{3 D^2 + 2 D F + 3 F^2}{6}  m_{K}^2   \left(1 + 3 \log \frac{\mu^2}{m_K^2}\right) 
-  \frac{m_s -\bar  m}{36 \bar m \varepsilon}    (D+ 3 F)^2 g_2(m_{K^0}, m_{K^+}) 
\right\}
\nonumber \\
& &  - 4 B  \bar m \varepsilon (b_D - b_F) \, \frac{D}{12 \pi^2 F_0^2}
\left\{ 
 (D-F) m_K^2 \left(1  + 3 \log \frac{\mu^2}{m_K^2} \right) \right. \nonumber \\  & & \left.
- \frac{m_s - \bar m}{6 \bar m \varepsilon}  (D - 3 F) g_2(m_{K^0}, m_{K^+})
\right\}\,\,\,,
\end{eqnarray}
where pion loops are proportional to the  combination $b_D + b_F$, which determines $\delta m_N$ at tree level, while kaon loops also give contributions proportional to $b_D - b_F$.
Only pion and kaon loops contribute to $\delta m_{N}^{(\textrm{a,e})}$. In the case of the $\eta$ and of  $\eta$ -- $\pi$ mixing, the  contributions of diagrams \ref{Diag3}(a) and \ref{Diag3}(e)
exactly cancel. 

Finally, diagram \ref{Diag3}(f) gives 
\begin{eqnarray}\label{N2LOf}
\delta m_{N}^{(\textrm{f})}  &=&  - 8 B \bar m \varepsilon\, (b_D + b_F) \frac{1}{32 \pi^2 F_0^2}
\left\{  m^2_{\pi}  \left(  1 +  \log \frac{\mu^2}{m_{\pi}^2}  \right)
+ \frac{ m_{\eta}^2}{3} \left(  1 + \log \frac{\mu^2}{m_{\eta}^2}  \right)  \right.  \nonumber \\
& & \left. - \frac{\bar m}{ ( m_s - \bar m) }\left(  m_{\pi}^2 - m_{\eta}^2 +  m_{\pi}^2 \log \frac{\mu^2}{m_{\pi}^2} - m_{\eta}^2 \log \frac{\mu^2}{m_{\eta}^2}  \right) \nonumber \right. 
\\ & & \left. + m_K^2 \left(  1  +  \log \frac{\mu^2}{m_K^2} \right) 
+ \frac{\bar m + m_s}{ 2 \bar m \varepsilon} g_1(m_{K^0}, m_{K^+})
\right\}\,\,\,.
\end{eqnarray}
The loop functions $g_1$ and $g_2$ are defined as
\begin{eqnarray}\label{f12}
g_{1}(m_{K^0}, m_{K^+}) &=& m_{K^0}^2 - m_{K^+}^2 + m_{K^0}^2 \log \frac{\mu^2}{m^2_{K^0}} - m_{K^+}^2 \log \frac{\mu^2}{m^2_{K^+}}\,\,\,,  \\
g_{2}(m_{K^0}, m_{K^+}) &=& m_{K^0}^2 - m_{K^+}^2 + 3 m_{K^0}^2 \log \frac{\mu^2}{m^2_{K^0}} - 3 m_{K^+}^2 \log \frac{\mu^2}{m^2_{K^+}}\,\,\,. 
\end{eqnarray}
Eqs. \eqref{N2LOabcd}, \eqref{N2LOae}, and \eqref{N2LOf} reproduce the results of Ref.~\cite{Frink:2004ic}.

We then consider N${}^2$LO  decuplet corrections  to the baryon octet mass splittings. The relevant diagrams are depicted in Fig. \ref{Diag3}(g,h). We give here the results for $\delta m_N$, while the contributions to the 
$\Xi$ and $\Sigma$ mass splittings can again be found in Appendix \ref{AppA}.

The  UV poles of the diagrams involving decuplet intermediate states have the form  
\begin{equation}
f^{\textrm{uv}}_2(m,\Delta) = - \left( 2\Delta^2 - m^2\right) \, L\,\,\,.
\end{equation}
The divergence proportional to $\Delta^2$ is absorbed by the $\mathcal O(\Delta^2)$  piece of the LECs $b_D$ and $b_F$,
while the divergence proportional to the quark mass is cancelled by the counterterms  $\tilde d_1$ and $\tilde d_2$.
After subtracting the UV poles, the decuplet contributions can be expressed in terms of the function
\begin{eqnarray}\label{fdecu}
 f_2(m,\Delta) &=&   - 2 \Delta^2  + ( m^2 - 2 \Delta^2 ) \log \frac{\mu^2}{m^2} - 4  \Delta \sqrt{m^2 - \Delta^2} \, \textrm{arccot} \frac{\Delta}{\sqrt{m^2 - \Delta^2}}\,\,\,,
\end{eqnarray}
and we define
\begin{equation}\label{fdecu2}
f^{\pm}_{2}(m_K,\Delta)   = (f_{2}(m_{K^0},\Delta) \pm f_{2}(m_{K^+},\Delta))\,\,\,.
\end{equation}
Diagram \ref{Diag3}(g) induces corrections to the mass splittings that are proportional to the LECs $b_0$, $b_D$, and $b_F$. 
\begin{eqnarray}\label{DecuPi}
\delta m^{(\textrm{g})}_{N} & =&  - \frac{ \mathcal C^2}{8 \pi^2 F_0^2} B \bar m \varepsilon\, (b_F + b_D) \,  \left(  
8 f_{2}(m_\pi,\Delta)
 +   f^+_{2}(m_{K},\Delta) \right) \nonumber  \\
 & & 
+ \frac{ \mathcal C^2}{24 \pi^2 F_0^2} B \left(  (m_s + 2 \bar m) b_0  +  \bar m ( b_D + b_F) + m_s (b_D- b_F) \right) f^-_{2}(m_{K},\Delta)  \,\,\,,
\end{eqnarray}
where we neglected relativistic corrections to the decuplet propagator. 
Diagram \ref{Diag3}(h) contains corrections that are induced by mass splittings of the decuplet, and are proportional to the LEC  $b_C$ in Eq. \eqref{eq:L2}
\begin{eqnarray}\label{DecuC}
\delta m^{(\textrm{h})}_{N} & =&      
- \frac{b_C\, \mathcal C^2}{36 \pi^2 F_0^2} B  \left(      2\bar m \varepsilon\, \left( 10 f_{2}(m_{\pi},\Delta) +  f^+_{2}(m_{K},\Delta) \right) - \frac{m_s + 2 \bar m }{2}  f^{-}_{2}(m_{K},\Delta) \right)\,\,\,.
\end{eqnarray}
Contributions proportional to $b_{\Delta}$ in Eq. \eqref{eq:L2} can be absorbed by a redefinition of $\Delta$ and we did not explicitly consider them.
If we neglect kaon loops, Eqs. \eqref{DecuPi} and \eqref{DecuC}  agree with the $SU(2)$ calculation of Ref. \cite{Tiburzi:2005na}.

\subsection{N${}^2$LO corrections to $\bar g_0$}\label{n2lo2}

\begin{figure}
\includegraphics[width=15cm]{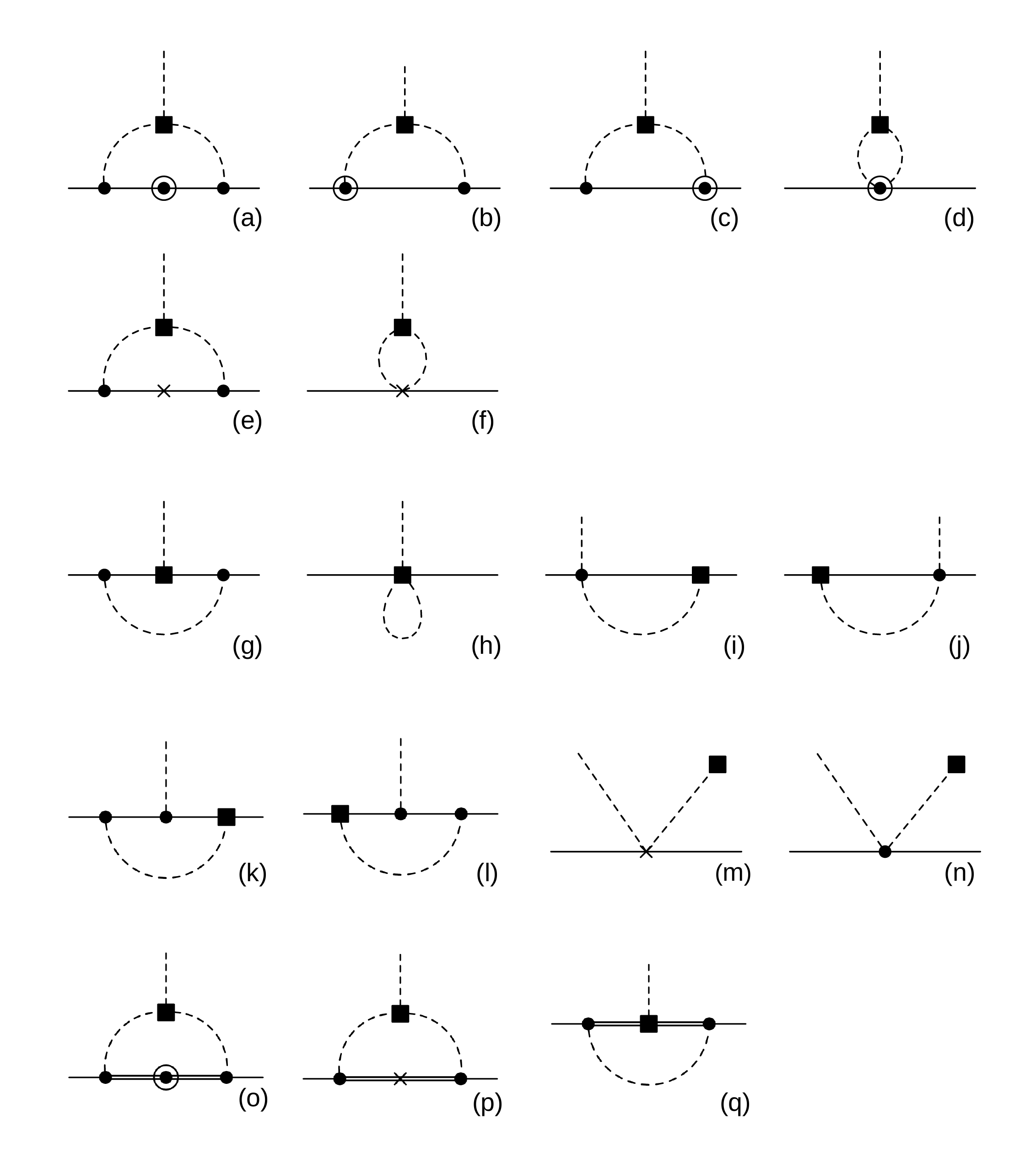}
\caption{N${}^2$LO contributions to $\slashT$ nucleon couplings. Squares denote $\slashT$ pNG and baryon-pNG couplings. Other notation is as in Figs.~\ref{Diag2}  and \ref{Diag3} . }\label{FigFF3}
\end{figure}

The N${}^2$LO corrections to $\bar g_0$ with octet and decuplet intermediate states are shown in  Fig.~\ref{FigFF3}.
In addition to these diagrams, we need the nucleon wave function renormalization $Z_N$, and the pion wave function renormalization, $Z_{\pi}$.
For the calculation of $\bar g_0$, it is sufficient to compute them in the isospin limit $m_{K^0} = m_{K^+} = m_K$ and $\phi  = 0$. Denoting
$\delta Z_N = Z_N - 1$ and $\delta Z_{\pi} = Z_{\pi} - 1$, we have
\begin{eqnarray}\label{Nwfr}
\delta Z_N & = &   3 (D+F)^2\frac{m^2_{\pi}}{64 \pi^2 F^2_0} \left( 1 + 3 \log \frac{\mu^2}{m^2_{\pi}} \right) +   ( 5 D^2 - 6 D F + 9 F^2)   \frac{m^2_{K}}{96 \pi^2 F^2_0} \left(1 + 3 \log \frac{\mu^2}{m^2_{K}}\right)
 \nonumber \\
& & + ( D - 3 F)^2\frac{m^2_{\eta}}{192 \pi^2 F^2_0} \left( 1 + 3 \log \frac{\mu^2}{m^2_{\eta}} \right) 
+ \frac{\mathcal C^2}{32 \pi^2 F^2_0} \left( f_2(m_K, \Delta) +  4 f_2(m_{\pi}, \Delta)\right)\,\,\,, \nonumber\\
\\
\delta Z_{\pi} &=&  - \frac{m^2_{\pi}}{24 \pi^2 F^2_0} \left(1 + \log \frac{\mu^2}{m_{\pi}^2}\right) - \frac{m_{K}^2}{48 \pi^2 F^2_0} \left( 1  +  \log \frac{\mu^2}{m_{K}^2} \right) 
- \frac{16 B}{F^2_0} \left( (m_s + 2 \bar m ) L_4 + \bar m L_5  \right)\,\,\,. \nonumber \\
\end{eqnarray}
Finally, at this order one has to consider the correction arising from expressing $F_0$ in the LO $\slashT$ pion-nucleon coupling in terms of $F_{\pi}$. 
At  N${}^2$LO, the relation between $F_0$ and $F_{\pi}$ is \cite{Gasser:1984gg}
\begin{eqnarray}
\delta F_{\pi} &  =& 1 - \frac{F_{\pi}}{F_0}  \nonumber \\
& = &  -  \frac{m^2_{\pi}}{16 \pi^2 F_0^2} \left(1 + \log \frac{\mu^2}{m_{\pi}^2}\right) - \frac{m_{K}^2}{32 \pi^2 F_0^2} \left(1 + \log \frac{\mu^2}{m_{K}^2}\right)
- \frac{8 B}{F^2_0} (  (m_s + 2 \bar m) L_4 + \bar m L_5 )\,\,\, . \nonumber\\ \label{FpiRen} 
\end{eqnarray}

We compute the diagrams in Fig.~\ref{FigFF3} with on-shell baryons, and an incoming pion with energy $v\cdot q$ and zero three-momentum $\vec q = 0$. The on-shell condition can be written as
\begin{eqnarray}
v\cdot K &=& \frac{1}{2} (v\cdot p + v\cdot p^{\prime}) = \frac{1}{2}(  m_i +  m_f  - 2 m_B  ) + \frac{\vec p^{\,\,2} + \vec p^{\,\,\prime\, 2}}{2 m_B}\,\,\,,\\
v\cdot q &=& (v\cdot p^{\prime} - v\cdot p^{}) =  m_f - m_i\,\,\, ,
\end{eqnarray}
where $m_{f,i}$ are the corrections to the masses of the baryons in the final and initial state, and $m_B$ is the common mass of the octet. 
For the diagrams in Fig.~\ref{FigFF3}, it is enough to use the tree-level expression of the baryon masses, which are given in Eqs. \eqref{tree} and \eqref{tree1}.

The counterterms are determined from Eq. \eqref{eq:L4}. Operators $d_1$ -- $d_6$ contribute in the same way to $\bar g_0$ and $\delta m_N$ and thus preserve the tree-level relation. Some of the remaining counterterms do spoil the $\bar g_0$ -- $\delta m_N$ relation. These corrections are discussed in Section \ref{sub}.

We move on to the loop diagrams, which we discuss in some detail. The contribution of  diagrams \ref{FigFF3}(a,b,c,d) to $\bar g_0$ can be written as 
\begin{equation}\label{g0N2LOabcd}
\bar g_0^{\textrm{(a,b,c,d)}} =   \delta m^{\textrm{(a,b,c,d)}}_N \, \frac{m_{*}}{\bar m \varepsilon} \bar\theta\,\,\,, 
\end{equation}
where we applied the LO expressions of the meson masses and mixing angle. In this expression  $\delta m_N^{\textrm{(a,b,c,d)}}$ is given in Eq. \eqref{N2LOabcd}. Both in Eq. \eqref{N2LOabcd} and in Eq. \eqref{g0N2LOabcd}, only the pieces of diagrams \ref{Diag3}(a) and \ref{FigFF3}(a) proportional to $1/m_B$ are considered. 

When we combine the contribution of diagram \ref{FigFF3}(a) proportional to $v \cdot K$, that on-shell becomes proportional to the average nucleon mass, the nucleon wave function renormalization $Z_N$, and diagrams \ref{FigFF3}(e,g), we obtain
\begin{eqnarray}\label{g0N2LOceg}
\bar g^{\textrm{(a,e,g)}}_{0} + \bar g^{(0)}_{0} \delta Z_N  =  \delta m^{\textrm{(a,e)}}_{N}  \, \frac{m_*}{\bar m \varepsilon} \bar\theta\,\,\,,
\end{eqnarray}
where $\bar g^{(0)}_{0}$ is given in Eq. \eqref{FFlo} and  $\delta m^{\textrm{(a,e)}}_{N}$  in Eq. \eqref{N2LOae}. 
Contributions from the $\eta$ meson in the loop diagrams \ref{FigFF3}(e,g) are cancelled by the sum of wave function renormalization and diagram \ref{FigFF3}(a), in the same way as it happens for the nucleon mass splitting. Diagrams \ref{FigFF3}(k,l) mutually cancel.

Then, we consider diagrams \ref{FigFF3}(f,h,i,j). These diagrams need to be combined
with one-loop corrections  to the pion wave function renormalization and to the  decay constant $F_{0}$. 
Considering these effects, we find
\begin{equation}\label{g0N2LOfhil}
\bar g_{0}^{\textrm{(f,h,i,j)}}   + \bar g^{(0)}_{0}\left( \frac{1}{2} \delta Z_{\pi}  - \delta F_{\pi} \right) =
  \delta m^{\textrm{(f)}}_{N} \, \frac{m_*}{\bar m \varepsilon} \bar\theta\,\,\,,  
\end{equation}
where $\delta m^{\textrm{(f)}}_{N}$ is in Eq. \eqref{N2LOf}.
Notice that the contribution of the LECs $L_4$ and $L_5$ cancels between the pion wave function renormalization and $\delta F_{\pi}$.

Finally we consider the decuplet corrections in Fig. \ref{FigFF3}(o,p,q). An explicit calculation shows  that diagrams \ref{FigFF3}(o) and \ref{FigFF3}(p,q) are in direct correspondence with  Fig. \ref{Diag3}(g) and \ref{Diag3}(h), respectively, once the decuplet corrections to the nucleon wave function renormalization are included. Thus N${}^2$LO decuplet corrections do not spoil the  $\bar g_0$ -- $\delta m_N$ relation.

The results in this section show that all UV-divergent contributions to $\bar g_0$ up to N${}^2$LO can be expressed in terms of the strong part of the nucleon mass splitting. However, at this order, we find some finite violations of the relation which we discuss now.

\subsubsection{N${}^2$LO violations of the relation between $\bar g_0$ and $\delta m_N$.}\label{sub}
There are three types of N${}^2$LO corrections   to $\bar g_0$ that cannot be written in terms of $\delta m_N$. The first of these corrections arises from additional counterterm contributions to $\bar g_0$. As discussed in the last section, the counterterms $d_1$ to $d_8$ in Eq. \eqref{eq:L4} conserve the relation,  while  $d_9$ -- $d_{16}$ can potentially spoil it. It can be seen that $d_9$ and $d_{12}$  do not contribute to $\slashT$ baryon-pNG couplings at this order, while $d_{15}$ and $d_{16}$ only contribute to $\bar g_1$.  The remaining correction can then be written as
\begin{equation}\label{ctviol}
\bar g^{\textrm{ct}}_0 = \delta m_N^{\textrm{ct}} \frac{m_* \bar\theta}{ \bar m \varepsilon}  - (4 B)^2\, \tilde d_{5} \, \left( 2 \bar m \right)  m_* \, \bar\theta\,\,\,  ,
\end{equation}
where $\tilde d_{5}$ is the combination
\begin{equation}
\tilde d_{5} =  2  d_{10} + 4 d_{11} + 3 d_{13} + 3 d_{14}\,\,\,.
\end{equation}
The combination of LECs $\tilde d_{5}$ is thus not related to mass splittings in the baryon spectrum, but could in principle be extracted from a precise analysis of nucleon-pion scattering. In practice, however, these LECs appear at too high order, and are not well constrained \cite{Hoferichter:2009gn}. This additional counterterm is present in $SU(2)$ $\chi$PT as well, where it also appears at N${}^2$LO \cite{Mereghetti:2010tp}. Both in $SU(2)$ and $SU(3)$, the additional contribution to $\bar g_0$ scales as $\bar m$ and not as $m_s$ in contrast to the terms in Eq.~\eqref{N2LOct}. Considering the good convergence of $SU(2)$ $\chi$PT, we expect these corrections to be of the expected size, $ m^2_{\pi}/\Lambda_\chi^2 $, of the order of a few percent.

The second type of contributions that violates the $\bar g_0$ -- $\delta m_N$ relation appears due to additional isospin violation. We have not calculated these contributions systematically, but give just one example. Diagram \ref{FigFF3}(a) induces, besides the component proportional to $v\cdot K$ discussed above, a contribution proportional to the energy transfer $v\cdot q$. Such a term gives rise to a correction to $\bar g_0$ that is quadratic in the quark mass splitting $\varepsilon^2$, which is not matched by an analogous correction to $\delta m_N$. This correction
is proportional to the tree-level pion-nucleon coupling, and we can write
\begin{equation}\label{dg0}
\frac{\delta \bar g^{\textrm{(c)}}_{0, v\cdot q}}{\bar g^{(0)}_0} =  \frac{ D^2 - 6 D F - 3 F^2}{256 \pi^2 F_0^2}   \frac{ m_{K^0}^4 - m_{K^+}^4 + 2 m_{K_0}^2 m_{K^+}^2 \log \frac{m^2_{K^+}}{m^2_{K^0}} }{m_{K^0}^2 - m_{K^+}^2}\,\,\,.
\end{equation}
Plugging in the values of the axial couplings and kaon masses,  Eq. \eqref{dg0} gives
\begin{equation}\label{dg0bis}
\frac{\delta \bar g^{\textrm{(c)}}_{0, v\cdot q}}{\bar g_0^{(0)}} \simeq -2\cdot10^{-6}\,\,\,,
\end{equation}
which is completely negligible. We do not expect significant corrections from the remaining $\mathcal O(\varepsilon^2)$ contributions that we did not compute.

Finally, the third type of violations arises from tadpole contributions, diagrams \ref{FigFF3}(m,n). In the $\mathcal O(q^4)$ meson Lagrangian, the operators $L_7$ and $L_8$
generate  pion and $\eta$ tadpoles. In addition, one has to consider one-loop diagrams with one external $\eta$ or $\pi_0$, and the $\slashT$ three-pNG vertex of Eq. \eqref{eq:2.7}.
Together with the two-pNG vertices from the baryon mass terms, the tadpoles generate contributions to $\bar g_0$ and $\bar g_1$.
Up to corrections of $\mathcal O(\varepsilon^2)$, the contribution to $\bar g_0$ arises only from $\eta$ tadpoles and is proportional to $\bar g^{(0)}_0$. We find
\begin{eqnarray}
\frac{\delta\bar g^{\textrm{(m,n)}}_0 }{\bar g_0^{(0)}}& = &    \frac{m^2_{\pi}}{3m_\eta^2} \left(  \frac{32 B}{F^2_0} ( L^r_8 + 3 L_7)( m_s -\bar m) 
+ \frac{1}{32 \pi^2 F_0^2} \left( 2 m^2_{K}  \log \frac{\mu^2}{m^2_K}  + m^2_{\eta} \log \frac{\mu^2}{m^2_\eta}    \right. \right. \nonumber \\ & & \left. \left. -3  m^2_{\pi}  \log \frac{\mu^2}{m^2_\pi}   \right)
\right) \,\,\,  ,  
\end{eqnarray}  
where here we have adopted the subtraction scheme of Ref. \cite{Gasser:1984gg} to define the renormalized coupling  $L_8^r$, while $L_7$ is not renormalized. 
Using the values of $L_7$ and $L_8$ discussed in Section \ref{Sec2Meson},  we can estimate the tadpole corrections to be
\begin{eqnarray}\label{dg0tad}
\frac{\delta \bar g^{\textrm{(m,n)}}_0}{\bar g_0^{(0)}} & = &  ( -0.5 \pm 1.3) \cdot 10^{-2} \,\,\, .
\end{eqnarray}
We thus expect the relation between $\bar g_0$ and $\delta m_N$ to hold up to a few percent. 

\subsection{N${}^2$LO corrections to $\bar g_1$}\label{n2log1}

We have shown that already at NLO no relations between $\bar g_1$ and baryon masses survive. However, in $SU(2)$ $\chi$PT, where $\bar g_1$ only appears at N${}^2$LO, it can be shown that part of $\bar g_1$ can be related to the pion mass splitting induced by the quark mass difference \cite{Mereghetti:2010tp}. At the same order, there appears an unknown direct contribution to $\bar g_1$ which has been estimated to be small by use of resonance saturation techniques in Ref.~\cite{Bsaisou:2012rg}. In this section, we study this relation in $SU(3)$ $\chi$PT and study the effects of the $SU(3)$ LO and NLO contributions to $\bar g_1$ that are missing in $SU(2)$. 

The $SU(2)$ relation of $\bar g_1$ to the strong component of the pion mass splitting can be recovered by studying the tadpole diagrams, Fig.~\ref{FigFF3} (m,n),
and, in particular, the contribution of the pion tadpole.
In $SU(3)$ $\chi$PT the pion tadpole receives contributions from the LECs $L_7$ and $L_8$, and from one-loop diagrams with insertion of the $\slashT$ three-pNG vertex in Eq. \eqref{eq:2.7}.
When these contributions are combined, the coupling of the pion to the vacuum given by
\begin{equation}
\mathcal L_{\mathrm{tad}} = -  2 B m_* \bar\theta\,  f_{\textrm{tad}}\, F_0 \pi^0\,\,\, ,
\end{equation}
where the function $f_{\textrm{tad}}$ is
\begin{eqnarray}\label{tadpi}
f_{\textrm{tad}} &=&  \frac{48 B}{F_0^2} \bar m \varepsilon \, (3 L_7 + L_8^r) + \frac{1}{32 \pi^2 F_0^2} \left(
m_{K^0}^2 \log \frac{\mu^2}{m_{K^0}^2} - m^2_{K^+}  \log \frac{\mu^2}{m_{K^+}^2}  \right)    \nonumber \\ & & 
 + \frac{\phi}{\sqrt{3}} \frac{1}{32 \pi^2 F_0^2} \left(2 m_K^2 \log \frac{\mu^2}{m_K^2} + 3 m^2_{\eta} \log \frac{\mu^2}{ m_{\eta}^2} - 5 m^2_{\pi} \log \frac{\mu^2}{m^2_{\pi}}\right)\,\,\,.
\end{eqnarray}
It is possible to show that $f_{\textrm{tad}}$ is related to the pion mass splitting in the large $m_s$ limit, more precisely 
\begin{equation}\label{tadmass}
\lim_{m_s \rightarrow \infty} f_{\textrm{tad}} =   -\lim_{m_s \rightarrow \infty} \frac{\delta m^2_{\pi}}{m^2_{\pi} \varepsilon}\,\,\,,
\end{equation}
where $\delta m_{\pi}^2$ denotes the component of the pion mass splitting induced by $m_d - m_u$. The expressions of the pion mass and mass splitting at one loop can be found in Refs. \cite{Gasser:1984gg,Gasser:1982ap}.

The pion tadpole induces a correction to $\bar g_1$, of the form
\begin{eqnarray}
\delta \bar g_1^{(\textrm{m,n})} &=& - 8 B (2 b_0 + b_D + b_F) f_{\textrm{tad}}   m_* \bar\theta\,\,\,,
\end{eqnarray}
which, using Eqs. \eqref{tree1} and \eqref{tadmass},  can be expressed in terms of the light quark sigma term and $\delta m_{\pi}^2$ as 
\begin{eqnarray} \label{g1su2}
\delta \bar g_1^{(\textrm{m,n})} &=&  - \left( \bar m \frac{d}{d \bar m} \Delta^{(0)} m_N \right)  \left. \frac{\delta m^2_{\pi}}{ m^2_{\pi} } \right|_{m_s \rightarrow \infty} \frac{1-\varepsilon^2}{\varepsilon}\,\,\, ,    
\end{eqnarray}
where we expanded also $m_*/\bar m$ in the large $m_s$ limit.
Eq. \eqref{g1su2} is exactly  what is found in $SU(2)$ \cite{Mereghetti:2010tp}. 
Notice, however, that in $SU(3)$  it is not possible to express $\bar g_1$ in terms of the full pion mass splitting, but only of its large $m_s$ limit. 

To estimate the pion tadpole correction to $\bar g_1$, we use the extraction of the light quark sigma term of Ref. \cite{Durr:2011mp}. 
In this paper, the $SU(3)$ $\chi$PT expressions of the  nucleon mass and sigma term are fitted to lattice data,  and the LO contribution to the light quark nucleon sigma term
is found to be
\begin{equation}\label{sigmaLO}
\bar m  \frac{d \Delta^{(0)} m_N}{  d\bar m} = 65 \pm 19 \, \textrm{MeV}\,\,\,.
\end{equation}
Using Eq. \eqref{sigmaLO}, together with $L_7$ and $L_8$ discussed in Section  \ref{Sec2Meson}, the pion tadpole contribution to $\bar g_1$ is 
\begin{equation}\label{g1tad}
\frac{ \delta \bar g_1^{(\textrm{m,n})} } {2 F_{\pi}} = - \left(  1.9 \pm 3.7 \right) \cdot 10^{-3}  \, \bar\theta\,\,\,.
\end{equation}
The central value is in reasonable agreement with the estimate of Ref. \cite{Bsaisou:2014zwa}, while the larger errors
stem mainly from the uncertainties on $L_7$ and $L_8$, and the partial cancellation between $L_7$ and $L_8$ in the combination $3L_7 + L_8$. 

Eq. \eqref{g1su2} is only a subset of the N${}^2$LO corrections to $\bar g_1$. In addition one should consider the loop diagrams in Fig. \ref{FigFF3}, 
the $\eta$ tadpole, contributions from corrections to $\eta$ -- $\pi$ mixing, and N${}^2$LO counterterms.  We have not computed these corrections, since,  
as we discuss now, they are not particularly instructive.

Comparing Eqs.  \eqref{g1tad}  and  \eqref{g1viol}, we see that the contribution of the pion tadpole, which is formally N${}^2$LO, is comparable with the LO and NLO pieces of $\bar g_1$.
This is not surprising, since the LO and NLO terms vanish in the limit $m_{s} \rightarrow \infty$, and thus are suppressed by powers of $\bar m/m_s$. On the other hand, at N${}^2$LO
$\bar g_1$ starts to receive contributions that are finite in the $m_s \rightarrow \infty$ limit, pion tadpoles being one such example. Eqs. \eqref{g1viol} and \eqref{g1tad} show that
for $\bar g_1$ the suppression due to inverse powers of $m_s$ or $\Lambda_\chi$ is similar. An analogous observation was made for $\delta m^2_{\pi}$ which is also determined at LO by the $\eta$ -- $\pi$ mixing angle \cite{Gasser:1984gg}. 
Thus, we conclude that the  $SU(3)$ $\chi$PT power counting does not provide a good organizational principle for $\bar g_1$.
Currently the best possible estimates are those based on $SU(2)$ \cite{Bsaisou:2014zwa}, which, however, are also affected by large  uncertainties,
leaving the determination of this important coupling in an unsatisfactory status. For clarity we repeat the $SU(2)$ estimate of Ref.~\cite{Bsaisou:2014zwa} here 
\begin{equation}\label{g1german}
\frac{  \bar g_1^{SU(2)} } {2 F_{\pi}} = -(3.4 \pm 1.5)\cdot 10^{-3}\, \tb\,\,\,,
\end{equation}
which is partially based on a resonance saturation estimate of an unknown N${}^2$LO LEC \cite{Bsaisou:2012rg} and on the tree-level relation $\delta m_\pi^2 = (\varepsilon^2/4) \mpi^4/(m_K^2 - \mpi^2)$. 
The error  in Eq. \eqref{g1german} is perhaps slightly underestimated. Including the uncertainty  on the pion mass splitting, either through the estimate of higher-order corrections \cite{Gasser:1984gg},
or using the error on the extraction of Ref. \cite{Amoros:2001cp} $\delta \mpi^2=(87\pm 55)\,$ MeV${^2}$, would raise the $45$\% uncertainty in Eq. \eqref{g1german} to about $70$\%. We stress that even with this large uncertainty, the $SU(2)$ determination of $\bar g_1$ is incompatible with the LO $SU(3)$ tree-level estimate in Eq.~\eqref{g1viol}. 
Finally, we note that Eq.~\eqref{g1german} does not include yet a higher-order contribution arising from the $\slashT$ three-pion vertex\footnote{In $SU(3)$ $\chi$PT this contribution already appears at NLO and is identified with the piece proportional to $(D+F)^2 \mpi$ in Eq.~\eqref{g11NLO}. }. This contribution enhances the estimate in Eq.~\eqref{g1german} by roughly $50\%$ \cite{deVries:2012ab, Bsaisou:2014zwa} making the tree-level estimate of $\bar g_1$ even less reliable. 

The large uncertainty on $\bar g_1$ dominates the uncertainty of light-nuclear EDMs when expressed in terms of $\bar\theta$ directly \cite{Dekens:2014jka}. It also affects the EDMs of diamagnetic atoms, but there the nuclear uncertainty associated with the complicated nuclear many-body problem is still larger. Nevertheless more theoretical work on the size of $\bar g_1$ could significantly increase the precision of EDM analyses.

\subsection{Other $\slashT$ nucleon couplings}\label{n2lo3}

The relations of $\bar g_{0\, \eta}$, $\bar g_{0\, N \Sigma K}$, and  $\bar g_{0\, N \Lambda K}$ to 
the nucleon  sigma term, the nucleon--$\Sigma$ and nucleon--$\Lambda$  mass splittings can be checked in a way analogous to what done in Sections \ref{n2lo1} and \ref{n2lo2}
for $\bar g_0$ and $\delta m_N$.
The expression for  $\Delta m_N$, $\Delta m_{\Sigma}$, and $\Delta m_{\Lambda}$ are given in Appendix \ref{AppB}. In addition, one needs the  
expressions of the baryon and meson wave function renormalizations, and the expression of $F_K$ and $F_{\eta}$ as function of $F_0$, which we also give in Appendix \ref{AppB}.
After all the ingredients are put together, it is possible to show that the loop contributions to the $\slashT$ couplings satisfy
\begin{eqnarray}
\sqrt{3}  \bar g^{(2)}_{0\, \eta} & = & \bar g^{\textrm{(loop)}}_{0\, \eta}  + \bar g_{0\, \eta}^{(0)} \left( \delta Z_N  + \frac{1}{2} \delta Z_{\eta}      - \delta F_\eta \right)  =  -  \left( \frac{\sigma^{(2)}_{N s}}{ m_s} - \frac{\sigma^{(2)}_{N l}}{2  \bar m} \right)\, 4 m_* \bar\theta \,\,\,,  \label{eta2} \\
\bar g_{0\, N \Sigma K}^{(2)} &=&   \bar g^{\textrm{(loop)}}_{0\, N \Sigma K}  + \bar g^{(0)}_{0\, N \Sigma K} \left( \frac{1}{2} \left(\delta Z_N + \delta Z_{\Sigma} + \delta Z_{K} \right)     - \delta F_K \right)   \nonumber\\
&=& - \left( \Delta^{(2)} m_{\Sigma} - \Delta^{(2)} m_{N} \right) \frac{2 m_* \bar\theta}{m_s - \bar m} \,\,\,,  \label{SigmaK2}\\
\sqrt{3} \bar g_{0\, N \Lambda K}^{(2)} &=&   \bar g^{\textrm{(loop)}}_{0\, N \Lambda K}  + \bar g^{(0)}_{0\, N \Lambda K} \left( \frac{1}{2} \left(\delta Z_N + \delta Z_{\Lambda} + \delta Z_{K} \right)     - \delta F_K \right)    \nonumber
\\ & =& - \left( \Delta^{(2)} m_{\Lambda} - \Delta^{(2)} m_{N} \right) \frac{6 m_* \bar\theta}{m_s - \bar m}\,\,\,. \label{LambdaK2}
\end{eqnarray}
Eqs. \eqref{eta2}, \eqref{SigmaK2} and \eqref{LambdaK2} include all the loop corrections, with the exception of the  $v \cdot q$ contribution to diagram \ref{FigFF3}(a).
In the case of  $\bar g_{0\, \eta}$, $v\cdot q = 0$, and this contribution vanishes. For $\bar g_{0\, N \Sigma K}$ and $\bar g_{0\, N \Lambda K}$, this contribution 
violates the tree-level relations, as in the case of $\bar g_0$. However, for $\bar g_0$ the breaking is proportional to $\varepsilon^2$, while for couplings involving the nucleon and the 
$\Sigma$ or $\Lambda$ baryons it goes as  $(m_s - \bar m)^2$, and thus is potentially larger.
We find
\begin{eqnarray}
\frac{\delta \bar g_{0\, N \Sigma K,\, v\cdot q}}{\bar g^{(0)}_{0\, N \Sigma K}} = 
& &   \frac{1}{512 \pi^2 F^2_{0}} 
\left(
(D^2  -18 D F  + 9 F^2) \frac{m_K^4 - m_\pi^4 - 2 m_K^2 m_{\pi}^2  \log\frac{m_K^2}{m_{\pi}^2}  }{m_K^2 - m_{\pi}^2} \nonumber \right. \\
& & \left.+ 9 (D-F)^2 \frac{m_K^4 - m_\eta^4 - 2 m_K^2 m_{\eta}^2  \log\frac{m_K^2}{m_{\eta}^2}  }{m_K^2 - m_{\eta}^2}
\right) \simeq - 0.015\,\,\,, \\
\frac{\delta \bar g_{0\, N \Lambda K,\, v\cdot q}}{\bar g^{(0)}_{0\, N \Lambda K}}   &=   & -   \frac{1}{512 \pi^2 F^2_{0}} 
\left(
3 (3 D^2  + 2 D F  + 3 F^2) \frac{m_K^4 - m_\pi^4 - 2 m_K^2 m_{\pi}^2  \log\frac{m_K^2}{m_{\pi}^2}  }{m_K^2 - m_{\pi}^2} \nonumber \right. \\
& & \left.+   (D+ 3F)^2 \frac{m_K^4 - m_\eta^4 - 2 m_K^2 m_{\eta}^2  \log\frac{m_K^2}{m_{\eta}^2}  }{m_K^2 - m_{\eta}^2}
\right) \simeq - 0.035\,\,\,.
 \end{eqnarray}
Therefore, also for $\bar g_{0\, N \Sigma K}$ and $\bar g_{0\, N \Lambda K}$, these violations are only a few percent.

The other contributions that violate the tree-level relations are the tadpole diagrams, \ref{FigFF3}(m,n), and the counterterms $d_{9}$ -- $d_{16}$.
For the nucleon--$\Sigma$ and nucleon--$\Lambda$ couplings, the tadpole contribution is proportional to the tree level, and we can write
\begin{eqnarray}\label{dg0tadK}
\frac{\delta \bar g^{\textrm{(m,n)}}_{0\, N \Sigma K }}{\bar g_{0\, N \Sigma K }^{(0)}} & = & \frac{\delta \bar g^{\textrm{(m,n)}}_{0\, N \Lambda K }}{\bar g_{0\, N \Lambda K }^{(0)}} =   
 -  \frac{15 m^2_\eta - 7 m^2_{\pi}}{48  m_\eta^2} \left(  \frac{32 B}{F^2_0} ( L^r_8 + 3 L_7)( m_s -\bar m) \right.  \nonumber  \\ & & 
\left. + \frac{1}{32 \pi^2 F_0^2} \left( 2 m^2_{K}  \log \frac{\mu^2}{m^2_K} + m^2_{\eta}  \log \frac{\mu^2}{m^2_\eta}   -3  m^2_{\pi} \log \frac{\mu^2}{m^2_\pi}   \right)
\right)\,\,\,,
\end{eqnarray}
where, again, we have used the subtraction scheme of Ref. \cite{Gasser:1984gg}.  
Using the values of $L_7$ and $L^r_8$ discussed in Section \ref{Sec2Meson},
we find that the tadpole corrections amount  to no more than $10\%$.

In the case of the $\eta$, the tadpole corrections are not proportional to the tree level 
\begin{eqnarray}
\frac{\delta \bar g^{\textrm{(m,n)}}_{0\, \eta }}{\bar g_{0\, \eta }^{(0)}} &=&
- \left( 1- \frac{m^2_{\pi}}{3 m^2_{\eta}} + \frac{2 b_0 + b_D + b_F}{b_D - 3 b_F}\right)
 \left(  \frac{32 B}{F^2_0} ( L^r_8 + 3 L_7)( m_s -\bar m) \right.  \nonumber  \\ & & 
\left. + \frac{1}{32 \pi^2 F_0^2} \left( 2 m^2_{K}  \log \frac{\mu^2}{m^2_K}  + m^2_{\eta}  \log \frac{\mu^2}{m^2_\eta}   -3  m^2_{\pi} \log \frac{\mu^2}{m^2_\pi}   \right)
\right)\,\,\,. 
\end{eqnarray}
In this case, the estimate of the tadpole corrections is affected by larger uncertainties. Nonetheless,  using the tree-level values of $b_D$
and $b_F$, $b_D = 0.068$ GeV$^{-1}$ and $b_F = -0.209$ GeV$^{-1}$,
and expressing $ 2 b_0 + b_D + b_F$ in terms of the LO contribution to the light quark nucleon sigma term, Eq. \eqref{sigmaLO},
we get that the tadpole corrections to $\bar g_{0\, \eta}$ come in at $30\%$.

Finally, the operators in the Lagrangian $\mathcal L^{(4)}$  can also violate the tree-level relations. 
The counterterms $d_1$ -- $d_8$, which are needed to absorb the divergences in the baryon mass, respect the relations, as one expects.
The operators $d_9$ -- $d_{11}$ only contribute to the $\slashT$ couplings, and we find 
\begin{eqnarray}
\frac{\sqrt{3} \delta \bar g^{(\textrm{ct})}_{0\, \eta}}{m_* \bar\theta} &=& -(4 B)^2    (  6 d_{10} - 4 d_{11} + 9 d_{13} - 3 d_{14})  \, \frac{2}{3}( 2 m_s + \bar m  )   \nonumber  \\
& &  + (4 B)^2  8 \left( \frac{4}{3} d_{11} +  2 d_{14} + 3 d_{15} + d_{16} \right) (m_s - \bar m)  \label{ctother1}
, \\
\frac{ \delta \bar{g}^{(\textrm{ct})}_{0\, N \Sigma K}}{m_* \bar\theta}  & =&    (4 B)^2    (  2 d_{10} - 4 d_{11} + 3 d_{13} - 3 d_{14}) \, (m_s + \bar m )  , \\
\frac{ \delta \bar{g}^{(\textrm{ct})}_{0\, N \Lambda K}}{m_* \bar\theta} &=&     (4 B)^2    (  6 d_{10} + 4 d_{11} + 9 d_{13} + 3 d_{14}) \, (m_s + \bar m ) .  \label{ctother3}
\end{eqnarray}
We cannot give a precise estimate of the  corrections in Eqs. \eqref{ctother1}-\eqref{ctother3}, due to the ignorance of the LECs $d_{10}$, $d_{11}$, and $d_{13}$ -- $d_{16}$. 
However, we see that the corrections to $\bar g_{0\, \eta}$ scale, at most, as  $m^2_{\eta}/\Lambda_\chi^2$, and to $\bar g_{0\, N \Sigma K}$ and $\bar g_{0 \, N \Lambda K}$ as $m_K^2/\Lambda_\chi^2$. 
Taking $\Lambda_\chi = 4 \pi F_{\pi}$, the counterterm corrections should be around $15\%$ -- $20\%$, of similar size as the tadpole contribution.

\section{Discussion}\label{Numerics}

\subsection{Best estimates of the $\slashT$ couplings}\label{estimates}
In this section we give the best estimates of the $\slashT$ couplings by using the relations in Eqs. ~\eqref{isorel1a} and \eqref{RelationsHeavy1}-\eqref{RelationsHeavy3}  that do not suffer from large $SU(3)$ $\chi$PT corrections.  

The coupling $\bar{g}_0$ is related to the strong contribution the nucleon mass splitting for which there are now several lattice QCD calculations~\cite{Beane:2006fk,Blum:2010ym,Horsley:2012fw,deDivitiis:2013xla,Borsanyi:2013lga,Borsanyi:2014jba}.
The first three calculations~\cite{Beane:2006fk,Blum:2010ym,Horsley:2012fw} were performed with only a single lattice spacing and with pion masses $m_\pi \gtrsim 250$~MeV.
According to the FLAG Lattice Averaging Group standards~\cite{Aoki:2013ldr}, these results would not be included in averages of lattice QCD predictions.  The next calculation~\cite{deDivitiis:2013xla} was performed with $m_\pi \gtrsim 283$~MeV and four lattice spacings.  According to the FLAG criterion, these results would be included in an average, but perhaps do not have all the systematic under complete control.  
The final two calculations~\cite{Borsanyi:2013lga,Borsanyi:2014jba}, although performed by the same group, are independent from each other, are performed with multiple lattice spacings and with pion masses at or near their physical value.  Both calculations include effects on the splitting from QED. In the first case,  QED was not included in the sea quarks, while in the second, the entire calculation included the effects of QED.  The strong contribution to the nucleon mass splitting from both of these calculations would receive the ``green star'' from FLAG.

There are only a small number of results that pass the FLAG criterion.
For a final determination of the $m_d - m_u$ contribution to $\delta m_N$ from lattice QCD, one must have more results.  In this case, the exclusion of the first three results mostly stems from the use of a single lattice spacing.  In the mass splitting, the leading discretization effects exactly cancel, since the lattice regulator used in those works respects flavor symmetry.  We therefore chose to include all the results~\cite{Beane:2006fk,Blum:2010ym,Horsley:2012fw,deDivitiis:2013xla,Borsanyi:2013lga,Borsanyi:2014jba} to construct a lattice average.  However, we assign a weight penalty to the first three calculations.  This follows the averaging scheme in Refs.~\cite{Walker-Loud:2014iea,Junnarkar:2013ac}: the weight factor is chosen to be
\begin{equation}
w_i = \frac{y_i}{\sigma_i^2}\,\,\,,
\end{equation}
where $\sigma_i$ are the given statistical and systematic uncertainties in the given lattice QCD calculation combined in quadrature and $y_i = 1$ for the first three calculations~\cite{Beane:2006fk,Blum:2010ym,Horsley:2012fw}, $y_i=2$ for Ref.~\cite{deDivitiis:2013xla}, and $y_i = 3$ for the most recent two calculations~\cite{Borsanyi:2013lga,Borsanyi:2014jba}, with these weights chosen somewhat arbitrarily.  This weighted average yields
\begin{equation}\label{deltamnLatt}
\delta m_N = 2.49 \pm 0.17 \textrm{ MeV}\,\,\,.
\end{equation}
Substituting this in  Eq. \eqref{isorel1a} together with $\varepsilon$, Eq.~(\ref{eq:mq_flag}), we obtain
\begin{equation}\label{bestg0}
\frac{\bar g_0}{2F_{\pi}} = (15.5  \pm 2.0 \pm 1.6 ) \cdot 10^{-3} \, \bar\theta\,\,\,.
\end{equation}
The first uncertainty comes from combining those on $\delta m_N$ and $\varepsilon$ in quadrature, while the second is a conservative estimate of the theoretical error associated to the  N${}^2$LO corrections  discussed in Section \ref{n2lo2}. This estimate agrees with the recent determination in Ref.~\cite{Bsaisou:2014zwa} based on $SU(2)$ $\chi$PT. Because the relation to the nucleon mass splitting is preserved, the only difference compared to $SU(2)$ is due to the $\bar m/m_s$ correction in Eq.~\eqref{eq:1.3}, which is tiny.  In addition, the error in Eq.~\eqref{bestg0} is slightly larger than in Ref.~\cite{Bsaisou:2014zwa}  due to inclusion of higher-order chiral corrections.

The fact that the relations of $\bar g_{0\, N \Sigma K}$ and $\bar g_{0\, N \Lambda K}$ to, respectively, $m_\Sigma - m_N$
and $m_\Lambda - m_N$ are only violated by finite N${}^2$LO corrections,  allows for a reliable estimate of these couplings. 
In this case, the electromagnetic contribution to the isospin averaged masses is relatively small, and we can use the experimentally observed baryon masses,
$\Delta m_{\Sigma} = 1193$ MeV, $\Delta m_{\Lambda} = 1116$ MeV, and $\Delta m_{N} = 939$ MeV, with negligible experimental uncertainties.  We use the ratio  $m_s/\bar m$ in Eq.~(\ref{eq:mq_flag}) to obtain 
\begin{eqnarray}
\frac{\bar g_{0\, N \Sigma K}}{2 F_K}  &=& -\left( 36 \pm 1 \pm 11 \right) \cdot 10^{-3} \, \bar\theta \,\,\,, \label{bestsigma}\\
\frac{\bar g_{0\, N \Lambda K}}{2 F_K} &=& -\left( 44 \pm 1 \pm 13 \right) \cdot 10^{-3} \, \bar\theta\,\,\,. \label{bestlambda}
\end{eqnarray}
The first error is given by the errors on $m_s/\bar m$ and $\varepsilon$, while the second estimates the effects of the finite terms that break the relations. In this case the breaking scales as $m_K^2/\Lambda_\chi^2$ which we estimate at the $30\%$ level, see the discussion in Section~\ref{n2lo3}.

Finally, $\bar g_{0\, \eta}$ is expressed in terms of the nucleon sigma terms.
For the light quark sigma term, the most precise value is determined from low-energy $\pi N$ scattering with the most recent determination from Ref.~\cite{Hoferichter:2015dsa},
\begin{equation}\label{eq:sigl_piN}
\sigma_{Nl} = 59.1 \pm 3.5 \textrm{ MeV}\,\,\, .
\end{equation}
This number is consistent with  earlier $\chi$PT analyses \cite{Alarcon:2011zs, Alarcon:2012kn,Ren:2014vea}.
This quantity can also be determined with lattice QCD.
However, there is significantly larger uncertainty from the lattice determination arising from a few systematic issues.  The primary means to determine this quantity is invoking the Feynman-Hellman theorem, Eq.~(\ref{eq:FH}), with a large spread of results, see Ref.~\cite{Young:2013nn} for a recent review.
There is a surprising ``phenomenological'' pion mass dependence of the nucleon mass found in lattice QCD calculations~\cite{WalkerLoud:2008pj,Walker-Loud:2013yua} yielding $m_N \simeq 800$~MeV  $+ m_\pi$ over a wide range of pion masses, including the physical point.  This in turns provides an estimate of $\sigma_{Nl} = 67\pm5$~MeV.
The best lattice QCD calculation, which also would receive a ``green star'' from FLAG was performed with pion masses as light as $m_\pi \sim 190$~MeV~\cite{Durr:2011mp} with the result
\begin{equation}
\sigma_{Nl} = 39_{-8}^{+18}\, \textrm{MeV}\,\,\, .
\end{equation}
It will likely be years before lattice QCD results can compete with the dispersive $\pi N$ scattering determination of Ref.~\cite{Hoferichter:2015dsa} in Eq.~(\ref{eq:sigl_piN}).

For the scalar strange content of the nucleon, there is no close second to the lattice QCD determination, although the results are not yet mature.
Ref.~\cite{Junnarkar:2013ac} compared all recent lattice QCD calculations of $\sigma_{Ns}$~\cite{Young:2009zb,Oksuzian:2012rzb,Engelhardt:2012gd,Freeman:2012ry,Durr:2011mp,Horsley:2011wr,Semke:2012gs,Shanahan:2012wh,Ren:2012aj} (where only Ref.~\cite{Durr:2011mp} evaluated all systematics) and found a systematically low value, as compared with prior estimates from $SU(3)$ baryon $\chi$PT.  An average value was determined
\begin{equation}\label{eq:sigs_latt}
	\sigma_{N s} = 40 \pm 10 \textrm{ MeV}\,\,\, .
\end{equation}
Thus our estimate for the coupling using inputs from Eqs.~(\ref{eq:sigl_piN}) and (\ref{eq:sigs_latt}) is
\begin{equation}
\frac{\bar g_{0\, \eta}}{2 F_{\eta}}  =  \left( 115 \pm 8  \pm 35 \right) \cdot 10^{-3} \bar\theta\,\,\, 
\end{equation}
where the first error is from the uncertainty in the sigma terms and the second from the N${}^2$LO corrections discussed in Section~\ref{n2lo3}.

\subsection{A comment on baryon EDMs}\label{bEDMs}
The $\tb$-induced EDMs of the baryon octet  in the framework of three-flavored $\chi$PT have been studied in great detail in Refs.~\cite{ Borasoy:2000pq, Ottnad:2009jw, Guo:2012vf,Akan:2014yha} (for a calculation of the nucleon EDM arising from the CKM phase, see Ref.~\cite{Seng:2014lea}). In these works $U(3)_L\times U(3)_R$ $\chi$PT is applied to calculate the EDMs of the whole baryon octet up to NLO in the chiral power counting. Compared to the $SU(3)\times SU(3)$ $\chi$PT framework used in this work, the main difference arises from the dynamical inclusion of the $\eta'$ meson. However, as shown in Ref.~\cite{Guo:2012vf}, these effects can be absorbed in a redefinition of a counterterm contributing to the EDMs of the charged baryons $p$, $\Sigma^{\pm}$, and $\Xi^-$. 

In the case of $\CP$ violation induced by the  $\bar\theta$ term,  the first contributions to baryon EDMs arise at $\mathcal O(q^2/\Lambda_\chi^3)$.
At this order, the baryon EDMs depend on only two combinations of counterterms in addition to one-loop diagrams involving the $\slashT$ nucleon-pNG vertices $\bar g_0$,  $\bar g_{0\, N \Sigma K}$, and $\bar g_{0\, N \Lambda K}$, and analogous couplings of the other octet baryons to pions and kaons. At NLO, no new counterterms appear but several additional one-loop diagrams contribute with no additional LECs. 
At NLO, one finds the first contributions from neutral mesons, and thus from $\bar g_{0\, \eta}$ and $\bar g_1$. 
Corrections induced by $\bar g_1$, being proportional to $\bar m \varepsilon/m_s$,  are small, and have been neglected in Refs.~\cite{Ottnad:2009jw, Guo:2012vf}.
In $SU(2)$ $\chi$PT, $\bar g_1$ only contributes at N${}^3$LO \cite{Mereghetti:2010kp}. 
The baryon-pNG vertices are related to $b_D$ and $b_F$ using the tree-level relations in Eqs. \eqref{FFlo}, \eqref{treeg0eta}-\eqref{treeg0K2} and values for $\{b_D,\,b_F\}=\{0.068,\,-0.209\}\mathrm{GeV}^{-1}$  are obtained 
by fitting the baryon masses to the tree-level expressions in Eq. \eqref{tree1}.  The two unknown counterterms are fitted to lattice data of the neutron and proton EDMs and the EDMs of the other baryons are predicted. It is found that NLO contributions are significant, in particular for nonphysical large pion masses that are typically used in lattice evaluations of the nucleon EDM.  

\begin{figure}
\center
\includegraphics[scale=.8]{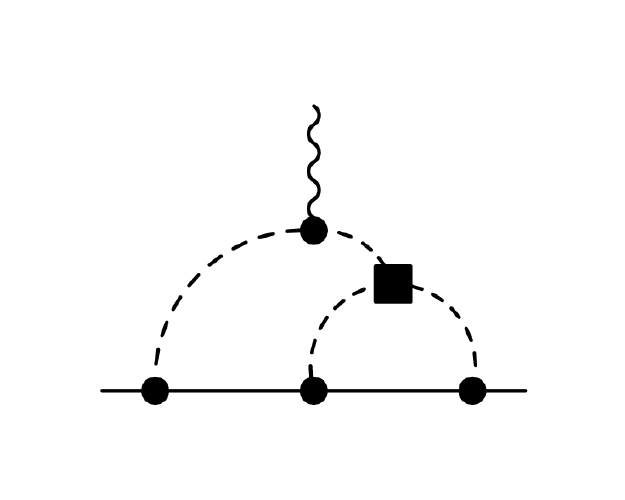}
\caption{An example of an NLO two-loop correction to baryon EDMs involving the $\slashT$ three-pNG vertex. Several other diagrams with different topologies appear at the same order. } \label{EDMcorrections}
\end{figure}

One class of diagrams has not been considered in these works. This class consists of diagrams involving the $\slashT$ three-pNG vertices. 
Despite being two-loop these diagrams are of order $\mathcal O(q^3/\Lambda_\chi^4)$, and thus contribute  to baryon EDMs at NLO\footnote{These diagrams are NLO because the three-pNG vertices appear at lower order than the tree-level $\slashT$ baryon-pNG vertices used in the LO one-loop diagrams. Although the extra loop comes with a suppression of $(q/\Lambda_\chi)^2$, the relative size of the three-pNG vertex brings in a factor $\Lambda_\chi/q$ making the two-loop diagrams genuine NLO corrections. }, and can be potentially large.
An example of such a two-loop diagram is shown in Fig.~\ref{EDMcorrections}. A two-loop calculation is beyond the scope of this work, but we note here that some of the two-loop diagrams involve a one-loop subpiece with the topology of Fig. \ref{Diag2}(c). Thus at least part of these corrections are taken into account by including one-loop corrections to $\bar g_0$, $\bar g_{0\, \eta}$, $\bar g_{0\, N \Sigma K}$, and $\bar g_{0\, N \Lambda K}$,
that is, they are  automatically taken  into account if the baryon-pNG couplings are obtained from the relation to baryon masses that survive higher-order corrections. 
Of course, a better assessment of the strangeness contribution to the nucleon EDM requires the full two-loop calculation.  In two-flavor $\chi$PT $\slashT$ three-pion couplings arise only at subleading order, implying that two-loop diagrams as the one depicted in Fig. \ref{EDMcorrections} only contribute at N${}^3$LO and can thus be neglected.

In Table \ref{comparison}, a comparison is given between the tree-level predictions in Eqs. \eqref{FFlo}, \eqref{treeg0eta}-\eqref{treeg0K2} using $\{b_D,\,b_F\}=\{0.068,\,-0.209\}\mathrm{GeV}^{-1}$  and the values obtained in Section~\ref{estimates}. 
The two predictions agree well for the couplings to kaons.   $\bar g_{0\, N \Sigma K}$ and $\bar g_{0\, N \Lambda K}$ are only  $10\%$ smaller, and the predictions agree within errors. 
The coupling to pions is more affected, being roughly $40\%$ smaller, with a smaller uncertainty (this point was already made in Ref.~\cite{Bsaisou:2012rg} based on LO $SU(3)$ $\chi$PT arguments). This can be easily understood, since using the values of $b_D$ and $b_F$ obtained by fits to the tree-level isospin-averaged octet masses is equivalent to use the relation of $\bar g_0$ to $\Delta m_{\Xi} - \Delta m_{\Sigma}$,
rather than the robust relation to $\delta m_N$.
Finally, using the nucleon sigma terms rather than the tree-level prediction in terms of $b_D$ and $b_F$ leads to a considerably larger $\bar g_{0\,\eta}$. This is a reflection of the poor convergence of the $SU(3)$
expansion for the nucleon sigma term. However, our best estimate is affected by a relatively large error due to unknown LECs that enter at N${}^2$LO.

We conclude that $SU(3)$ corrections to the baryon-pNG couplings  moderately alter the tree-level predictions. We do not expect that the $40\%$ shift in $\bar g_0$ significantly affects the nucleon EDM extractions performed in Ref.~\cite{Guo:2012vf,Akan:2014yha}, as the lattice data are not yet sensitive to non-zero values of $\slashT$ nucleon-pNG couplings  \cite{Mereghetti:2015rra}. (Notice, however, that the analysis of Ref.~\cite{Mereghetti:2015rra} did not include the more recent results of Ref. \cite{Guo:2015tla}.) 
For future extractions based on more precise lattice data with pion masses closer to the physical point, we recommend the values (and uncertainties\footnote{A part of the uncertainties on the couplings arise from N${}^2$LO corrections and thus are formally higher order than considered in the baryon EDM calculation. They were estimated in Refs.~\cite{Guo:2012vf,Akan:2014yha} by varying the renormalization scale appearing in the chiral logarithms.}) of the $\slashT$ couplings given in the third column of Table \ref{comparison}. 

Finally we discuss decuplet corrections to the nucleon EDMs, which have not been calculated in the literature. 
The leading $\slashT$ Lagrangian induced by $\bar\theta$, Eq. \eqref{eq:L2}, contains non-derivative $\slashT$ decuplet-pNG and octet-pNG couplings, 
but no $\slashT$ decuplet-octet-pNG couplings. Such couplings require, in order to conserve angular momentum, at least two derivatives, and thus are suppressed in the $\chi$PT power counting.
The lack of a $\slashT$ nucleon-decuplet-pNG vertex at the order  we work implies that there are no LO one-loop contributions to the nucleon EDM. 
In three-flavor $\chi$PT, two-loop contributions appear at NLO due to diagrams with similar topology as Fig.~\ref{EDMcorrections}, but with an internal decuplet propagator. However, as argued above, these corrections are partially taken into account if the $\slashT$ baryon-pNG couplings are inferred from the protected relations. We therefore do not expect decuplet corrections to play an important role in the study of baryon EDMs. We note that contributions from the decuplet EDMs to octet EDMs only appear at N${}^2$LO. 

\begin{table}[t]

\begin{center}\footnotesize
\begin{tabular}{||c||c|c||}
\hline
    \rule{0pt}{2.5ex}& Tree-level values & Values obtained here \\
& $[\times10^{-3} \,\bar{\theta}]$ & $[\times10^{-3}\, \bar{\theta}]$ \\
     \hline
    $\bar g_0/(2F_\pi)$ & $\phantom{-}26  $ & $\phantom{-}15.5\pm 2.5  $\\
     \hline
    $\bar g_{0\, \eta}/(2F_\eta)$ &$\phantom{-}56  $ & $\phantom{-}115\pm 37  $ \\
     \hline
    $\bar g_{0\, N \Sigma K}/(2F_K)$ &$-41  $ & $-36\pm11$\\
     \hline
    $\bar g_{0\, N \Lambda K}/(2F_K)$ &$-48  $ & $-44\pm13  $\\

 \hline
\end{tabular}
\end{center}
\caption{Comparison between tree-level predictions for $\slashT$ nucleon-pNG couplings using $\{b_D,\,b_F\}=\{0.068,\,-0.209\}\,\mathrm{GeV}^{-1}$ (see Refs.~\cite{Ottnad:2009jw, Guo:2012vf}), and the predictions from Section~\ref{estimates}. All values are in units of $10^{-3} \tb$. }

\label{comparison} 
\end{table}

\subsubsection{The nucleon Schiff moment}
An alternative way of extracting the $\slashT$ nucleon-pNG couplings is to study the momentum dependence of the nucleon electric dipole form factor (EDFF).
While the EDM gets both long-range and short-range contributions, the momentum dependence of the EDFF at NLO  is finite, and determined purely by loop diagrams. Short range effects enter only at N${}^2$LO. 
The full momentum dependence of the EDFF  is given in Refs.~\cite{Ottnad:2009jw, Guo:2012vf}, with the omission of  two-loop diagrams as the one depicted in Fig.~\ref{EDMcorrections}. 
Denoting the nucleon (proton) EDFF as $F_n(\vec q^{\,\, 2})$  ($F_{p} (\vec q^{\, \,2}) $)\,\,\,,
the Schiff moment  is defined as
\begin{equation}
S_{n,p} =  - \left. \frac{d F_{n,p}(\vec q^{\, 2})}{d \vec q^{\, 2}} \right|_{\vec q^{\, 2} = 0}\,\,\,,
\end{equation}
and is given at NLO by 
\begin{eqnarray}
S_n &=&   - \frac{e}{(4\pi F_{\pi})^2 } \left[  \frac{\bar g_0 g_A}{6 m^2_{\pi}} \left(1 -  \frac{5 \pi m_{\pi}}{4 m_N} \right) 
- \frac{\bar g_{0\, N \Sigma K} (D - F)}{6 m_K^2}
\left(1 - \frac{5 \pi m_K}{4 m_N} - \pi \frac{\Delta m_{\Sigma} - \Delta m_N}{2 m_K}\right)
\right]\,\,\,, \nonumber  \label{sn}\\
\\
S_p &=&     \frac{e}{(4\pi F_{\pi})^2 } \left[  \frac{\bar g_0 g_A}{6 m^2_{\pi}} \left(1 -  \frac{5 \pi m_{\pi}}{4 m_N} \right) 
+ \frac{\bar g_{0\, N \Sigma K} (D - F)}{12 m_K^2}
\left(1 - \frac{5 \pi m_K}{4 m_N} - \pi \frac{\Delta m_{\Sigma} - \Delta m_N}{2 m_K}\right)
\right . \nonumber \\ & &  \left. 
- \frac{\bar g_{0\, \Lambda N \Sigma}(D+ 3F)}{12 \sqrt{3} m_K^2}
\left(1 - \frac{5 \pi m_K}{4 m_N} - \pi \frac{\Delta m_{\Lambda} - \Delta m_N}{2 m_K}\right)
\right]\,\,\,, \label{sp}
\end{eqnarray}
where we expressed the results in Refs.~\cite{Ottnad:2009jw, Guo:2012vf} in terms of $\bar g_0$, $\bar g_{0\, N \Sigma K}$, and  $\bar g_{0\, N \Lambda K}$, 
and of the nucleon-$\Sigma$ and nucleon-$\Lambda$ mass splittings, and we used $D+F = g_A$. Considering only pion loops, these results agree with the $SU(2)$ calculations of Refs.~\cite{Thomas:1994wi,Hockings:2005cn,Mereghetti:2010kp}.
In $SU(3)$ $\chi$PT, the neutron and proton Schiff moments have two components. The contribution of pion loops is isovector and its typical scale is 
determined by the pion mass.  In addition, both neutron and proton receive contributions from kaon loops, which vary on the scale of the kaon mass. 
The neutron only receives contributions from loops involving the $\Sigma$ baryon, while the proton from both $\Sigma$ and $\Lambda$ intermediate states.
One can immediately see that the kaon contributions receive large NLO corrections, indeed larger than the LO, being  $\pi m_K/m_N \sim 2$. 
However, the neutron and proton Schiff moments receive their largest contributions from pion loops, so that in this case the poor convergence is not a big issue quantitatively.

Using the values of the $\slashT$ couplings in Table  \ref{comparison}, we find
\begin{eqnarray}\label{schiff}
S_0 &=& \frac{S_n + S_p}{2} =  - \left( 0.5  \pm 0.7 \right)   \cdot 10^{-5} \bar\theta \,  \, e \,\textrm{fm}^3\,\,\,,  \nonumber \\  
S_1 &=& \frac{S_p - S_n}{2} =  ( \left(7.6 \pm 1.4 \right)   - \left(2.4 \pm  0.6 \right) )\cdot 10^{-5} \bar\theta \,  \, e \,\textrm{fm}^3\,\,\, ,
\end{eqnarray}
where the isoscalar Schiff moment is given purely by kaon loops. For $S_1$ we have listed separately the contributions of pion and kaon loops.
The uncertainties only include the errors from the $\slashT$ couplings, while we do not give an estimate of the theoretical error from higher-order corrections.
Since the isovector Schiff moment is very sensitive to $\bar g_0$, using our best estimate in Eq. \eqref{bestg0} results in nucleon Schiff moments that are smaller by roughly a factor $2$ than found in Ref.~\cite{Ottnad:2009jw}. 
It is interesting  that even in $SU(3)$ $\chi$PT the Schiff moment is predominantly isovector, due to accidental cancellations between the pieces proportional to $\bar g_{0\, N \Sigma K }$ and
$\bar g_{0\, N \Lambda K }$. 

While a measurement of the nucleon EDFF is not going to happen in the foreseeable future, the predictions \eqref{sn}, \eqref{sp}, and \eqref{schiff} can be compared to lattice evaluations of the EDFFs.
In particular, lattice calculations performed at several values of light quark masses could provide enough information to disentangle the contributions from the pion-nucleon and kaon-nucleon $\slashT$ couplings,
which have different dependence on $\bar m$ and $m_s$. This would provide a method to check the values of the $\slashT$ couplings.

\subsection{A few comments on the $\slashT$ nucleon-nucleon potential}\label{potential}

\begin{figure}
\center
\includegraphics[scale=.8]{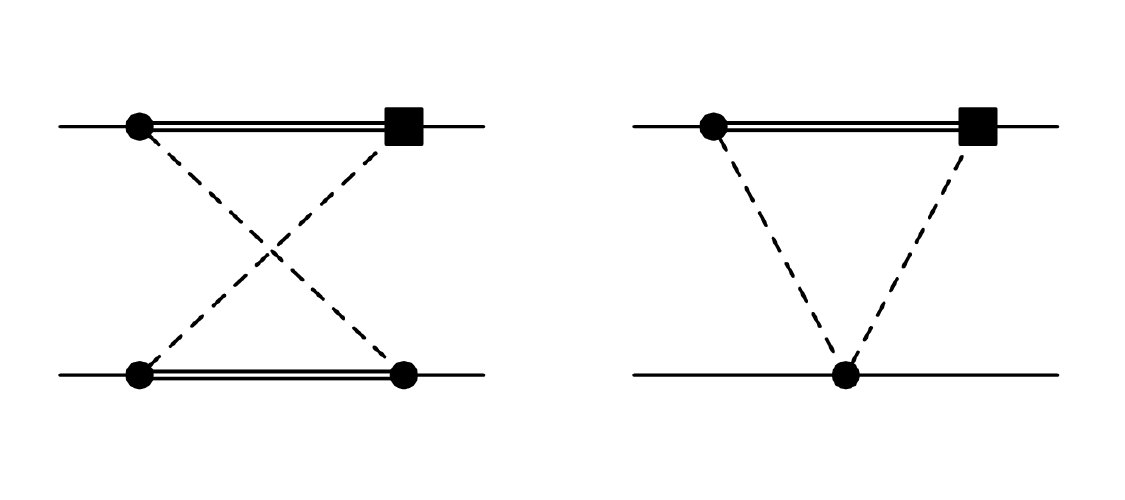}
\caption{Contribution of $\bar g_{0\, N \Sigma K}$ and $\bar g_{0\, N \Lambda K}$ to the nucleon-nucleon $\slashT$ potential. Single lines denote nucleon external states. Double lines denote $\Lambda$ or $\Sigma$ baryon propagators. 
Only one possible ordering per topology is shown.} \label{pot1}
\end{figure}

The EDMs of light nuclei and diamagnetic atoms obtain important contributions from the $\slashT$ nucleon-nucleon potential. In case of the QCD $\tb$ term, this potential is expected to be dominated by one-pion-exchange diagrams. In particular, because $\bar g_1$ is suppressed with respect to $\bar g_0$, often only the latter is taken into account. However, we stress that for certain $\slashT$ quantities such as the EDMs of the light nuclei ${}^2$H, ${}^6$Li, and ${}^9$Be \cite{Khriplovich:1999qr,Liu:2004tq,Yamanaka:2015qfa}, the ${}^{225}$Ra Schiff moment \cite{Engel:2013lsa}, or the ${}^{181}$Ta magnetic quadrupole moment \cite{Skripnikov:2015eia}, the nuclear matrix element for $\bar g_1$ is significantly larger than that for $\bar g_0$. In these cases $\bar g_1$ should be included in the analysis.  
The value of $\bar g_0$ extracted in Section \ref{estimates} can be immediately used in the existing calculations of EDMs that use the lowest-order chiral $\slashT$ potential
induced by the QCD $\bar\theta$ term \cite{Maekawa:2011vs}. On the other hand, as discussed in Section \ref{n2log1}, at the moment chiral symmetry only allows a determination of $\bar g_1$, and thus of the $\slashT$ isospin-breaking potential, with a relatively large uncertainty.

Nuclear EDMs can obtain important contributions from $\slashT$ nucleon-nucleon contact interactions. In case of strong $\CP$ violation, such interactions appear at N${}^2$LO only and are thus expected to be small. The corresponding potential is of the form \cite{Maekawa:2011vs}
\begin{equation}
V_{\textrm{SR}}(\vec q\,) = - \frac{i}{2} \left( \bar C_1  + \bar C_2 \boldtau^{(1)} \cdot \boldtau^{(2)} \right) \left( \vec\sigma^{(1)} - \vec\sigma^{(2)} \right) \cdot \vec q\,\,\,,
\end{equation}
where $\vec \sigma^{(i)}$ and $\boldtau^{(i)}$ denote the spin and isospin of nucleon $i$, and $\vec q$ is the momentum transfer $\vec q = \vec p - \vec p^{\,\prime}$, with $\vec p$ ($\vec p^{\, \prime}$) the center-of-mass momentum of the incoming (outgoing) nucleons.
$\bar C_{1,2}$ are LECs that in $SU(2)$ $\chi$PT scale as $\bar C_{1,2} = \mathcal O( \bar\theta  m^2_{\pi}/F_{\pi}^2 \Lambda_\chi^3)$.
$\bar C_2$ is needed to absorb the divergences in two-pion exchange diagrams (TPE), while $\bar C_1$ is not renormalized at this order \cite{Maekawa:2011vs}. 
The TPE contribution can be used to give a rough estimate of $\bar C_2$ \cite{Bsaisou:2014zwa},
\begin{equation}\label{C2est}
|\bar C_2| = \mathcal O\left( \frac{\bar g_0 g_A^3}{F^2_{\pi} (4 \pi F_{\pi})^2} \right) \simeq 2 \cdot 10^{-3} \,\bar\theta \, \textrm{fm}^3\,\,\,, 
\end{equation}
and $\bar C_1$ is expected to be of similar size. Clearly, this estimate is not very precise and below we derive an independent estimate of the sizes of $\bar C_{1,2}$.

For nuclear physics applications, the typical momentum transfer is  smaller than the kaon and $\eta$ masses. We can estimate the size of $\bar C_{1,2}$ by calculating contributions from $\bar g_{0\,\eta}$, $\bar g_{0\, N \Sigma K}$,
and $\bar g_{0\, N \Lambda K}$ to the potential, and expanding them in powers of  $| \vec q\,|/m_{K,\eta}$, where $|\vec q\,| \sim \mpi$. In principle the same can be done for decuplet corrections but these only appear at higher order because of the absence of a $\slashT$ nucleon-decuplet-pNG vertex (see the discussion at the end of Sect.~\ref{bEDMs}).

The coupling to $\eta$, $\bar g_{0\, \eta}$, contributes to the potential at tree level, providing a finite piece to $\bar C_1$. We obtain 
\begin{equation}\label{c1eta}
\bar C_1 = \frac{\bar g_{0\, \eta}}{2F^2 _{\eta}} \frac{D-3F}{\sqrt{3}} \frac{1}{m_\eta^2} = - 8 \cdot 10^{-3} \, \bar\theta \, \textrm{fm}^3  \,\,\,,
\end{equation}
which is in reasonable agreement with Eq.~\eqref{C2est} considering the uncertainty of that estimate. 

The couplings  $\bar g_{0\, N \Sigma K}$ and $\bar g_{0\, N \Lambda K}$ contribute to the nucleon potential only at one loop and are thus formally suppressed by two powers in the chiral counting compared to the one-pion-exchange contribution. Nevertheless, because the $\slashT$ coupling to kaons are somewhat larger than $\bar g_0$, their contribution might be sizeable. The corresponding diagrams are shown in Fig.~\ref{pot1}. They are similar to the TPE
diagrams studied in Ref. \cite{Maekawa:2011vs}, with the exception that only triangle and crossed diagrams are possible because  box diagrams are forbidden by strangeness conservation. 
We computed the diagrams in the limit $|\vec q\, | \ll m_K$, while keeping the mass difference between the nucleon and the strange baryons in the baryon propagator.
The diagrams contribute both to $\bar C_1$ and to $\bar C_2$. After cancelling the UV poles,  diagrams involving the $\slashT$ nucleon-$\Sigma$ coupling give
\begin{eqnarray}
\bar C_{1\, \Sigma} &=&   \frac{9 \bar g_{0\, N \Sigma K} \, (D-F) }{8 F^2_{K} (4\pi F_{K})^2}  \left(  
\left( 1 - \frac{3}{2} (D-F)^2 \right) v(m_K,\Delta_{\Sigma})  - \frac{(D+3F)^2}{6}  v(m_K,\Delta_{\Sigma \Lambda})  + \frac{2}{3}\right)\,\,\,, \nonumber \\
\bar C_{2\, \Sigma} &=&  - \frac{ \bar g_{0\, N \Sigma K} \, (D-F) }{8 F^2_{K} (4\pi F_{K})^2 } \left(   
\left( 1 + \frac{3}{2} (D-F)^2 \right) v(m_K,\Delta_{\Sigma})  - \frac{(D+3F)^2}{2}  v(m_K,\Delta_{\Sigma \Lambda})  + \frac{2}{3} \right)\,\,\,, \nonumber\\
\end{eqnarray}
while those with the $\slashT$ nucleon-$\Lambda$ coupling
\begin{eqnarray}
\bar C_{1\, \Lambda} &=&  - \frac{3 \bar g_{0\, N \Lambda K} \, (D+ 3 F) }{8 \sqrt{3} F^2_{K} (4\pi F_{K})^2}  \left(  
\left( 1 -  \frac{(D+3F)^2}{6} \right) v(m_K,\Delta_{\Lambda})  - \frac{3}{2}   (D-F)^2 v(m_K,\Delta_{\Sigma \Lambda})    + \frac{2}{3}\right)\,\,\,, \nonumber \\
\bar C_{2\, \Lambda} &=&  - \frac{ \bar g_{0\, N \Lambda K} \, (D+ 3F)  }{8  \sqrt{3} F^2_{K} (4\pi F_{K})^2}  \left( 
\left( 1 - \frac{(D+3F)^2}{2} \right) v(m_K,\Delta_{\Lambda})   + \frac{3}{2} (D-F)^2 v(m_K,\Delta_{\Sigma \Lambda })  + \frac{2}{3}\right)\,\,\,. \nonumber\\
\end{eqnarray}
The function $v(m_K,\Delta)$ is defined as
\begin{equation}
v(m_K,\Delta) = \left(\log \frac{\mu^2}{m_K^2}  - \frac{2}{3} - 2 \frac{\Delta }{\sqrt{m_K^2 - \Delta^2}}
\textrm{arccot} \frac{\Delta}{\sqrt{m_K^2 - \Delta^2}} \right)\,\,\,,
\end{equation}
and $\Delta_{\Sigma} = \Delta m_{\Sigma} - \Delta m_N$, $\Delta_{\Lambda} = \Delta m_{\Lambda} - \Delta m_N$.
The first diagram in Fig. \ref{pot1} can involve $\Sigma$ and $\Lambda$ intermediate states at the same time, 
and thus is a function of $\Delta_{\Sigma}$ and $\Delta_\Lambda$. In this case, we find that the loop function is well approximated by 
evaluating  $v$ in
$ \Delta_{\Sigma \Lambda} = (\Delta m_{\Sigma} + \Delta m_{\Lambda} - 2\Delta m_N)/2$, and ignoring contributions proportional to $\Delta m_{\Sigma} - \Delta m_{\Lambda}$.

We can estimate the contributions to the potentials by setting the scale $\mu = m_N = 939$ MeV, and using the values for the $\slashT$ couplings in Eqs. \eqref{bestsigma} and \eqref{bestlambda}
\begin{eqnarray}
\bar C_{1\, \Sigma} &=& -0.5 \cdot 10^{-3} \, \bar\theta \, \textrm{fm}^3, \, \qquad \bar C_{2\, \Sigma} = \hspace{10pt}  0.1 \cdot 10^{-3} \, \bar\theta \, \textrm{fm}^3\,\,\,, \\
\bar C_{1\, \Lambda} &=&  \hspace{10pt}  1.0 \cdot 10^{-3} \, \bar\theta \, \textrm{fm}^3, \, \qquad \bar C_{2\, \Lambda} = - 0.5 \cdot 10^{-3} \, \bar\theta \, \textrm{fm}^3\,\,\,. 
\end{eqnarray}
Thus, the contributions of the nucleon-kaon couplings to the $\slashT$ potentials are below the NDA estimate of $\bar C_{1,2}$ in Eq.~\eqref{C2est},
and significantly smaller than the contribution of $\bar g_{0\, \eta}$.

The contributions of $\bar C_{1,2}$  to the EDMs of $^3$He and $^3$H were studied in Refs. \cite{deVries:2011an,Bsaisou:2014zwa}.  
Unfortunately, these operators are very sensitive to the choice of the $T$-conserving strong-interaction potential and results vary by about an order of magnitude. 
Here we use the results of Ref. \cite{Bsaisou:2014zwa} obtained with the N${}^2$LO chiral potential  \cite{Epelbaum:2004fk},
which gives the largest dependence on $\bar C_{1,2}$. The EDM of  $^3$He 
is, ignoring uncertainties, given by
\begin{equation}
d_{^3\textrm{He}} =  0.9 d_n - 0.03 d_p +  \left( - 0.11 \frac{\bar g_0}{2 F_{\pi}}  -0.40  \, \bar C_1  F_{\pi}^3  + 0.88 \, \bar C_2 F_{\pi}^3 \right) \, e\,  \textrm{fm}\,\,\,.
\end{equation}
Focusing on the pieces proportional to $\bar g_0$ and $\bar C_1$, and using the estimates in Eqs. \eqref{bestg0}
and \eqref{c1eta}, we find
\begin{equation}
d_{^3\textrm{He}}-0.9 d_n + 0.03 d_p =   \left ( - 1.8  + 0.3  \right) \cdot 10^{-3} \bar\theta\; e \,\textrm{fm}\,\,\,,
\end{equation}
where the first (second) number is the contribution of $\bar g_0$ ($\bar C_1$).
The short-range potential provides a $15\%$ correction compared to the one-pion-exchange contribution\footnote{A similar conclusion can be drawn by inserting the value of the $\bar g_{0 \eta}$ coupling in the results of Ref.~\cite{Song:2012yh}  where $\eta$ exchange was considered explicitly.}, well within the $\mathcal O(20\%)$ uncertainties of the nuclear calculation of the latter (see Ref.~\cite{Bsaisou:2014zwa} for a more detailed discussion). It seems safe to neglect the short-range contributions to, at least, light-nuclear EDMs in case of the QCD $\tb$ term. This conclusion is in line with Refs.~\cite{deVries:2011an, Bsaisou:2014zwa}.

\section{Conclusion}\label{conclusion}

In this work we have investigated 
higher-order $SU(3)$ flavor-breaking corrections to the relations between $\slashT$ meson-nucleon couplings and baryon masses 
in the framework of $SU(3)$ $\chi$PT. 
For each isospin-invariant $\slashT$ nucleon-meson coupling induced by the QCD $\bar\theta$ term, we have identified one relation to baryon mass splittings or sigma terms which is not spoiled by loop corrections up to next-to-next-to-leading order.  
The determination of these couplings from spectroscopy is therefore independent of the convergence issues of $SU(3)$ baryon $\chi$PT through this order. 
In Sect.~\ref{estimates} we have used the conserved relations to derive precise values for $\slashT$ couplings. We recommend these values and corresponding uncertainties in future lattice QCD extractions of the nucleon EDM and in studies of nuclear $\slashT$ quantities such as EDMs, Schiff moments, and magnetic quadrupole moments.

The most important of these relations is the one linking the pion-nucleon coupling $\bar g_0$ to the nucleon mass splitting induced by $m_d - m_u$, the quark mass difference.
We find that all loop corrections, with the exception of small terms quadratic in $m_d - m_u$, affect $\bar g_0$ and $\delta m_N$ in the same way,
so that at N${}^2$LO we can express $\bar g_0$ in terms of $\delta m_N$, plus corrections that are not enhanced by chiral logarithms, and are not proportional to $m_s$.
We stress that, at this order, the effects of strangeness on $\bar g_0$ are completely buried in $\delta m_N$, and thus accounted for when using lattice calculations of $\delta m_N$  
with dynamical strange quarks.
We conservatively  estimate the impact of terms violating the relation between $\bar g_0$ and $\delta m_N$ to be about $10\%$. Thus, available lattice calculations of $\delta m_N$ allow 
to determine $\bar g_0$ with $15\%$ uncertainty.
Further reduction of the errors will require improvements of the lattice calculations of the nucleon mass splitting, but also the determination of the unknown LECs that enter at $\mathcal O(q^4)$, Eq. \eqref{ctviol}, which are not related to baryon masses. 
Using the tree-level relation to $\Delta m_{\Xi} - \Delta m_{\Sigma}$, as often done in the literature, overestimates $\bar g_0$ by about $50\%$.
It will be interesting to see if a direct extraction of $\bar g_0$ from the lattice, e.g. from the momentum dependence of the nucleon electric dipole form factor, will give a value compatible with Eq. \eqref{bestg0}.

Similarly, the $\slashT$ couplings of the nucleon to the $\eta$ meson, $\bar g_{0\,\eta}$, and the couplings involving kaons, $\bar g_{0\, N \Sigma K}$ and $\bar g_{0\, N \Lambda K}$,
are determined, respectively, by the nucleon sigma term, and by the mass differences of the nucleon and $\Sigma$ and $\Lambda$ baryons. 
In this case, the importance of the terms breaking the relation is larger, since they scale as $m_K^2/\Lambda_\chi^2$ rather than $m_\pi^2/\Lambda_\chi^2$, and the corrections are estimated to be $30\%$. These couplings contribute to the nucleon EDM at LO, but, as discussed in Section \ref{potential}, they do not considerably  affect 
the $\slashT$ nucleon-nucleon potential induced by the QCD $\tb$ term, and therefore play a minor role in the calculation of $\slashT$ nuclear observables. 

For the phenomenologically interesting coupling $\bar g_1$, all LO relations to baryon mass splittings and sigma terms obtain $\mathcal O(100\%)$ corrections already at next-to-leading order and higher-order corrections are even larger. We see no pattern of convergence and conclude that $SU(3)$ $\chi$PT does not provide a reliable method to extract a value of $\bar g_1$. This coupling plays an important role in many interesting $\slashT$ observables such as the deuteron EDM and the ${}^{225}$Ra Schiff moment, such that the lack of a robust relation to the baryon spectrum is unfortunate. In this case, the $SU(2)$ $\chi$PT extraction discussed in Sect.~\ref{n2log1} is more reliable but nevertheless suffers from a large uncertainty. This uncertainty could be reduced by more precise  evaluations of the pion mass splitting induced by the quark mass difference. 

As a byproduct of our study, we have obtained expressions for the octet baryon masses and mass splittings at N${}^2$LO in $SU(3)$ $\chi$PT.
For diagrams involving  octet intermediate state, our results reproduce the findings of Ref.~\cite{Frink:2004ic}. We also included the effects of the decuplet baryons on the mass splittings of the nucleon, $\Xi$, and $\Sigma$ baryons.
The N${}^2$LO expressions of the octet masses and mass splittings depend on several LECs, which cannot be determined purely from experimental data. 
Given the poor convergence/lack of convergence of $SU(3)$ baryon $\chi$PT~\cite{WalkerLoud:2008bp,Jenkins:2009wv,Torok:2009dg,WalkerLoud:2011ab}, it is not clear they can meaningfully be determined from a comparison with lattice QCD either.

In this work we have focused on strong $\CP$ violation, but it would be interesting to extend the study to higher-dimensional $\CP$-violating operators. In many scenarios of beyond-Standard-Model (BSM) physics, large nucleon and nuclear EDMs are induced by light-quark chromo-electric dipole moments (qCEDMs). However, the sizes of the nucleon EDMs and  $\slashT$ pion-nucleon couplings $\bar g_{0,1}$ are poorly known \cite{Pospelov:2001ys, Engel:2013lsa}, leading to large uncertainties in the analysis of EDM constraints on BSM physics (see for instance Ref.~\cite{Inoue:2014nva}). Just as for the $\tb$ term, it is possible to derive leading-order relations between qCEDM-induced $\slashT$ pion-nucleon couplings and baryon mass splittings induced by $\CP$-even quark chromo-magnetic dipole moments \cite{deVries:2012ab}, the chiral partners of the qCEDMs. The baryon mass splittings   can be evaluated on the lattice providing a method to accurately evaluate $\bar g_{0,1}$. However, the relations have only been studied at leading order and they might suffer from large higher-order corrections \cite{Chien}. Finally, a recent evaluation \cite{Fuyuto:2012yf} of the neutron EDM in $SU(3)$ $\chi$PT  found a much larger dependence on the strange qCEDM than previous studies based on QCD sum rules \cite{Pospelov:2000bw, Hisano:2012sc}. As qCEDMs typically scale with the quark mass, this would strongly impact neutron EDM constraints on BSM scenarios. However, the analysis of Ref.~\cite{Fuyuto:2012yf} is based on leading-order $SU(3)$ $\chi$PT and, as demonstrated in this work,  higher-order corrections might strongly affect the results. 

\section*{Acknowledgments}
We thank V. Cirigliano, W. Dekens, F.-K. Guo, C.~Hanhart, A.~Ritz, C.~Y.~Seng, U.~van~Kolck, and A.~Wirzba for useful discussions. We are grateful to F.-K. Guo, Ulf-G. Mei{\ss}ner, C.~Y.~Seng, and U.~van~Kolck for comments on the manuscript.
This work (JdV) is supported in part by the DFG and the NSFC
through funds provided to the Sino-German CRC 110 ``Symmetries and
the Emergence of Structure in QCD'' (Grant No. 11261130311).
The research of EM was supported by the LDRD program at 
Los Alamos National Laboratory.
The work of AWL is supported in part by the U.S. Department of Energy (DOE) contract DE-AC05-06OR23177, under which Jefferson Science Associates, LLC, manages and operates the Jefferson Lab and by the U.S. DOE Early Career Award contract DE-SC0012180.

\appendix 

\phantom{asd}

\section{N${}^2$LO corrections to mass splittings of the $\Sigma$ and $\Xi$ baryons}\label{AppA}

In this appendix we give the expression of the loop contributions to  $\delta m_{\Xi}$ and $\delta m_{\Sigma}$ at N${}^2$LO.
The counterterm contributions are given in Eq. \eqref{N2LOct}, and the violation to the Coleman-Glashow relation in Eq. \eqref{CGviolation}.
The loop functions  $g_1$ and $g_2$ are defined in Eq. \eqref{f12},
while  $f_2(x,y)$ and $f^{\pm}_{2}(x,y)$  in Eqs. \eqref{fdecu} and \eqref{fdecu2}. 
The octet contributions agree with Ref. \cite{Frink:2004ic}, while the decuplet corrections are new results.

\begin{itemize} 
\item $\Xi$
\end{itemize}
The contribution of the relativistic corrections and of the two-pion couplings $b_{1}$, $b_2$ and $b_3$ is
\begin{eqnarray}
\delta m^{\textrm{(a,b,c,d)}}_\Xi  & = &  - \left(b_1 - b_2 + b_3\right)\frac{1}{8  \pi^2 F^2_0 } \left( m_{K^0}^4 - m_{K^+}^4  + m_{K^0}^4 \log \frac{\mu^2}{m^2_{K^0}} - m_{K^+}^4  \log \frac{\mu^2}{m^2_{K^+}} \right) \nonumber \\ & &
+ \left( 3 b_1 - b_2 - b_3\right) \frac{\phi}{\sqrt{3}}  \frac{1}{4  \pi^2 F^2_0 } \left( m_{\pi}^4 - m_{\eta}^4  + m_{\pi}^4 \log \frac{\mu^2}{m^2_{\pi}} - m_{\eta}^4  \log \frac{\mu^2}{m^2_{\eta}} \right) \nonumber \\ & &
 + \frac{D^2 + 6 D F - 3 F^2}{96 \pi^2 F^2_0 m_B} \left(   m_{K^0}^4 \log \frac{\mu^2}{m^2_{K^0}} -  m_{K^+}^4  \log \frac{\mu^2}{m^2_{K^+}}  \right) \nonumber 
 \\ & & 
  - \frac{(D - F)(D + 3 F)}{16 \pi^2 F^2_0 m_B} \frac{\phi}{\sqrt{3}} \left(   m_{\pi}^4 \log \frac{\mu^2}{m^2_{\pi}} -  m_{\eta}^4  \log \frac{\mu^2}{m^2_{\eta}}  \right) 
 \label{relativistic.2}\,\,\,. 
\end{eqnarray} 
Loop corrections involving the operators $b_D$ and $b_F$ give
\begin{eqnarray}
\delta m_{\Xi}^{(\textrm{a,e})}  &=&  - 8 B (b_F - b_D) \bar m \varepsilon\, \frac{1}{16 \pi^2 F^2_0}
\left\{ 
(D-F)^2 m_{\pi}^2  \left( 1+  3 \log \frac{\mu^2}{m_{\pi}^2} \right) \right.  \nonumber\\
& & \left.+ \frac{3 D^2 - 2 D F + 3 F^2}{6} m_K^2  \left( 1 + 3 \log \frac{\mu^2}{m_K^2}\right)
- \frac{m_s - \bar m}{36 \bar m \varepsilon} (D- 3 F)^2\, g_2(m_{K^0}, m_{K^+})\right\} \nonumber \\
& &+ 4 B  \bar m \varepsilon\, (b_D + b_F)\, \frac{D}{12 \pi^2 F^2_{\pi}} \left\{ (D+F) \left( 1 + 3 \log \frac{\mu^2}{m_K^2}\right)
\right. \nonumber \\ && \left.- \frac{m_s - \bar m}{6 \bar m \varepsilon}(D + 3 F) g_2(m_{K^0}, m_{K^+})
\right\}\,\,\,, \\
\delta m_{\Xi}^{(\textrm{f})}  &=&  - 8 B (b_F - b_D) \bar m \varepsilon\,   \frac{1}{32 \pi^2 F^2_0}
\left\{  
 m_{\pi}^2 \left(1 + \log \frac{\mu^2}{m_{\pi}^2}\right) + \frac{m_{\eta}^2}{3} \left(1 + \log \frac{\mu^2}{m_{\eta}^2}\right) \right. \nonumber 
 \\
 & &  \left. - \frac{ \bar m }{m_s - m}   \left(  m_{\pi}^2 - m_{\eta}^2 +  m_{\pi}^2 \log \frac{\mu^2}{m_{\pi}^2} - m_{\eta}^2 \log \frac{\mu^2}{m_{\eta}^2}  \right) \right. \nonumber  \\ && \left.
+ m_K^2 \left(1 + \log \frac{\mu^2}{m_K^2}\right)  + \frac{m_s + \bar m}{2 \bar m \varepsilon} g_1(m_{K^0}, m_{K^+})
\right\}\,\,\,.
\end{eqnarray}
The decuplet contribution to the $\Xi$ mass splitting is
\begin{eqnarray}
\delta m^{(\textrm{g,h})}_{\Xi} & =&   B (b_D - b_F) \bar m \varepsilon\,  \frac{\mathcal C^2}{4 \pi^2 F^2_0} \left(  f_{2}(m_\pi,\Delta) + f_{2}(m_\eta,\Delta)  + \frac{3}{2} f^+_{2}(m_{K},\Delta) 
\right ) 
\nonumber \\
& & + \frac{\mathcal C^2}{2 \pi^2 F^2_0}  B \left(  b_0 (m_s +2 \bar m ) +  b_D (\bar m +  m_s) +  b_F ( m_s - \bar m) \right)   \nonumber\\
& & \times \left(
( f_{2}(m_\pi,\Delta) - f_{2}(m_\eta,\Delta) ) \frac{\phi}{\sqrt{3}} - \frac{7}{12} f^-_{2}(m_{K},\Delta) \right)
\nonumber \\
&& +   \frac{ b_C\, \mathcal C^2}{36 \pi^2 F^2_0}  B \bar m \varepsilon\, \left( f_{2}(m_{\pi},\Delta) - 3 f_{2}(m_{\eta},\Delta) - 2 f^+_{2}(m_{K},\Delta)  \right) \nonumber \\ 
&& + \frac{b_C \, \mathcal C^2 }{6 \pi^2 F^2_0} B \left (  ( 2 m_s + \bar m) \frac{\phi}{\sqrt{3}} \left( f_{2}(m_{\pi},\Delta) -  f_{2}(m_{\eta},\Delta) \right) 
-  \frac{2 \bar m + 19 m_s}{12} f^-_{2}(m_{K},\Delta) 
\right)\,\,\,. \nonumber\\
\end{eqnarray}

\begin{itemize}
\item $\Sigma$
\end{itemize}

The recoil corrections to $D$ and $F$ and the couplings $b_1$, $b_2$ and $b_3$ give
\begin{eqnarray} 
\delta m^{\textrm{(a,b,c,d)}}_\Sigma  & = &  - 2 b_2 \frac{1}{8  \pi^2 F^2_0 } \left( m_{K^0}^4 - m_{K^+}^4  + m_{K^0}^4 \log \frac{\mu^2}{m^2_{K^0}} - m_{K^+}^4  \log \frac{\mu^2}{m^2_{K^+}} \right) \nonumber \\
& &  +  b_2 \frac{1}{ 2 \pi^2 F^2_0 } \frac{\phi}{\sqrt{3}}\left( m_{\pi}^4 - m_{\eta}^4  + m_{\pi}^4 \log \frac{\mu^2}{m^2_{\pi}} - m_{\eta}^4  \log \frac{\mu^2}{m^2_{\pi}} \right) 
\nonumber \\
 & & - \frac{  D F }{8 \pi^2 F^2_0 m_B} \left(  m_{K^0}^4 \log \frac{\mu^2}{m^2_{K^0}} -  m_{K^+}^4  \log \frac{\mu^2}{m^2_{K^+}} \right) \nonumber \\
  & & + \frac{  D F }{4 \pi^2 F^2_0 m_B} \frac{\phi}{\sqrt{3}}\left(  m_{\pi}^4 \log \frac{\mu^2}{m^2_{\pi}} -  m_{\eta}^4  \log \frac{\mu^2}{m^2_{\eta}}  \right)\,\,\,. \label{relativistic.3} 
 \end{eqnarray}
Loop corrections involving the operators $b_D$ and $b_F$ are
\begin{eqnarray}
\delta m_{\Sigma}^{(\textrm{a,e})}  &=&    B   \bar m \varepsilon\, \frac{1}{12 \pi^2 F_0^2} 
\left\{ 
4 ( 2 b_D D F  +  b_F (  D^2 + 3 F^2)     )\, m^2_{\pi} \left( 1 + 3 \log \frac{\mu^2}{m_{\pi}^2} \right)  \right. \nonumber \\ & &  
\left. +  \left(  2 b_D  D F +  b_F (  D^2 + F^2   )  \right) \left( 6 m_K^2 \left(1 + 3 \log \frac{\mu^2}{m_K^2}\right)   
- 3 g_2(m_{K^0}, m_{K^+}) \frac{ m_s  - \bar m}{\bar m \varepsilon} \right) 
\right\}\,\,\,, \nonumber\\
\delta m_{\Sigma}^{(\textrm{f})}  &=&    B b_F  \bar m \varepsilon\,  \frac{1}{2 \pi^2 F_0^2}
\left\{  m_{\pi}^2 \left(  1+   \log \frac{\mu^2}{m_{\pi}^2} \right) + \frac{m_\eta^2}{3} \left( 1 + \log \frac{\mu^2}{m_{\eta}^2} \right) \right. \nonumber \\ & & \left.
- \frac{\bar m}{m_s - \bar m}  \left(  m_{\pi}^2 - m_{\eta}^2 +  m_{\pi}^2 \log \frac{\mu^2}{m_{\pi}^2} - m_{\eta}^2 \log \frac{\mu^2}{m_{\eta}^2}  \right) \right.  \nonumber \\ && \left.
+ m_K^2 \left(1 + \log\frac{\mu^2}{m_K^2} \right) + \frac{m_s + \bar m}{2 \bar m \varepsilon} g_1(m_{K^0}, m_{K^+})
\right\}\,\,\,.
\label{SigmaK1} 
\end{eqnarray}

The decuplet corrections are
\begin{eqnarray}
\delta m^{(\textrm{g,h})}_{\Sigma} & =&  \frac{\mathcal C^2}{ 6 \pi^2 F^2_0} B b_F  \bar m \varepsilon\, \left( 2 f_{2}(m_\pi,\Delta) +3 f_{2}(m_\eta,\Delta) + 5 f^+_{2}(m_{K},\Delta)  \right) \nonumber \\ 
& & - \frac{\mathcal C^2}{ 2 \pi^2 F^2_0 } \, B \left( b_0 ( m_s + 2 \bar m) + 2 \bar m b_D \right) \left(  \frac{\phi}{\sqrt{3}}   ( f_{2}(m_\pi,\Delta) - f_{2}(m_\eta,\Delta) )
- \frac{1}{2} f^-_{2}(m_{K},\Delta) \right) 
\nonumber \\
& &  + \frac{ b_C\, \mathcal C^2}{18 \pi^2 F^2_0}  B \bar m \varepsilon\, \left( f_{2}(m_{\pi},\Delta) + 3 f_{2}(m_{\eta},\Delta) + \frac{11}{2}  f^+_{2}(m_{K},\Delta) \right) \nonumber \\
& & - \frac{b_C\, \mathcal C^2 }{6 \pi^2 F^2_0} B  
\left( 
(m_s + 2 \bar m) \frac{\phi}{\sqrt{3}} \left( f_{2}(m_{\pi},\Delta) -  f_{2}(m_{\eta},\Delta) \right)
- \frac{7 \bar m + 2 m_s}{6}  f^-_{2}(m_{K},\Delta)
\right) \,\,\,.
\end{eqnarray}

\section{N${}^2$LO corrections to the octet average masses}\label{AppB}

In this Appendix, we give the corrections to the nucleon, $\Xi$, $\Sigma$ and $\Lambda$ average masses.
In order to verify the relations involving the couplings $\bar g_{0\, \eta}$, $\bar g_{0\, N \Sigma K}$
and $\bar g_{0\, N \Lambda K}$,  in addition to $\Delta^{(2)} m_N$, $\Delta^{(2)} m_\Sigma$ and $\Delta^{(2)} m_{\Lambda}$,
one needs the octet baryon wave function renormalization, the kaon and $\eta$ meson wave function renormalization,
and  the corrections to $F_K$ and $F_{\eta}$.

The corrections to the mesons wave function renormalization and decay constants are  \cite{Gasser:1984gg}
\begin{eqnarray}
\delta Z_K &=& - \frac{m_\pi^2}{64 \pi^2 F_0^2} \left( 1 + L_{\pi}\right) - \frac{m_K^2}{32 \pi^2 F_0^2} \left( 1 + L_K\right) - \frac{m_{\eta}^2}{64 \pi^2 F_{0}^2} \left(
1 + L_{\eta} \right) \nonumber  \\
& & - \frac{16 B}{F_0^2}\left( (m_s + 2 \bar m ) L_4  + \frac{1}{2} (m_s + \bar m) L_5\right)\,\,\,, \\
\delta Z_{\eta} &=& - \frac{m_K^2}{16 \pi^2 F^2_0} \left(1 + L_K \right) - \frac{16 B}{F_0^2}\left( L_4  (m_s + 2 \bar m )+ \frac{1}{3} (2 m_s + \bar m) L_5\right)\,\,\,,  \\
\delta F_{K} & =& - \frac{3 m_\pi^2}{128 \pi^2 F_0^2} \left(1 + L_{\pi}\right) - \frac{3 m_K^2}{64 \pi^2 F_0^2} \left(1 + L_K\right) - \frac{3 m_{\eta}^2}{128\pi^2 F_0^2}
\left(1 + L_{\eta}\right) \nonumber
\\
& &  - \frac{8 B}{F_0^2} \left( (m_s + 2 \bar m) L_4  + \frac{1}{2} (m_s + \bar m) L_5\right)\,\,\,, \\
\delta F_{\eta} & =& - \frac{3 m_K^2}{32 \pi^2 F_0^2} \left(1 +L_K\right)  - \frac{8 B}{F_0^2} \left( (m_s + 2 \bar m) L_4  + \frac{1}{3} (2 m_s + \bar m) L_5\right)\,\,\,,
\end{eqnarray}
where we  have introduce the shorthand $L_{i} = \log \mu^2/m_i^2$, for $i = \{ \pi, K, \eta\}$, in order to make the formulae in this Appendix more compact.

Then, we give the corrections to the baryon octet masses and wave function renormalization.
The result of diagrams \ref{Diag3}(a) -- \ref{Diag3}(f),  which involve octet intermediate states, agree with Ref. \cite{WalkerLoud:2004hf,Frink:2004ic}. 
The decuplet corrections agree with the results of Ref.  \cite{WalkerLoud:2004hf}. We also agree with \cite{Jenkins:1991ts}, after we expand in $\Delta$, and 
set the light quark mass $\bar m$ to zero, as was done in Ref. \cite{Jenkins:1991ts}.

\begin{itemize}
\item nucleon
\end{itemize}

The nucleon wave function renormalization is given in Eq. \eqref{Nwfr}.
The loop contributions to the nucleon mass, including decuplet corrections, are
\begin{eqnarray}
 \Delta^{} m_N &=& \frac{1}{96 \pi^2 F^2_0  m_B}  \left( ( 5 D^2 - 6 D F + 9 F^2) m_K^4 L_{K} + \frac{9}{2} (D + F)^2  m^4_{\pi} L_{\pi} 
+  (D - 3 F)^2 \frac{m^4_{\eta}}{2} L_{\eta}   \right)  \nonumber \\
 && +  \frac{1}{48\pi^2F_{0}^2}\bigg[6 (3b_1-b_2+3b_3+4 b_8)\, m_K^4 \left(1+L_K\right)\nonumber\\
&&+ (9b_1-3b_2+ b_3+6 b_8)\, m_\eta^4 \left(1+L_{\eta}\right)
+ 9 (b_1+b_2+b_3+2 b_8)\, m_\pi^4 \left(1+L_{\pi}\right)\bigg]  \nonumber \\
& & - B  (m_s - \bar m)\frac{m_K^2}{72 \pi^2 F^2_0} \left( (13 D^2 -30 D F + 9 F^2) b_D - 3   (5 D^2 - 6 D F + 9 F^2) b_F  \right) \nonumber \\ 
& & \times \left(1 + 3 L_K \right)  - B  (m_s + \bar m)\frac{m_K^2}{8 \pi^2 F^2_0} (4 b_0 +3 b_D - b_F) \left(1 +  L_K\right)   \nonumber \\
& &  - 3 B  \bar m \frac{m_\pi^2}{8 \pi^2 F^2_0} (2 b_0 +  b_D + b_F) \left(1 +  L_{\pi}\right) 
 \nonumber \\ 
& &- B \frac{m_\eta^2}{24 \pi^2 F^2_0} ( 4 m_s (b_0 + b_D - b_F) + \bar m (2 b_0 + b_D + b_F)) \left(1 +  L_{\eta}\right) \nonumber \\
& & - B \frac{\mathcal C^2}{8 \pi^2 F^2_0} ( (2 b_0 + b_D + b_F) \bar m + (b_0 + b_D - b_F) m_s ) ( 4 f_2(m_{\pi},\Delta) + f_2(m_{K},\Delta)  ) \nonumber \\
& & - B \frac{b_C \, \mathcal C^2}{24 \pi^2 F^2_0} ( (2 \bar m + m_s) f_2(m_K,\Delta) + 12 \bar m f_2(m_\pi,\Delta) ) \,\,\,.
\end{eqnarray}
The counterterm operators in $\mathcal L^{(4)}$ give 
\begin{eqnarray}
\frac{\Delta^{(\textrm{ct})} m_N}{(4 B)^2}  &=& - \left( (d_1-d_2+d_3-d_5+d_6+d_7+d_8) m_s^2  - (2 d_1 - 2 d_3 + d_5- 3 d_6-4 d_7) \bar m m_s \right. \nonumber  \\ & & \left. + (d_1+d_2+d_3 +2 d_5+ 2 d_6+4 d_7+2 d_8) \bar m^2  \right)\,\,\,.
\end{eqnarray}

\begin{itemize}
\item $\Xi$
\end{itemize}
The wave function renormalization of the $\Xi$ field, in the isospin limit, is
\begin{eqnarray}
\delta Z_{\Xi} &=&  ( 5 D^2 + 6 D F + 9 F^2)   \frac{m^2_{K}}{96 \pi^2 F^2_0} \left(1 + 3 L_K \right) +   ( D + 3 F)^2\frac{m^2_{\eta}}{192 \pi^2 F^2_0} \left( 1 + 3 L_{\eta} \right)
 \nonumber \\
& & +  3 (D-F)^2\frac{m^2_{\pi}}{64 \pi^2 F^2_0} \left( 1 + 3 L_{\pi} \right)  + \frac{\mathcal C^2}{32 \pi^2 F_0^2} \left( 3 f_2(m_K,\Delta) + f_2(m_\eta,\Delta) + f_2(m_{\pi},\Delta)  \right)\,\,\,. \nonumber \\
\end{eqnarray}
The loop contributions to the $\Xi$ mass, including decuplet corrections, are
\begin{eqnarray}
\Delta^{} m_\Xi &=& \frac{1}{96 \pi^2 F^2_0 m_B}  \left( ( 5 D^2 + 6 D F + 9 F^2) m_K^4 L_K 
+ \frac{9}{2} (D - F)^2 m^4_{\pi} L_{\pi} 
   +  (D +  3 F)^2 \frac{m^4_{\eta}}{2} L_{\eta}    \right)  \nonumber \\
 && + \frac{1}{48\pi^2F_{0}^2}\bigg[6 (3b_1+ b_2+3b_3+4 b_8) m_K^4 \left(1+L_K\right)\nonumber\\
&&+ (9b_1+ 3b_2+ b_3+6 b_8) m_\eta^4 \left(1+L_{\eta}\right) + 9 (b_1- b_2+b_3+2 b_8)m_\pi^4 \left(1+L_{\pi}\right)\bigg] \nonumber \\
&& - B  (m_s - \bar m)\frac{m_K^2}{72 \pi^2 F^2_0} \left( (13 D^2 + 30 D F + 9 F^2) b_D + 3   (5 D^2 + 6 D F + 9 F^2) b_F  \right) \nonumber \\ & & \times \left(1 + 3 L_K\right) 
- B  (m_s + \bar m)\frac{m_K^2}{8 \pi^2 F^2_0} (4 b_0 +3 b_D + b_F) \left(1 +  L_K\right) \nonumber \\
& & - 3 B  \bar m \frac{m_\pi^2}{8 \pi^2 F^2_0} (2 b_0 +  b_D - b_F) \left(1 +  L_\pi\right) \nonumber \\ 
& &- B \frac{m_\eta^2}{24 \pi^2 F^2_0} ( 4 m_s (b_0 + b_D + b_F) + \bar m (2 b_0 + b_D - b_F)) \left(1 +  L_{\eta}\right) \nonumber \\
& & -  B \frac{\mathcal C^2}{8 \pi^2 F^2_0} ( (2 b_0 + b_D - b_F) \bar m + (b_0 + b_D + b_F) m_s ) \nonumber \\ & & \times  (  f_2(m_{\pi},\Delta) + f_2(m_{\eta},\Delta) + 3 f_2(m_{K},\Delta)  ) \nonumber \\
& & - B \frac{b_C\, \mathcal C^2}{24 \pi^2 F^2_0}  ( ( \bar m + 2 m_s) (f_2(m_\pi,\Delta) + f_2(m_\eta,\Delta))+  ( 7 m_s +  2 \bar m) f_2(m_K,\Delta) ) .
\end{eqnarray}
The counterterm operators in $\mathcal L^{(4)}$ give 
\begin{eqnarray}
\frac{\Delta^{(\textrm{ct})} m_{\Xi}}{(4 B)^2}  &=& -  \left( (d_1 + d_2+d_3 + d_5+d_6+d_7+d_8) m_s^2 - (2 d_1 - 2 d_3 - d_5- 3 d_6-4 d_7) \bar m m_s \right. \nonumber  \\ & & \left. + (d_1-d_2+d_3 -2 d_5+ 2 d_6+4 d_7+2 d_8) \bar m^2  \right)\,\,\,.
\end{eqnarray}

\begin{itemize}
\item $\Sigma$
\end{itemize}
The wave function renormalization of the $\Sigma$ field, in the isospin limit, is
\begin{eqnarray}
\delta Z_{\Sigma} &=&  (  D^2 + F^2)   \frac{m^2_{K}}{16 \pi^2 F^2_0} \left(1 + 3 L_{K}\right)  + (D^2 + 6 F^2)\frac{m^2_{\pi}}{48 \pi^2 F^2_0} \left( 1 + 3 L_{\pi} \right) 
  \nonumber \\
& & +  D^2 \frac{m^2_{\eta}}{48 \pi^2 F^2_0} \left( 1 + 3 L_{\eta} \right)  + \frac{\mathcal C^2}{32 \pi^2 F_0^2} \left( \frac{10}{3} f_2(m_K,\Delta) + f_2(m_\eta,\Delta) + \frac{2}{3} f_2(m_{\pi},\Delta)  \right)\,\,\,. \nonumber\\
\end{eqnarray}
The loop contributions to the $\Sigma$ mass, including decuplet corrections, are
\begin{eqnarray}
\Delta^{} m_\Sigma &=& \frac{1}{48 \pi^2 F^2_0 m_B}  \left( 3(  D^2 + F^2) m_K^4 L_K   + (D^2 + 6 F^2) m^4_{\pi} L_{\pi}  + D^2 m^4_{\eta} L_{\eta}   \right)  \nonumber \\
&& + \frac{1}{24\pi^2F_{0}^2}\bigg[6 ( b_1 + b_3 + 2 b_8) m_K^4 \left(1+L_K \right) + ( 2 b_3 + 3 b_8) m_\eta^4 \left(1+L_{\eta}\right)\nonumber\\
&&+ 3 ( 4 b_1+ 2 b_3 + 3 b_8)m_\pi^4 \left(1+L_{\pi}\right)\bigg] \nonumber \\
& & +
B  (m_s - \bar m)\frac{m_\pi^2}{9 \pi^2 F^2_0}  b_D D^2   \left(1 + 3 L_{\pi}\right) \nonumber \\ & &
+ B  (m_s - \bar m)\frac{m_K^2}{4 \pi^2 F^2_0} \left( (D^2 + F^2) b_D + 2 D F b_F  \right)  \left(1 + 3 L_K\right) \nonumber \\ & & 
- B  (m_s + \bar m)\frac{m_K^2}{4 \pi^2 F^2_0} (2 b_0 + b_D) \left(1 +  L_K\right) 
 - 3 B  \bar m \frac{m_\pi^2}{4 \pi^2 F^2_0} ( b_0 +  b_D ) \left(1 +  L_{\pi}\right) \nonumber \\ 
& &- B \frac{m_\eta^2}{12 \pi^2 F^2_0} (  \bar m b_D  + b_0 (2 m_s + \bar m)) \left(1 +  L_{\eta}\right) \nonumber \\
& &  -  B \frac{\mathcal C^2}{24 \pi^2 F^2_0} ( 2 ( b_0 + b_D ) \bar m + b_0  m_s ) (  2 f_2(m_{\pi},\Delta) + 3 f_2(m_{\eta},\Delta) + 10 f_2(m_{K},\Delta)  ) \nonumber \\
& & - B \frac{b_C\, \mathcal C^2}{72 \pi^2 F^2_0} ( ( 2 \bar m +  m_s) (2 f_2(m_\pi,\Delta) +3 f_2(m_\eta,\Delta))+  2 ( 2 m_s +  13 \bar m) f_2(m_K,\Delta) )\,\,\, . \nonumber \\
\end{eqnarray}
The counterterm operators in $\mathcal L^{(4)}$ give 
\begin{eqnarray}
\frac{\Delta^{(\textrm{ct})} m_{\Sigma}}{(4B)^2}  &=& -  \left( (d_7 + d_8) m_s^2 + 2 (d_6 + 2 d_7) \bar m m_s + (4 d_3 + 4 d_6 +4 d_7 +2 d_8) \bar m^2  \right)\,\,\,.
\end{eqnarray}

\begin{itemize}
\item $\Lambda$
\end{itemize}

The wave function renormalization of the $\Lambda$ field, in the isospin limit, is
\begin{eqnarray}
\delta Z_{\Lambda} & = &
(  D^2 + 9 F^2)   \frac{m^2_{K}}{48 \pi^2 F^2_0} \left(1 + 3 L_K \right) + D^2 \frac{m^2_{\pi}}{16 \pi^2 F^2_0} \left( 1 + 3 L_{\pi} \right) +  D^2 \frac{m^2_{\eta}}{48 \pi^2 F^2_0} \left( 1 + 3 L_{\eta} \right)
 \nonumber \\
& & 
+ \frac{\mathcal C^2}{32 \pi^2 F_0^2} \left( 2 f_2(m_K,\Delta)  +  3 f_2(m_{\pi},\Delta)  \right)\,\,\,.  \nonumber\\
\end{eqnarray}
The loop contributions to the $\Lambda$ mass, including decuplet corrections, are
\begin{eqnarray}
\Delta^{} m_\Lambda &=& \frac{1}{48 \pi^2 F^2_0 m_B}  \left( (  D^2 + 9 F^2) m_K^4 L_K
 + 3  D^2  m^4_{\pi} L_{\pi}
+ D^2 m^4_{\eta} L_{\eta}   \right) \nonumber \\  
&&+  \frac{1}{24\pi^2F_{0}^2}\bigg[ 2 ( 9 b_1 + b_3 + 6 b_8)m_K^4 \left(1+L_K\right)+ 3 ( 2 b_3 +  b_8) m_\eta^4 \left(1+L_{\eta}\right) \nonumber\\
&& + 3 ( 2 b_3 + 3 b_8)  m_\pi^4 \left(1+L_{\pi}\right)\bigg] \nonumber \\
& &
- B  (m_s - \bar m)\frac{m_\pi^2}{3 \pi^2 F^2_0} \left( b_D D^2  \right)  \left(1 + 3 L_{\pi} \right) \nonumber \\ & &
- B  (m_s - \bar m)\frac{m_K^2}{36 \pi^2 F^2_0} \left( (D^2 + 9 F^2) b_D + 18 D F b_F  \right)  \left(1 + 3 L_K\right) \nonumber \\ & & 
- B  (m_s + \bar m)\frac{m_K^2}{12 \pi^2 F^2_0} (6 b_0 +5 b_D) \left(1 +  L_K\right)   -  B  \bar m \frac{m_\pi^2}{4 \pi^2 F^2_0} ( 3 b_0 +  b_D ) \left(1 +  L_{\pi}\right) \nonumber \\ 
& &- B \frac{m_\eta^2}{36 \pi^2 F^2_0} (  (\bar m + 8 m_s) b_D  + 3 b_0 (2 m_s + \bar m)) \left(1 +  L_{\eta}\right) \nonumber \\
& & - B \frac{\mathcal C^2}{24 \pi^2 F^2_0} ( 3 b_0 ( 2 \bar m + m_s) + 2 b_D (\bar m  + 2 m_s) ) (  3 f_2(m_{\pi},\Delta) + 2 f_2(m_{K},\Delta)  ) \nonumber \\
& & - B \frac{b_C\, \mathcal C^2}{24 \pi^2 F^2_0} ( ( 2 \bar m +  m_s)  3 f_2(m_\pi,\Delta) +  2 ( 2  m_s +   \bar m) f_2(m_K,\Delta) ) \,\,\,. \nonumber \\
\end{eqnarray}
The counterterm operators in $\mathcal L^{(4)}$ give 
\begin{eqnarray}
\frac{\Delta^{(\textrm{ct})} m_{\Lambda}}{(4 B)^2}  &=& - \frac{1}{3} \left( 
(8 d_3 + 2 d_4 + 4 d_6 + 
      3 (d_7 + d_8)) m_s^2 + 2 (-2 d_4 + 5 d_6 + 6 d_7) \bar m m_s \right. \nonumber \\  & &  
\left.
+ 2 (2 d_3 + d_4 + 2 d_6 + 6 d_7 + 3 d_8) \bar m^2 
\right)\,\,\,.
\end{eqnarray}

\bibliography{su3}

\end{document}